\documentclass[10pt]{article}
\usepackage{fancyhdr}
\usepackage{extramarks}
\usepackage{amsmath}
\usepackage{amsthm}
\usepackage{amsfonts}
\usepackage{siunitx}
\usepackage{tikz}
\usepackage{algpseudocode}
\usepackage{multirow}
\usepackage{booktabs}
\usepackage{palatino}
\usepackage{graphicx}
\usepackage{subfigure}
\usepackage[colorlinks,linkcolor=black,anchorcolor=black,citecolor=black,urlcolor=blue]{hyperref}
\usepackage{amsmath,bm}
\usepackage{booktabs}
\usepackage{mathtools}
\usepackage{amssymb}
\usepackage{caption}
\usepackage{capt-of}
\usepackage{mciteplus}
\usepackage{cite}
\usepackage{mathrsfs}
\usepackage[title,titletoc,toc]{appendix}
\usepackage{xr}
\usepackage{parskip}
\usepackage{textcomp}
\usepackage[colaction]{multicol}
\usepackage[switch]{lineno}
\usepackage{lipsum}
\usepackage{etoolbox}
\usepackage{longtable}
\usepackage{array}
\usepackage{float} 
\usepackage{tablefootnote}
\usepackage[ruled,vlined]{algorithm2e}
\newcolumntype{C}[1]{>{\centering\arraybackslash}p{#1}}
\usetikzlibrary{automata,positioning}
\usetikzlibrary{automata,positioning}

\begin{document}
	
\title{Methodology-centered review of molecular modeling, simulation, and prediction of SARS-CoV-2}

\author{Kaifu Gao$^1$, Rui Wang$^1$\footnote{Kaifu Gao and Rui Wang contributed equally.}, Jiahui Chen$^1$, Limei Cheng$^2$, Jaclyn Frishcosy$^1$,\\ Yuta Huzumi$^1$, Yuchi Qiu$^1$, Tom Schluckbier$^1$, and Guo-Wei Wei$^{1,3,4}$\footnote{
Corresponding author.		Email: wei@math.msu.edu} \\
$^1$ Department of Mathematics, \\
Michigan State University, MI 48824, USA.\\
$^2$ Clinical Pharmacology and Pharmacometrics,\\ 
Bristol Myers Squibb, Princeton, NJ 08536, USA\\
$^3$ Department of Electrical and Computer Engineering,\\
Michigan State University, MI 48824, USA. \\
$^4$ Department of Biochemistry and Molecular Biology,\\
Michigan State University, MI 48824, USA. \\
}

\date{\today} 
	
\maketitle

\begin{abstract}
The deadly coronavirus disease 2019 (COVID-19) pandemic caused by severe acute respiratory syndrome coronavirus 2 (SARS-CoV-2) has gone out of control globally. Despite much effort by scientists, medical experts, healthcare professions, and the society in general, the slow progress on drug discovery and antibody therapeutic development, the unknown possible side effects of the existing vaccines, and the high transmission rate of the SARS-CoV-2, remind us the sad reality that  our current understanding of the transmission, infectivity, and evolution of SARS-CoV-2 is unfortunately very limited. The major limitation is the lack of mechanistic understanding of viral-host cell interactions, the viral regulation of host cell functions and immune systems, protein-protein interactions, including antibody-antigen binding, protein-drug binding, host immune response, etc. This limitation will likely haunt the scientific community for a long time and have a devastating consequence in combating COVID-19 and other pathogens. Notably, compared to the long-cycle, highly cost, and safety-demanding molecular-level experiments, the theoretical and computational studies are economical, speedy and easy to perform. There exists a tsunami of the literature on molecular modeling, simulation, and prediction of SARS-CoV-2 that has become impossible to fully be covered in a review. To provide the reader a quick update about the status of molecular modeling, simulation, and prediction of SARS-CoV-2, we present a comprehensive and systematic methodology-centered narrative in the nick of time. Aspects such as molecular modeling,  Monte Carlo (MC) methods, structural bioinformatics, machine learning, deep learning, and mathematical approaches are included in this review. This review will be beneficial to researchers who are look for ways to contribute to SARS-CoV-2 studies and those who are assessing the current status in the field. 

\end{abstract}


\newpage

{\setcounter{tocdepth}{4} \tableofcontents}

\clearpage \pagebreak \setcounter{page}{1}
\renewcommand{\thepage}{{\arabic{page}}}

\section{Introduction}
Since its first case was identified in Wuhan, China, in December 2019, coronavirus disease 2019 (COVID-19)  caused by severe acute respiratory syndrome coronavirus 2 (SARS-CoV-2) has expeditiously spread to as many as 218 countries and territories worldwide, and led to over 100 million confirmed cases and over 2 million fatalities as of January 20, 2021. This pandemic has also brought a massive economic recession globally.

\begin{figure}[ht]
	\centering
	\includegraphics[width = 1\textwidth]{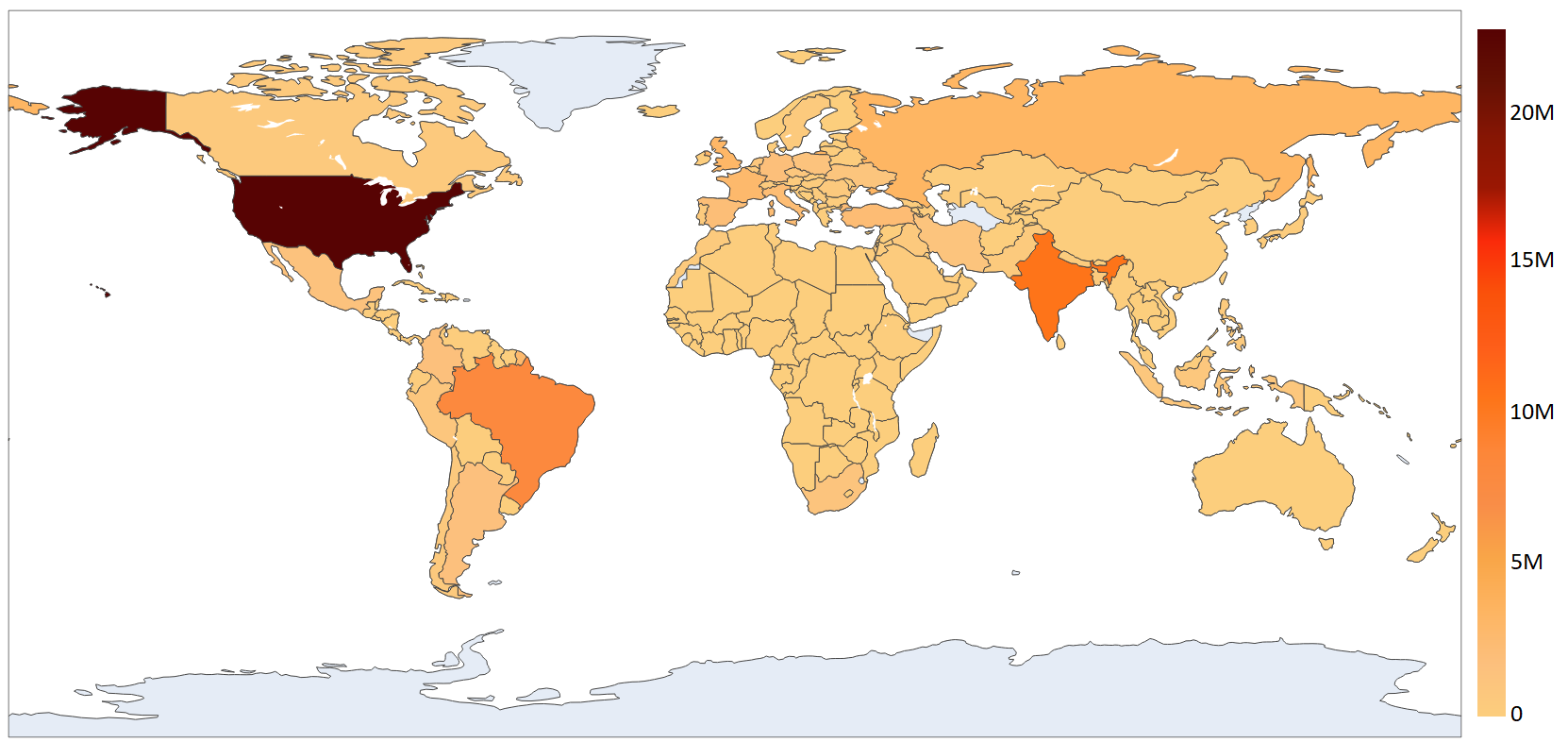}
	\caption{Confirmed cases all around the world until January 20, 2021.}  
	\label{fig:SARS2-dis}
\end{figure}

Although 10 types of SARS-CoV-2 vaccines are already approved or in the status of emergency use worldwide (https://www.nytimes.com/interactive/2020/science/coronavirus-vaccine-tracker.html), effective therapy against the virus has not been discovered yet. Considering SARS-CoV-2's unprecedentedly high infection rate, high prevalence rate, long incubation period \cite{shin2020covid}, asymptomatic transmission \cite{day2020covid,long2020clinical,wang2020decoding2}, potential seasonal pattern \cite{kissler2020projecting}, and emergence of new variants \cite{yin2020genotyping,wang2020mutations,wang2020characterizing},  unveil that the understanding of  virus's molecular mechanism \cite{wang2020host}, tracking genetic evolution \cite{wang2020decoding}, and developing specific antiviral drugs or antibody therapies are still of paramount importance.

\begin{figure}[ht]
	\centering
	\includegraphics[width = 1\textwidth]{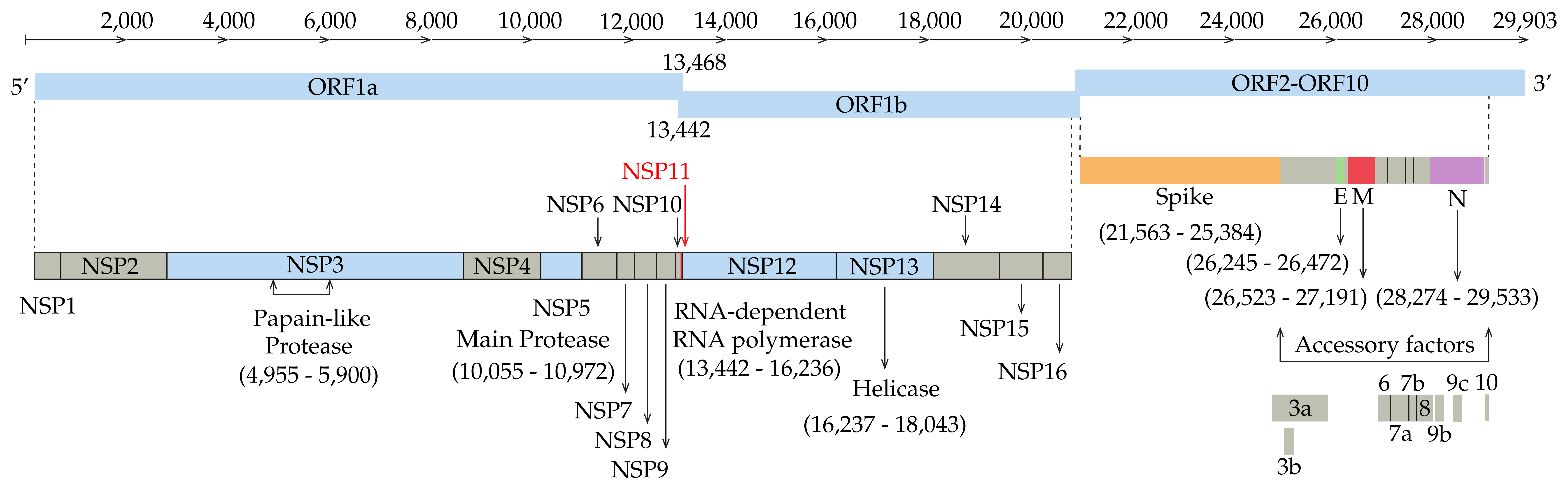}
	\caption{Genomics organization of SARS-CoV-2.  }  
	\label{fig:Whole genome}
\end{figure}

Belonging to the $\beta$-coronavirus genus and coronaviridae family, SARS-CoV-2 is an unsegmented positive-sense single-stranded RNA virus with a compact 29,903 nucleotide-long genome and the diameter of each SARS-CoV-2 virion is about 50-200 nm \cite{chen2020epidemiological}. In the first 20 years of the 21st century, $\beta$-coronaviruses have triggered three major outbreaks of deadly pneumonia: SARS-CoV (2002), Middle East respiratory syndrome coronavirus (MERS-CoV) (2012), and SARS-CoV-2 (2019) \cite{lu2020genomic}. Like SARS-CoV and MERS-CoV, SARS-CoV-2 also causes respiratory infections, but at a much higher infection rate \cite{walls2020structure,wrapp2020cryo}. The complete genome of SARS-CoV-2 comprises 15 open reading frames (ORFs), which encode 29 structural and non-structural proteins, as illustrated in \autoref{fig:Whole genome}. The 16 non-structural proteins NSP1-NSP16 get expressed by protein-coding genes ORF1a and ORF1b, while four canonical 3' structural proteins: spike (S), envelope (E), membrane (M), and nucleocapsid (N) proteins, as well as accessory factors, are encoded by other four major ORFs, namely ORF2, ORF4, ORF5, and ORF9 (See Figure \ref{fig:Whole genome}) \cite{michel2020characterization,helmy2020covid,naqvi2020insights,mu2020sars}.

\begin{figure}[ht!]
	\centering
	\includegraphics[width = 1\textwidth]{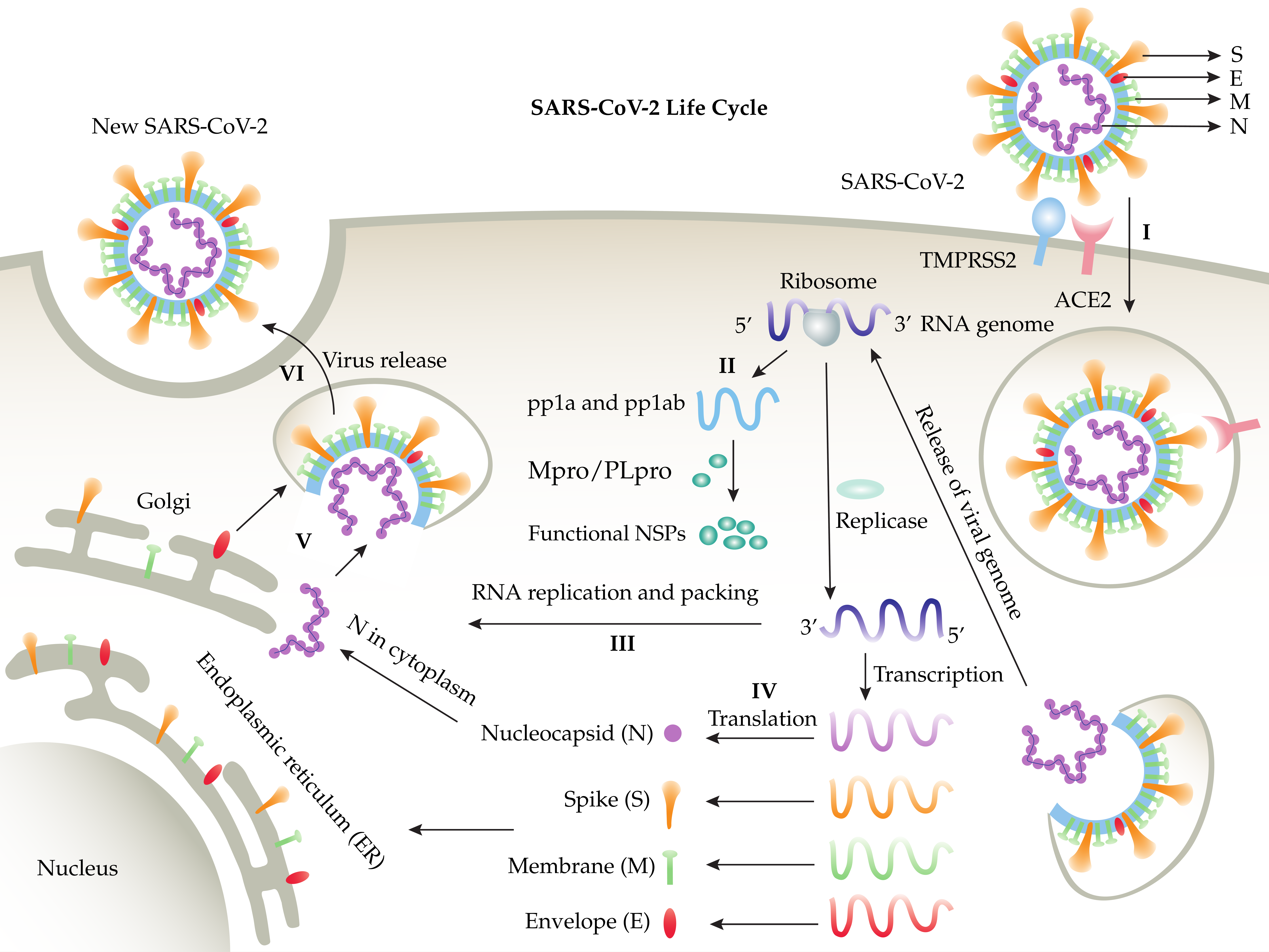}
	\caption{Six stages of the SARS-CoV-2 life cycle. Stage {\bf I}: Virus entry. Stage {\bf II}: Translation of viral replication. Stage {\bf III}: Replication. Here, NSP12 (RdRp) and NSP13 (helicase) cooperate to perform the replication of the viral genome. Stage {\bf IV}: Translation of viral structure proteins. Stage {\bf V}: Virion assembly. Stage {\bf VI}: Release of virus. }  
	\label{fig:Life cycle}
\end{figure}

\begin{figure}[ht!]
	\includegraphics[width=1.0\textwidth]{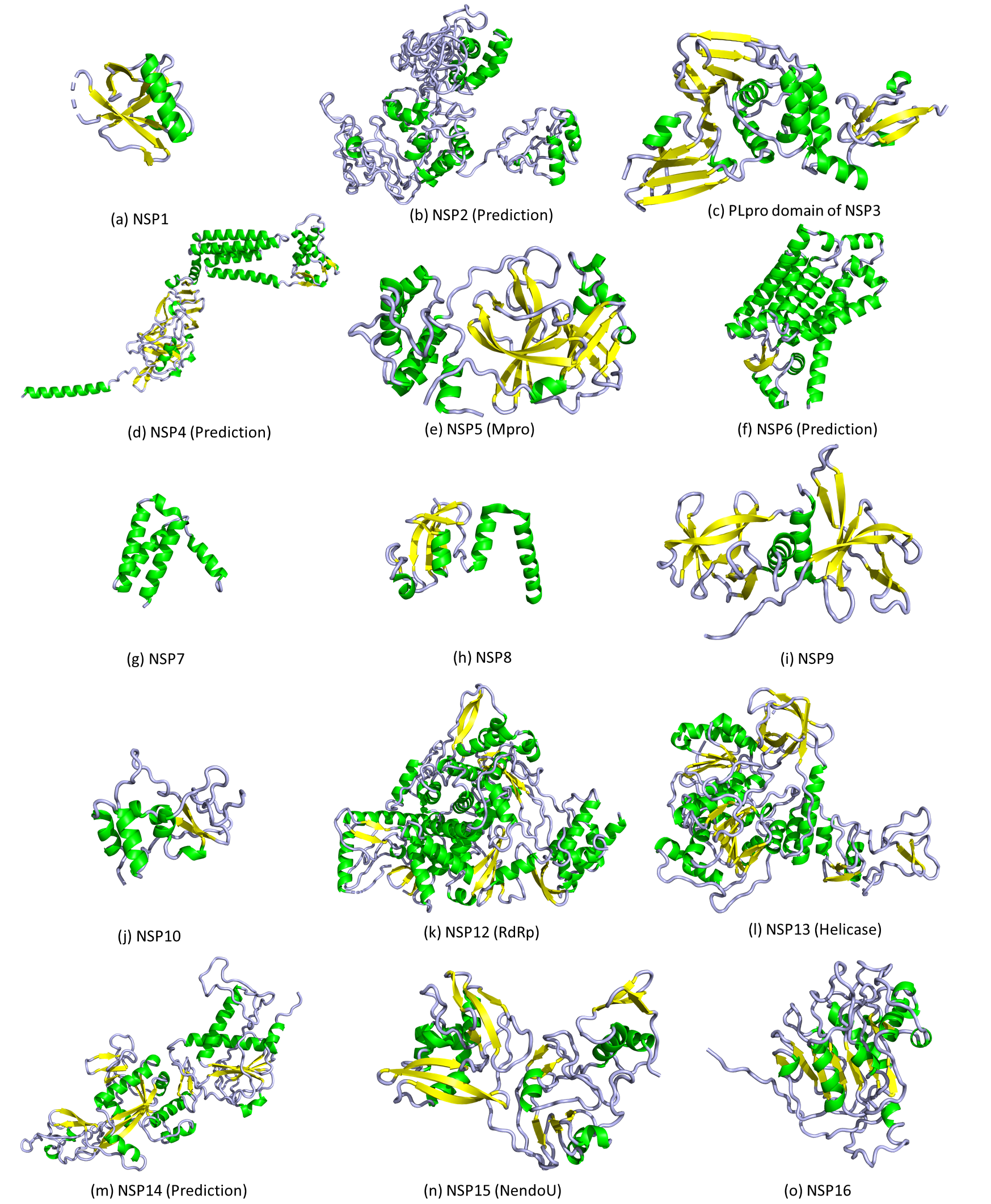}
	\caption{3D conformations of the SARS-CoV-2 nonstructural proteins.}
	\label{fig:sars2-nsp-proteins}
\end{figure}

\begin{figure}[ht!]
	\includegraphics[width=1.0\textwidth]{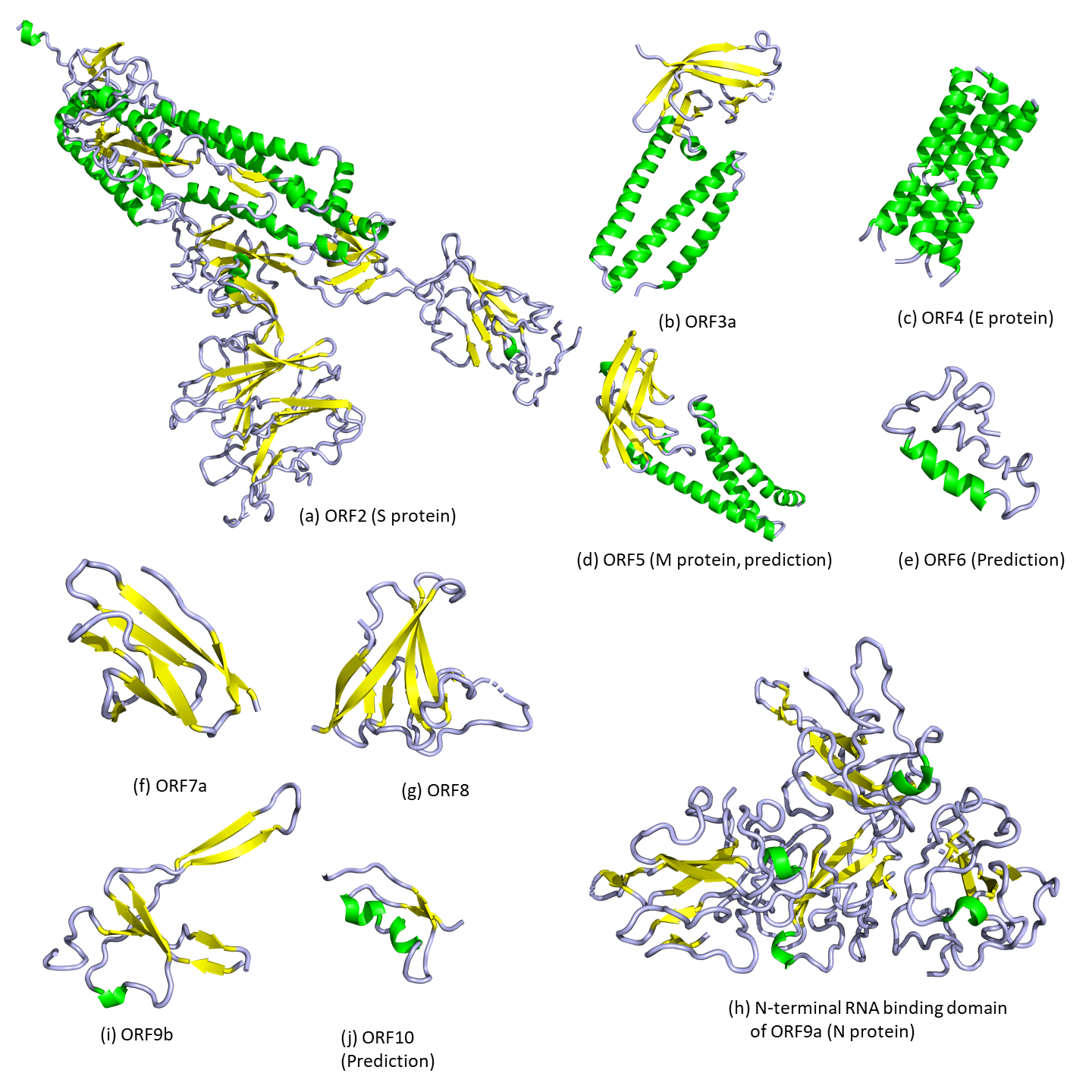}
	\caption{3D conformations of the ORF2-ORF10 proteins.}
	\label{fig:sars2-orf-proteins}
\end{figure}

The viral structure of SARS-CoV-2 can be found in the upper right corner of Figure \ref{fig:Life cycle}. This structure is formed by the four structural proteins: the N protein holds the RNA genome, and the S, E, and M proteins together construct the viral envelope \cite{wu2020analysis}. The studies on SARS-CoV-2 as well as previous SARS-CoV and other coronaviruses have mostly identified the functions of these structural proteins, nonstructural proteins as well as accessory proteins, which are summarized in Table \ref{table:sars2-proteins}; their 3D structures are also largely known from experiments or predictions, which are shown in Figures \ref{fig:sars2-nsp-proteins} and \ref{fig:sars2-orf-proteins}.

\begin{center}
	\begin{longtable}{|p{2.5cm } |p{11cm}|p{2.5cm}|}
		\hline
		Protein & Functions & 3D structure availability \\ \hline		
		NSP1 & NSP1 (180 residues) likely inhibits host translation by interacting with the 40S ribosomal subunit. Its C terminus binds to and obstructs the ribosomal mRNA entry tunnel, thereby inhibiting antiviral response triggered by innate immunity or interferons. The NSP1-40S ribosome complex further induces an endonucleolytic cleavage near the 5'UTR of host mRNAs, targeting them for degradation. By suppressing host gene expression, NSP1 facilitates efficient viral gene expression in infected cells and evasion from host immune response \cite{jauregui2013identification}. & Experiment (PDB ID: 7k3n, etc.) \\ \hline
		NSP2 & NSP2 (638 residues) may play a role in the modulation of host cell survival signaling pathway by interacting with the host factors, prohibitin 1 and prohibitin 2, which are involved in maintaining the functional integrity of the mitochondria and protecting cells from various stresses. It appears that NSP2 could change the intracellular milieu and perturb host intracellular signaling \cite{cornillez2009severe}. & Prediction \cite{zhang_predict} \\ \hline
		NSP3 (Contains PLpro) & NSP3 (1945 residues) includes the papain-like protease (PLpro) and some multi-pass membrane proteins. PLpro is responsible for cleaving and releasing NSP1, NSP2, and NSP3; PLpro also possesses a deubiquitinating/deISGylating activity and processes both ``Lys-48''- and ``Lys-63''-linked polyubiquitin chains from cellular substrates. It cleaves preferentially ISG15 from substrates in vitro, which can play a role in host ADP-ribosylation by binding ADP-ribose. In addition, NSP3 participates together with NSP4 in the assembly of virally-induced cytoplasmic double-membrane vesicles necessary for viral replication, and antagonizes innate immune induction of type I interferon by blocking the phosphorylation, dimerization and subsequent nuclear translocation of host IRF3; it also prevents host NF-kappa-B signaling \cite{baez2015sars}. & Partially available from experiments: Residues 1570-1877 (PLpro, PDB ID: 7kol, etc.); Residues 819-929 (PDB ID: 7kag, etc.); Residues 1024-1192 (PDB ID: 6wcf \cite{michalska2020crystal}, etc.)\\ \hline		
		NSP4 & NSP4 (500 residues) is a multi-pass membrane protein. Together with NSP3, it participates in the assembly of virally-induced cytoplasmic double-membrane vesicles, which is necessary for viral replication. \cite{sakai2017two}. & Prediction \cite{deepmind_predict} \\ \hline
		NSP5 (Mpro) & NSP5 (306 residues) is the main protease (3CL protease) of the SARS-CoV-2. It takes charge of cleaving and releasing NSP4-NSP16. Additionally, it recognizes substrates containing the core sequence [ILMVF]-Q-|-[SGACN] and is also able to bind an ADP-ribose-1"-phosphate (ADRP). Moreover, it plays a role in NSP maturation \cite{wu2020analysis}. & Experiment (PDB ID: 5r84 \cite{alic2020mpro}, etc.)\\ \hline
		NSP6 & NSP6 (290 residues) is a multi-pass membrane protein, working with NSP3 and NSP4, it induces double-membrane vesicles (autophagosomes) in infected cells from their reticulum endoplasmic. It also limits the expansion of these autophagosomes that are no longer able to deliver viral components to lysosomes \cite{angelini2013severe,cottam2014coronavirus}. & Prediction \cite{deepmind_predict} \\ \hline
		NSP7 & NSP7 (83 residues) plays a role in viral RNA synthesis. It forms a hexadecamer with NSP8 that may participate in viral replication by acting as a primase. Alternatively, may synthesize substantially longer products than oligonucleotide primers \cite{kirchdoerfer2019structure}. & Experiment (PDB ID: 6m5i, etc.) \\ \hline
		NSP8 & NSP8 (198 residues) plays a role in viral RNA synthesis. It forms a hexadecamer with NSP7 that may participate in viral replication by acting as a primase. Alternatively, it may synthesize substantially longer products than oligonucleotide primers \cite{kirchdoerfer2019structure, snijder2016nonstructural}. & Experiment (PDB ID: 6m5i, etc.)         \\ \hline
		NSP9 & NSP9 (113 residues) functions in viral replication as a dimeric ssRNA-binding protein.  \cite{snijder2016nonstructural} & Experiment (PDB ID: 6w4b, etc.)\\ \hline		
		NSP10 & NSP10 (139 residues) plays a pivotal role in viral transcription. It forms a dodecamer and interacts with both NSP14 and NSP16 to stimulate their respective 3'-5' exoribonuclease and 2'-O-methyltransferase activities in viral mRNAs cap methylation \cite{snijder2016nonstructural}. & Experiment (PDB ID: 6zct \cite{rogstam2020crystal}, etc.)      \\ \hline		
		NSP11 & NSP11 (13 residues) is a pp1a cleavage product at the NSP10/11 boundary. For pp1ab, it is a frameshift product that becomes the N-terminal of NSP12. Its function, if any, is currently unknown \cite{snijder2016nonstructural}. & No  \\ \hline		
		NSP12 (RdRp) & NSP12 (932 residues) is the RNA-dependent RNA polymerase (RdRp) performing both replication and transcription of the viral genome. Specifically, it catalyzes the synthesis of the RNA strand complementary to a given RNA template. The RdRp of SARS-CoV-2 can be inhibited by the nucleoside analogue Remdesivir \cite{snijder2016nonstructural}. & Experiment (PDB ID: 6m71\cite{gao2020structure}, etc.)                  \\ \hline		
		NSP13 (Helicase) & NSP13 (601 residues)  is a multifunctional superfamily 1 helicase capable of using both dsDNA and dsRNA as substrates with 5’-3’ polarity. In addition to working with NSP12 in viral genome replication, it is also involved in viral mRNA capping. It associates with nucleoprotein in membranous complexes \cite{jang2020high}. & Experiment (PDB ID: 5rlh\cite{gao2020structure}, etc.)                    \\ \hline
		NSP14 & NSP14 (527 residues) possesses two different activities: (1) An exoribonuclease activity on both ssRNA and dsRNA in a 3' to 5' direction; (2) A N7-guanine methyltransferase (viral mRNA capping) activity. It acts as a proofreading exoribonuclease for RNA replication, thereby lowering the sensitivity of the virus to RNA mutagens \cite{ferron2018structural}. It always interacts with NSP10 \cite{snijder2016nonstructural}. & Prediction \cite{zhang_predict}       \\ \hline
		NSP15 (NendoU) & NSP15 (346 residues) is the nidoviral RNA uridylate‐specific endoribonuclease (NendoU) that favors the cleavage of RNA at the 3’-ends of uridylates, loss of NSP15 affects both viral replication and pathogenesis. It is also required for the evasion of host cell dsRNA sensors \cite{deng2017coronavirus}.   & Experiment (PDB ID: 5s72, etc.)                   \\ \hline
		NSP16 & NSP16 (298 residues) is activated by and interacts with NSP10. Its 2’-O-methyltransferase activity mediates mRNA cap 2'-O-ribose methylation to the 5'-cap structure of viral mRNAs. Since N7-methyl guanosine cap is a prerequisite for binding of NSP16, it plays an essential role in viral mRNAs cap methylation which is essential to evade the immune system. It may also work against host cell antiviral sensors \cite{snijder2016nonstructural}.   & Experiment (PDB ID: 6w4h \cite{rosas2020crystal}, etc.)                  \\ \hline		
		ORF2 (Spike (S) protein) & The S protein (1273 residues) may down-regulate host tetherin (BST2) by lysosomal degradation, thereby counteracting its antiviral activity. It can be cleaved into two subunits, S1 and S2.  S1 attaches the virion to the cell membrane by interacting with the host receptor, initiating the infection. Binding to human ACE2 receptor and internalization of the virus into the endosomes of the host cell induces conformational changes in the S protein. The stalk domain of S contains three hinges, giving the head unexpected orientational freedom. The S protein uses human TMPRSS2 for priming in human lung cells, which is an essential step for viral entry. S2 mediates fusion of the virion and cellular membranes by acting as a class I viral fusion protein. Under the current model, the protein has at least three conformational states: pre-fusion native state, pre-hairpin intermediate state, and post-fusion hairpin state. During viral and target cell membrane fusion, the coiled coil regions (heptad repeats) assume a trimer-of-hairpins structure, positioning the fusion peptide in close proximity to the C-terminal region of the ectodomain. The formation of this structure appears to drive apposition and subsequent fusion of viral and target cell membranes.\cite{hoffmann2020sars}.            & Experiment (PDB ID: 7c2l \cite{chi2020neutralizing}, etc.)         \\ \hline
		ORF3a & ORF3a (275 residues) is a multi-pass membrane protein that forms homotetrameric potassium sensitive ion channels (viroporin). It upregulates expression of fibrinogen subunits FGA, FGB, and FGG in host lung epithelial cells, induces apoptosis in cell culture, and downregulates the type 1 interferon receptor by inducing serine phosphorylation within the IFN alpha-receptor subunit 1 (IFNAR1) degradation motif and increasing IFNAR1 ubiquitination. More importantly, it activates both NF-kB and NLRP3 inflammasome and contributes to the generation of the cytokine storm. it may also modulate viral release \cite{siu2019severe}.   & Experiment (PDB ID: 6xdc \cite{kern2020cryo})                  \\
		\hline
		ORF3b & Along with nucleocapsid protein and ORF6, ORF3b (22 residues) appears to block induction of IFN-I. This 22-residue variant is also present in SARS-CoV-2-related viral genomes in bats and pangolins \cite{alam2020functional}.   &  No             \\ \hline
		ORF4 (Envelope (E) protein) & The E protein (75 residues) is a single-pass type III membrane protein playing a central role in virus morphogenesis and assembly, it acts as a viroporin and self-assembles in host membranes forming pentameric protein-lipid pores that allow ion transport. It is also involved in the induction of apoptosis. \cite{schoeman2019coronavirus}. & Partially available from experiment: Residues 8-38 (PDB ID:7k3g \cite{mandala2020structure})                   \\ \hline
		ORF5 (Membrane (M) protein) & The M protein (222 residues) is the most abundant structural component of the virion, and very conserved. It mediates morphogenesis, assembly, and budding of viral particles through the recruitment of other structural proteins to the ER-Golgi-intermediate compartment (ERGIC). It also interacts with N for RNA packaging into virion \cite{voss2009studies}.  &  Prediction \cite{deepmind_predict}                   \\ \hline		
		ORF6 & ORF6 (61 residues) appears to be a virulence factor. It disrupts cell nuclear import complex formation by tethering karyopherin alpha 2 and karyopherin beta 1 to the membrane. Retention of import factors at the ER/Golgi membrane leads to a loss of transport into the nucleus, thereby preventing STAT1 nuclear translocation in response to interferon signaling and thus blocking the expression of interferon stimulated genes (ISGs) that display multiple antiviral activities. \cite{huang2017phage}.   & Prediction \cite{zhang_predict}       \\ \hline
		ORF7a & ORF7a (121 residues) is a type I membrane protein that plays a role as an antagonist of bone marrow stromal antigen 2 (BST-2), disrupting its antiviral effect. As BST-2 tethers virions to the host’s plasma membrane, ORF7a binding inhibits BST-2 glycosylation and interferes with this restriction activity. ORF7a may suppress small interfering RNA (siRNA) and also may bind to host ITGAL, thereby playing a role in attachment or modulation of leukocytes. \cite{taylor2015severe}.   & Partially available from experiment: Residues 16-82 (PDB ID:6w37)        \\ \hline
		ORF7b & ORF7b (43 residues) is a type III integral transmembrane protein in the Golgi apparatus. In SARS-CoV-2, it appears to be a viral attenuation factor. It may be involved in human infectivity of SARS-CoV-2 \cite{pfefferle2009reverse}.  & No  \\ \hline
		ORF8 & ORF8 (121 residues) might be a luminal ER membrane-associated protein. It may trigger ATF6 activation and affect the unfolded protein response (UPR). Like ORF7b, it may be involved in human infectivity of SARS-CoV-2 \cite{pfefferle2009reverse,muth2018attenuation,sung20098ab}.  & Experiment (PDB ID:7jtl \cite{flower2020structure}, etc.)      \\ \hline
		ORF9a (Nucleocapsid (N) protein) & The N protein (419 residues) packages the positive strand viral genome RNA into a helical ribonucleocapsid (RNP) and plays a fundamental role during virion assembly through its interactions with the viral genome and membrane protein M. It also plays an important role in enhancing the efficiency of subgenomic viral RNA transcription as well as viral replication. It may modulate transforming growth factor-beta signaling by binding host smad3 \cite{mu2020sars}. & Partially available from experiments: Residues 41-174 (PDB ID:6m3m\cite{kang2020crystal}, etc.); Residues 247-364 (PDB ID:6zco\cite{zinzula2020high}, etc.);      \\ \hline
		ORF9b & ORF9b (97 residues) plays a role in the inhibition of the host's innate immune response by targeting the mitochondrial-associated adapter MAVS. Mechanistically, it usurps the E3 ligase ITCH to trigger the degradation of MAVS, TRAF3, and TRAF6. In addition, it can cause mitochondrial elongation by triggering ubiquitination and proteasomal degradation of dynamin-like protein 1/DNM1L\cite{shi2014sars}.  & Experiment (PDB ID:6z4u)                   \\ \hline
		ORF9c & ORF9c (70 residues), located in the N coding region, interacts with various host proteins including Sigma receptors, implying involvement in lipid remodeling and the ER stress response. It might also target NF-kB signaling \cite{gordon2020sars}.  &  No       \\ \hline
		ORF10 & ORF10 (38 residues) interacts with factors in the CUL2 RING E3 ligase complex and thus may modulate ubiquitination \cite{gordon2020sars}.  & Prediction \cite{zhang_predict}       \\ \hline		
	\end{longtable}
	\label{table:sars2-proteins}
\end{center}


With these SARS-CoV-2 proteins, the intracellular viral life cycle of SARS-CoV-2 can be realized \cite{kim2020architecture}. This life cycle has six stages as shown in  \autoref{fig:Life cycle}. The first stage is the entry of the virus. SARS-CoV-2 enters the host cell either via endosomes or plasma membrane fusion. In both ways, the S protein of SARS-CoV-2 first attaches to the host cell-surface protein, angiotensin-converting enzyme 2 (ACE2). Then, the cell's protease, transmembrane protease serine 2 (TMPRSS2), cuts and opens the S protein of the virus, exposing a fusion peptide in the S2 subunit of the S protein \cite{matsuyama2020enhanced}. After fusion, an endosome forms around the virion, separating it from the rest of the host cell. The virion escapes when the pH of the endosome drops or when cathepsin, a host cysteine protease, cleaves it. The virion then releases its RNA into the cell \cite{hoffmann2020sars}. After the RNA releasing, the polyproteins pp1a and pp1ab are translated. Notably, facilitated by viral Papain-like protease (PLpro), NSP1, NSP2, NSP3, and the amino terminus of NSP4 from the pp1a and pp1ab are released. Moreover, NSP5-NSP16 are also cleaved proteolytically by the main protease (Mpro) \cite{v2020coronavirus}. The next stage of the life cycle is the replication process, where the NSP12 (RdRp) and NSP13 (helicase) cooperate to perform the replication of the viral genome. Stages IV and V are the translation of viral structural proteins and the virion assembly process. In these stages, the structural proteins S, E, and M are translated by ribosomes and then present on the surface of the endoplasmic reticulum (ER), which will be transported from the ER through the golgi apparatus for the preparation of virion assembly. Meanwhile, multiple copies of N protein package the genomics RNA in cytoplasm, which interacts with the other 3 structural proteins to direct the assembly of the virions. Finally, virions will be secreted from the infected cell through exocytosis. 

Since the first outbreak of the COVID-19, the raging pandemic caused by SARS-CoV-2 has lasted over a year. Despite much effort by scientists, medical professions, and the society in general, there is still no effective control, prevention, and cure for the deadly disease at present. The discovery of small-molecular drugs and antibody therapies has been progressing slowly. We do have many promising vaccines, but they might have side effects and their full side effects, particularly, long-term side effects, remain unknown. To make things worse, near 30,000 unique mutations have been recorded for SARS-CoV-2 as shown by Mutation Tracker (\url{https://users.math.msu.edu/users/weig/SARS-CoV-2_Mutation_Tracker.html}).  All of these reveal a sad reality that our current understanding of life science, virology, epidemiology, and medicine is severely limited. This limitation hinders our understanding of the transmission, infectivity, and evolution of SARS-CoV-2. Ultimately, the root of the challenge is the lack of the molecular mechanistic understanding of coronavirus RNA proofreading, virus-host cell interactions, antibody-antigen interactions, protein-protein interactions, protein-drug interactions, viral regulation of host cell functions, including autophagocytosis and apoptosis, and irregular host immune response behavior such as cytokine storm and antibody-dependent enhancement. Molecular-level experiments on SARS-CoV-2 are expensive and time-consuming, and require to take heavy safety measures. However, the advances in computer power, the accumulation of molecular data, the availability of artificial intelligence (AI) algorithms, and the development of new mathematical tools have paved the road for mechanistic understanding from molecular modeling, simulation, and prediction. A gigantic literature for molecular modeling, simulation, and prediction of SARS-CoV-2 has been published or available online. It has become impossible for senior exports, not to mention junior researchers, to go through this literature at present. Therefore, it is time to present a methodology-centered review so that a reader can still grasp the current status of SARS-CoV-2 modeling, simulation, and prediction without having to read the entire literature.  In this review, we cover the studies of molecular modeling, Monte Carlo (MC) methods,   structural bioinformatics, machine learning \& deep learning, and mathematical approaches,  in the combating of COVID-19. Comments are given in the discussion section, while future perspectives are presented in the concluding remarks.

\section{Methods and Approaches}

\subsection{Molecular modeling}

\subsubsection{Molecular docking}

\begin{figure}[ht!]
	\includegraphics[width=1.0\textwidth]{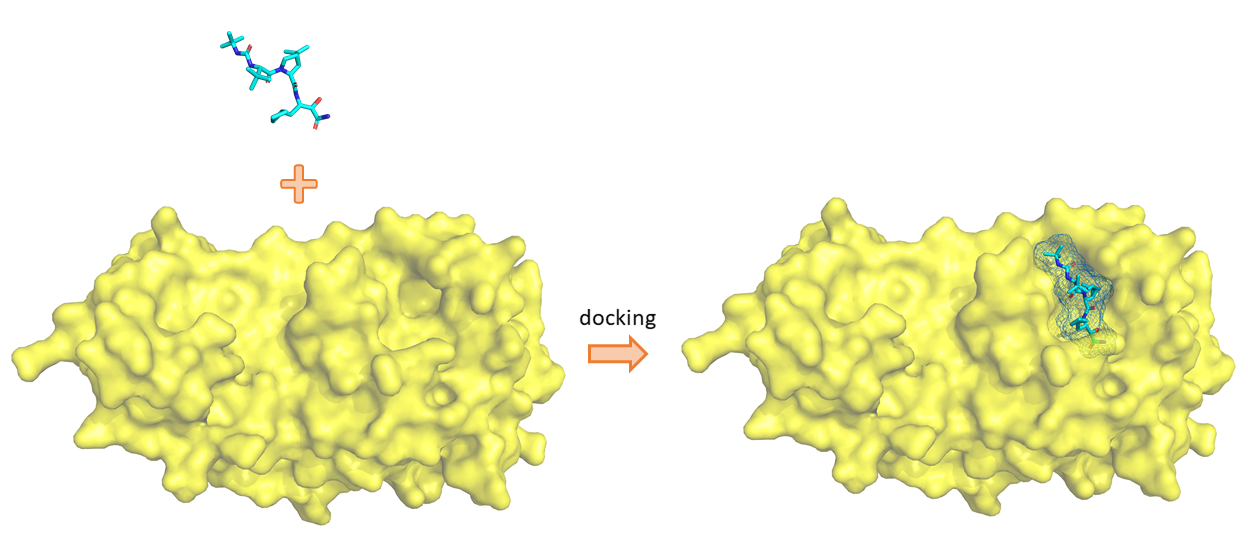}
	\caption{The molecular docking procedure targeting the SARS-CoV-2 main protease.}
	\label{fig:docking}
\end{figure}

In the field of molecular modeling, docking is a method to predict the preferred orientation of one molecule to a second when bound to each other to form a stable complex as shown in Figure \ref{fig:docking}. Since binding behavior plays a vital role in the rational design of drugs, molecular docking, which can provide binding conformations of ligands to specific binding sites, is one of the most popular methods in structure-based drug design\cite{thomsen2006moldock,kitchen2004docking}. A docking program includes two key components: a scoring function to evaluate the energies of different conformations and a search algorithm to sample the conformational degrees of freedom and locate the global energy minimum from all the sampled conformations \cite{brooijmans2003molecular}. In addition to regular docking, ensemble docking \cite{amaro2018ensemble} docks a ligand to an ensemble of receptor conformations (often generated by molecular dynamics simulation) and picks up the optimal binding pose. Molecular docking is well-established in early-stage drug discovery and is, as a result, widely applied to many SARS-CoV-2 proteins. 

\paragraph{Targeting the SARS-CoV-2 main protease.} 
One important source for SARS-CoV-2 treatment is existing drugs. Chen et al. \cite{chen2020docking_prediction} implemented a quite extensive drug-repurposing work: they docked and predicted the binding affinities of 7173 purchasable drugs, 4574  unique compounds  as well as  their stereoisomers to the main protease. As a result, disosmin,  hesperidin, and MK-3207 with an affinity of -10.1 kcal/mol, were suggested as the most potent inhibitors. Sencanski et al. \cite{sencanski2020drug} and Gurung et al. \cite{gurung2020silico} both screened about 1400 FDA (the United States Food and Drug Administration)-approved drugs through docking, predicting that dihydroergotamine has a promising affinity (-9.4 kcal/mol). Eleftheriou et al. \cite{eleftheriou2020silico} used docking to evaluate the potency of around 100 approved protease inhibitors and suggested that  faldaprevir has the strongest binding affinity of -11.15 kcal/mol. Many others  \cite{joshi2020molecular,li2020prioritization,mamidala2020silico,verma2020potential,marinho2020virtual,gandhi2020ayurvedic,anandamurthy2020anti,wafa2020molecular,usman2020molecular,hadni2020silico,das2020investigation,hosseini2020simeprevir,kumar2020silico_2, shamsi2020glecaprevir} have also docked and repurposed existing drugs against the SARS-CoV-2 main protease. 

Another type of potential inhibitors is natural products. Chandel et al. \cite{kumar2020silico_1} screened 1000 active phytochemicals from Indian medicinal plants by molecular docking. Among them, rhein and aswagandhanolide were predicted to have binding affinities over -8.0 kcal/mol. Mazzini et al.\cite{mazzini2020putative} docked more than 100 natural and nature-inspired products from an in-house library to the main protease, predicting leopolic acid A to have with the highest affinity, -12.22 kcal/mol. Vijayaraj et al. \cite{vijayaraj2020bioactive} studied some bioactive compounds from marine resources, suggesting that an ethyl ester (-8.42 kcal/mol) from Marine Sponges Axinella cf. corrugata is the most potent against SARS-CoV-2 among them. Moreover, a variety of natural products from aloe vera, Moroccan medicinal plants, fungal metabolite, millet, tannins, neem leaves, nigella sativa, etc., \cite{mishra2020antiviral,mpiana2020identification,aanouz2020moroccan,mishra2020millet,enmozhi2020andrographolide,singh2020identification,sampangi2020molecular,rehman2020natural,khalifa2020tannins,ojo2020novel,abd2020inhibition,das2020naturally,khaerunnisa2020potential,taofeek2020molecular,subramanian2020some, shaghaghi2020molecular,bouchentouf2020identification,gonzalez2020theoretical,santosidentification} were also investigated by docking-based virtual screening.

Other works focus on small compounds from other sources. Tsuji et al. \cite{tsuji2020potential} virtually screened the compounds in ChEMBL database \cite{gaulton2012chembl} against the main protease, suggesting that the compound CHEMBL1559003 with a binding affinity of -10.6 kcal/mol as the most potent. Udrea et al. \cite{udrea2020laser} predicted the binding affinities of 15 phenothiazines, reporting the compound SPZ (-10.1 kcal/mol) as the most effective. Ghaleb et al. \cite{ghaleb2020silico} also studied some pyridine N-oxide compounds, indicating the most potent one with a predicted pIC50 value of 5.294 (about -7.22 kcal/mol). More similar works are described in  Refs. \cite{kumar2020development, utami2020molecular, cherrakpotential, agrawal2020molecular, owis2020molecular, mohamedmolecular, chhetri2020exploration, liu2020potent, garza2020copper, ahmad2020molecular_1, bhaliya2020identification, bhaliya2020silico}.

\paragraph{Targeting the S protein.} Many researchers  have analyzed the binding interactions between the S protein and human ACE2. For example, Ortega et al. \cite{ortega2020role} used docking calculations to compare the binding affinities of the SARS-CoV-2 S protein and the SARS-CoV S protein to ACE2. Their results suggested more residue interactions are between the SARS-CoV-2 S protein and ACE2, leading to a higher binding affinity, which is consistent with recent experimental research \cite{wrapp2020structural}. 

More investigations were about drug or compound repurposing. To repurpose existing drugs against the SARS-CoV-2 S protein, Miroshnychenko et al.\cite{miroshnychenko2020combined} systematically docked 248 drugs to the S protein and found that amentoflavone and ledipasvir (-8.5 kcal/mol and -8.4 kcal/mol) were the top two inhibitors. Since their binding domains were predicted to be different, the combined use of them to treat SARS-CoV-2 was suggested by the authors. The potency of lopinavir and ubrogepant against the S protein was also investigated \cite{shaikh2020lopinavir, omotuyi2020disruption}. 

For natural products, Subbaiyan et al. \cite{subbaiyan2020silico} performed virtual screening on 12 compounds from indigenous food additives and herbal constituents. Among them, epigallocatechin gallate was predicted as the top compound with a binding affinity of -9.2 kcal/mol. Basu et al. \cite{basu2020computational} revealed that hesperidin had the highest binding affinity of -8.99 kcal/mol among the 6 phytochemicals from Indian medicinal plants. Other natural products from tea flavonoids and triterpenoid were also studied against the S protein \cite{maiti2020epigallocatechin,jomhori2020investigation,goswami2020molecular}. 

Inhibitors from other sources were repurposed to inhibit the S protein as well. For example, Fakih et al. \cite{fakih2020dermaseptin} performed docking on dermaseptin-based antiviral peptides. Zegheb et al. \cite{zegheb2020n} evaluated N-ferrocenylmethyl derivatives against the S-protein. Abo-zeid et al. \cite{abo2020molecular} virtually screened some FDA-approved iron oxide nanoparticles to target the S-protein receptor-binding domain (RBD).

\paragraph{Targeting the RdRp.} 
Through docking, Ahmad et al. \cite{ahmad2020sars} systematically assessed 7922 approved or experimental drugs against SARS-CoV-2 RdRp and suggested that Nacartocin has the highest binding affinity of -13.943 kcal/mol. Beg et al. \cite{beg2020anti} screened 70 anti-HIV (human immunodeficiency virus) or anti-HCV (hepatitis C virus) drugs, reporting the top drug paritaprevir with a -10 kcal/mol binding affinity.  Aftab et al. \cite{aftab2020analysis} studied 10 antiviral drugs and revealed Remdesivir's docking score was the highest (-14.06 kcal/mol). Other RdRp drug-repurposing works include Refs. \cite{parvez2020prediction, elfiky2020ribavirin, heydargoy2020investigation, mukhoty2020silico}.

Lung et al. \cite{lung2020potential} retrieved 83 traditional Chinese medicinal compounds as well as their similar structures from the ZINC15 database, evaluated their potency against the RdRp by docking, and reported the binding affinity of theaflavin (-9.11 kcal/mol) as the highest among them. Singh et al. \cite{singh2020potential} virtually screened over 100 phytochemical inhibitors and predicted that withanolide E, with a binding affinity of -9 kcal/mol, was the most potent. Pandeya et al. \cite{pandeya2020natural} also investigated some biologically active alkaloids of argemone mexicana.

\paragraph{Other targets.} 
Through docking, Mohideen et al.\cite{mohideenmolecular} found that the binding affinity of the  natural product thymoquinone to the E protein was -9.01 kcal/mol. Borgio et al. \cite{borgio2020state} screened 23 FDA-approved drugs to target the helicase of SARS-CoV-2 and reported vapreotide with a binding affinity of -11.58 kcal/mol as the most potent.

\paragraph{Targeting multiple proteins.} 
Maurya et al. \cite{maurya2020evaluation} assessed phytochemicals and active pharmacological agents present in Indian herbs against 7 different proteins of SARS-CoV-2 (The main protease, N protein, NendoU, NSP3, NSP9, and S protein) and human proprotein convertase (furin) through docking. Deshpande et al.\cite{deshpande2020silico} also docked 11 antiviral drugs to the main protease, S-protein, PLpro, NSP10, NSP16, and NSP9, and calculated their binding affinities. Nimgampalle et al. \cite{nimgampalle2020screening} virtually screened chloroquine, hydroxychloroquine, and their derivatives against multiple SARS-CoV-2 protein drug targets: the main protease, RdRp, S-protein, ADP-ribose-1 monophosphatase (in NSP3), and NSP9. Da Silva et al.\cite{da2020essential} studied the potency of essential oil components against the main protease, PLpro, NendoU, ADP-ribose-1 phosphatase, RdRp, S protein, and human ACE2. Notably, there is a controversy about using human ACE2 as a target (See Section \ref{ace2_targeting}). Some limonoids and triterpenoids were evaluated by Vardhan et al. \cite{vardhan2020silico} against the main protease, PLpro, S protein, RdRp, and human ACE2. Laksmiani et al. \cite{laksmiani2020active} performed docking on some medicinal plants against the main protease, PLpro, RdRp, human cellular transmembrane protease serine 2 (TMPRSS2), and ACE2. Khan et al. \cite{khan2020identification_2} tested some dietary molecules by docking against the main protease, S protein, HR2 domain and post fusion core S2 subunit of the S protein, and NendoU. Thurakkal et al. \cite{thurakkal2020silico} docked organosulfur compounds against the main protease, PLpro, S-protein, RdRp, and helicase. Yu et al. \cite{yu2020computational}, targeting the main protease, PLpro, RdRp, and S protein, evaluated the potency of five FDA-approved drugs and some Chinese traditional drugs.  

Vijayakumar et al. \cite{vijayakumar2020silico} assessed the potency of natural flavonoids and synthetic indole chalcones against the main protease, S protein, and RdRp. On the same set of targets,  Parvez et al. \cite{parvez2020virtual} studied some plant metabolites, Maurya et al. \cite{maurya2020evaluation_1} investigated yashtimadhu (glycyrrhiza glabra) active phytochemicals, and Alexpandi et al. \cite{alexpandi2020quinolines} simulated quinoline-based inhibitors. Targeting the main protease, RdRp, and PLpro, Hosseini et al. \cite{hosseini2020computational} calculated some drug candidates; Chowdhury et al. \cite{chowdhury2020silico_1} reported that an arsenic-based approved drug darinaparsin had docking binding affinities over -7.0 kcal/mol. Additionally, Iftikhar et al. \cite{iftikhar2020identification} reported results targeting the main protease, RdRp, and helicase.

More works studied the potency of compounds inhibiting two different proteins. Elmezayen et al. \cite{elmezayen2020drug} virtually screened 4500 approved or experimental drugs against the main protease and human TMPRSS2, finding ZINC000103558522 had the highest binding affinity to the main protease (-12.36 kcal/mol), and ZINC000012481889 had the highest binding affinity to the TMPRSS (-12.14 kcal/mol). Chandel et al. \cite{chandel2020structure} 
repurposed about 2000 FDA-approved compounds targeting S protein and NSP9, reporting that Tegobuvir was the most potent to S protein (-8.1 kcal/mol) and Conivaptan was the most potent to NSP9 (-8.4 kcal/mol). Targeting the main protease and S protein, Narkhede et al. \cite{narkhede2020molecular}, Durdagi et al. \cite{durdagi2020screening}, and Cubuk et al. \cite{cubuk2020comparison} evaluated tens of existing drugs, Kamaz et al. \cite{kamaz2020screening} and Tallei \cite{tallei2020potential} tested some plant products, Moreover, Maiti et al. \cite{maiti2020active} docked Nigellidine to these targets. Targeting the main protease and RdRp, Al-Masoudi et al. \cite{almolecular} focused on some antiviral and antimalarial drugs, Rono et al. \cite{rono2020azole} studied azole derivatives,  and Lakshmanan et al. \cite{lakshmanan2020screening} and Suresh et al. \cite{suresh2020screening} investigated Kabasura Kudineer Chooranam and Maramanjal Kudineer Churnam, respectively. Against the S protein and NenDoU, Sinha et al. \cite{sinha2020silico} reported that saikosaponin V was potent to both targets. 

\paragraph{Targeting human ACE2 or related targets.} \label{ace2_targeting}

Some docking studies involved targeting human proteins such as the ACE2 \cite{guler2020investigation, joshi2020silico, abdelli2020silico, hakmi2020silico, zhang2020biological, sharma2020natural, rana2020virtual, da2020essential, laksmiani2020active, braz2020silico, vardhan2020protein} and glucose regulated protein 78 (GRP78) \cite{palmeira2020preliminary}, which are related to SARS-CoV-2 binding. However, according to some reports, it is controversial to design SARS-CoV-2 inhibitors targeting human ACE2 or related proteins. ACE2 is an important enzyme attached to cell membranes in  lungs, arteries, heart, kidney, and intestines. It is critical for lowering blood pressure in a human body \cite{keidar2007ace2}. It is unclear whether drugs to inhibit ACE2 or related targets are more beneficial than harmful. Further investigation is needed \cite{fang2020antihypertensive}.

\subsubsection{Molecular dynamics (MD) simulation}
Biomolecules are not static. X-ray crystallography and nuclear magnetic resonance (NMR) have already revealed that even a same molecule can adopt multiple conformations \cite{gerstein1998database, goh2004conformational}. Conformational change plays a significant role in biomolecular function, such as enzyme catalytic cycles \cite{gao2015molecular, gao2017network}. While X-ray crystallography and NMR can only provide static structures, MD simulation is a feasible way to investigate biomolecular conformational changes \cite{karplus2002molecular, hospital2015molecular}. Furthermore, thanks to high-performance computing platforms such as graphical processing units (GPUs), current MD simulation can reveal conformation changes of biomacromolecules such as proteins, DNA, and RNA in the time scale of milliseconds (ms) \cite{shaw2010atomic}.

\begin{figure}[ht!]
	\centering
	\includegraphics[width=0.6\textwidth]{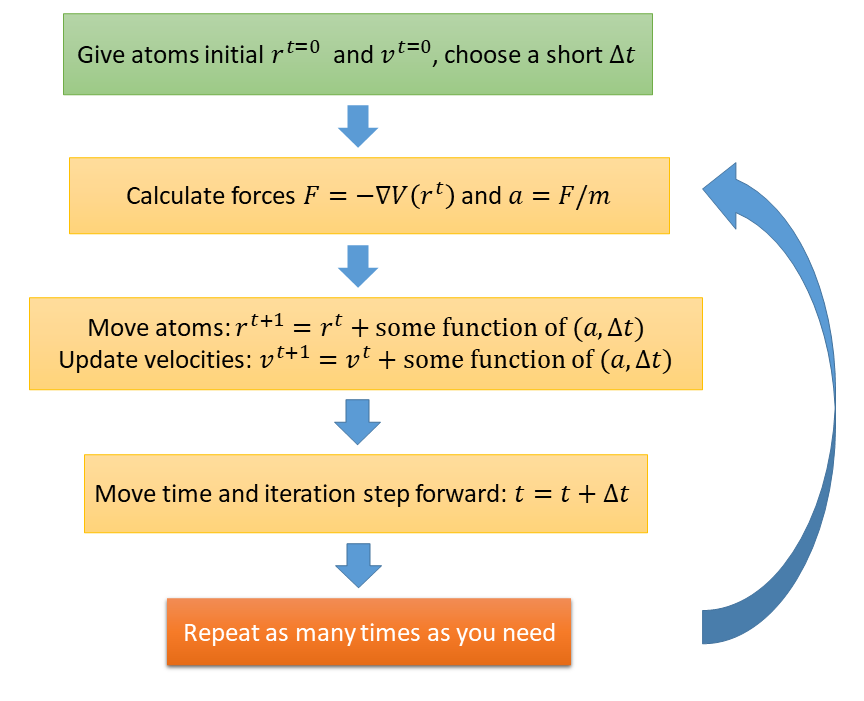}
	\caption{The workflow of MD simulations.}
	\label{fig:md-flow}
\end{figure}

MD simulation is a commonly used computational method for understanding the physical movement of atoms (or particles) in molecules (see Figure \ref{fig:md-flow}). The motions driven by  force fields are determined using Newton's second law \cite{adcock2006molecular}. In MD simulations, the interactions between atoms are described by force fields. Most force fields in chemistry are empirical, which consist of a summation of bonded forces associated with chemical bonds, bond angles, bond dihedrals, and non-bonded forces associated with van der Waals and electrostatic forces \cite{allen2004introduction}. The functional form of a typical force field such as AMBER looks like \cite{cornell1995second}: 

\begin{equation}\label{eq:force_field}
\begin{split}
V(r^N)=\sum_{\mathrm{bonds}}k_{r}(r-r_0)^2+\sum_{ \mathrm{angles}}k_{\theta}(\theta-\theta_0)^2 \\
+\sum_{ \mathrm{torsions}}\frac{1}{2}V_n[1+\mathrm{cos}(n\omega_i-\gamma_i)] \\
+\sum_{j=1}^{N-1}\sum_{i=j+1}^{N}\left[{\frac{A_{ij}}{R_{ij}^{12}}}-{\frac{B_{ij}}{R_{ij}^6}}+\frac{q_iq_j}{\epsilon R_{ij}}\right],
\end{split}
\end{equation}
where $k_{r}$ and $k_{\theta}$ are the force constants for bond lengths and bond angles, respectively. Here, $r$ and $\theta$ are a bond length and a bond angle, $r_0$ and $\theta_0$ are the equilibrium bond length and bond angle, $\omega_i$ is the dihedral angle, $V_n$ is the corresponding force constant, $n$ is the multiplicity, and phase angle $\gamma_i$ takes values of either $0^\circ$ or $180^\circ$. The non-bonded part of the potential is represented by the Lennard-Jones  repulsive $A_{ij}$ and attractive  $B_{ij}$ terms, Coulomb interactions between partial atomic charges ($q_i$ and $q_j$). Here, $R_{ij}$ is the distance between atoms $i$ and $j$. Finally, $\epsilon$ is the dielectric constant that considers the medium effect that is not explicitly represented and usually equals 1.0 in a typical solvated environment where solvent is represented explicitly. The non-bonded terms are calculated for atom pairs that are either separated by more than three bonds or not bonded. 

To tackle systems with an excessive number of atoms, coarse-grained models are also developed. In these models, a group of atoms is represented by a ``pseudo-atom'', so the number of atoms is largely reduced \cite{kmiecik2016coarse}. Popular coarse-grained models are the G{\=o} model \cite{ueda1978studies}, MARTINI force field \cite{marrink2007martini}, UNRES force field \cite{maisuradze2010investigation}, et al.

Since MD simulations can provide many samplings, one can calculate free energy change between different states from these samplings. Typical binding free energy calculation methods based on MD simulations are the molecular mechanics energies combined with Poisson–Boltzmann or generalized Born and surface area continuum solvation (MM/PBSA and MM/GBSA) \cite{massova1999computational,genheden2015mm}, free energy perturbation (FEP) \cite{lu2001accuracy}, thermodynamic integration \cite{straatsma1988free}, 
and metadynamics \cite{laio2002escaping}. Recently, a method that is more efficient method than normal-mode analysis, called WSAS \cite{wang2012develop}, was developed to estimate the entropic effect in the free energy calculation.

\paragraph{MD simulations revealing conformational changes.} MD simulations can be applied to investigate the dynamical properties of SARS-CoV-2 proteins and the interactions between proteins and inhibitors. Multiscale coarse-grained model was employed to understand the behavior of the SARS-CoV-2 virion \cite{yu2020multiscale}.  
Multiscale simulations were designed to examine glycan shield effects on drug binding to influenza neuraminidase \cite{seitz2020multiscale}.

\subparagraph{1. The main protease.} 
Grottesi et al.\cite{grottesi2020computational} analyzed 2-$\mu$s MD trajectories of the apo form of the SARS-CoV-2 main protease and indicated that the long loops,  which connect domain II and III and provide access to the binding site and the catalytic dyad, carried out large conformational changes. Bz{\'o}wka et al.\cite{bzowka2020structural} applied MD simulations to compare dynamical properties of the SARS-CoV-2 main protease and SARS-CoV main protease, which suggests that the SARS-CoV main protease has a larger binding cavity and more flexible loops. Yoshino et al.\cite{ryunosuke2020identification} used MD simulations to reveal the key interactions and pharmacophore models between the main protease and its inhibitors. 

\subparagraph{2. The S protein.} Bernardi et al. \cite{bernardi2020development} built a glycosylated molecular model of ACE2-Fc fusion proteins with the SARS-CoV-2 S protein RBD and used MD simulations to equilibrate it. Veeramachaneni et al.\cite{veeramachaneni2020structural} ran 100-ns MD simulations of the complexes of human ACE2 and S protein from SARS-CoV-2 and SARS-CoV. Their simulations showed that the SARS-CoV-2 complex was more stable. Gur et al.\cite{gur2020conformational} carried out steered MD to simulate the transition between closed and open states of S protein, a semi-open intermediate state was observed. Han et al. \cite{han2020computational} applied MD simulation to design and investigate peptide S-protein inhibitors extracted from ACE2. Oliveira \cite{oliveira2020simulations} used MD simulation to support his hypothesis that the SARS-CoV-2 S protein can interact with a nAChRs inhibitor.
All atom molecular dynamics was used to understand the interactions between the S protein and ACE2 \cite{casalino2020beyond,barros2020flexibility}.

\subparagraph{3. Other SARS-CoV-2 proteins.}
MD simulations were also used to investigate other SARS-CoV-2 proteins' conformational changes.  Henderson et al.\cite{henderson2020assessment} performed pH replica-exchange CpHMD simulations to estimate the pKa values of Asp/Glu/His/Cys/Lys sidechains and assessed possible proton-coupled dynamics in SARS-CoV, SARS-CoV-2, and MERS-CoV PLpros. They also suggested a possible conformational-selection mechanism by which inhibitors bind to the PLpro.

\paragraph{The combination of docking and MD simulation.}

Much effort combines docking and MD simulation. For example, molecule docking predicts binding poses, and MD simulation further optimizes and stabilizes the conformations of complexes. Some researchers rescore the optimized complexes by docking programs, or follow an ensemble-docking procedure to dock compounds to multiple conformations of the protein extracted from MD simulations. 

An ensemble docking of the SARS-CoV-2 main protease was performed by Sztain et al.\cite{sztain2020elucidation}. They docked almost 72,000 compounds to over 80 conformations of the main protease generated from MD simulations and screened these compounds through the ensemble docking strategy. To obtain extensive conformational samplings of the main protease, a Gaussian accelerated MD simulation \cite{miao2015gaussian} was run. Another ensemble docking work of the main protease was implemented by Koulgi \cite{koulgi2020drug}. They carried out long-time MD simulations  on the apo form of the main protease. Sixteen representative conformations were collected from these MD simulations by clustering analysis and Markov state modeling analysis \cite{chodera2014markov}. Targeting these 16 conformations, ensemble docking was performed on some FDA-approved drugs and other drug leads, suggesting some potent candidates such as Tobramycin. Additionally, Stoddard et al. \cite{stoddard2020optimization} docked inhibitors to two different crystal structures of the main protease. A similar scheme was also applied to RdRp by Elfiky et al. \cite{elfiky2020sars}. They extracted 8 conformations from clustering analysis of MD simulation and docked 31 drugs and other compounds to these conformations of RdRp.
Many other investigations focused on docking and then optimizing by MD simulations.

\subparagraph{1. Targeting the main protease.} 
Odhar et al. \cite{odhar2020molecular} applied docking and MD simulation to systematically investigate the binding affinities and interactions of 1615 FDA-approved drugs to the main protease, suggesting some potential repurposed drugs such as Perampanel with a predicted binding affinity of -8.8 kcal/mol. Baildya et al. \cite{baildya2020inhibitory} tested Hydroxychloroquine, reporting a binding affinity of -6.3 kcal/mol. Other existing drugs such as lopinavir, oseltamivir, ritonavir, atazanavir, darunavir, tetracyclines, flaviolin, Hydroxyethylamine Analogs, buriti oil (mauritia flexuosa L.) like inhibitors, etc. were also investigated specifically by docking and MD simulations \cite{kumar2020discovery, muralidharan2020computational, huynh2020silico,kumar2020molecular, khan2020comparative, kumar2020silico, mahanta2020potential, bharadwaj2020computational, fintelman2020atazanavir, razzaghi2020identification, costa2020constituents, rao2020proposing, de2020silico, shree2020targeting, kadil2020silico}.

Another important inhibitor source is natural products. Following the workflow of ligand docking, MD optimization, and rescoring,  Gentile et al. \cite{gentile2020putative} screened the library of marine natural products (MNP), which includes 14064 marine natural products. The best one, heptafuhalol A, was predicted to have a docking score as high as -18.0 kcal/mol. Khan et al. \cite{khan2020marine} also investigated some marine products. Qamar et al. \cite{ul2020structural} used docking and MD simulation to screen a medicinal plant library containing 32,297 potential anti-viral phytochemicals/traditional Chinese medicinal compounds. Potent inhibitors such as 5,7,3\',4\'-Tetrahydroxy-2\'-(3,3-dimethylallyl) isoflavone with a docking score of -16.35 kcal/mol were predicted. Virtual screening was also performed towards other natural products such as Indian medicinal herbs \cite{islam2020molecular, rao2020reckoning, ul2020structural, krupanidhi2020screening, chowdhury2020silico, umesh2020identification, mazzini2020putative, yepes2020investigating,peele2020molecular}.

Other small molecules were also screened to inhibit the SARS-CoV-2 main protease. Ton et al.\cite{ton2020rapid} identified potential main protease inhibitors by docking 1.3 billion compounds, and suggested that compound ZINC000541677852 had the highest binding affinity of -11.32 kcal/mol. Jiménez-Alberto et al. \cite{jimenez2020virtual} performed docking and MD simulations to test 4384 molecules from the Zinc dataset \cite{sterling2015zinc} and some of them are FDA-approved drugs. Among them, the best one was Bisoctrizole, which has a docking score of -10.25 kcal/mol. Besides, the prototypical-ketoamide inhibitors, HIV protease inhibitors, Leucoefdin, some nutraceuticals, and other compounds from literature were also studied \cite{liang2020interaction,cardoso2020molecular, battisti2020computational, kodchakorn2020molecular, stroylov2020computational, singh2020leucoefdin, lokhande2020molecular}. Notably, Mohammad et al. \cite{mohammad2020identification} optimized some complexes of the main protease with ligands in the protein data bank (PDB) and re-scored them by AutoDock Vina. The best PDB structure is 6m2n with a predicted binding affinity of -8.3 kcal/mol. 

\subparagraph{2. Targeting the S protein.} 
Trezza et al. \cite{trezza2020integrated} ran docking and MD simulations to identify potential inhibitors against SARS-CoV-2 S protein from 1582 FDA-approved drugs. Lumacaftor was predicted to have the highest binding affinity (-9.4 kcal/mol). Moreover, in order to evaluate the binding interactions between the S protein and compounds, steered MD simulations \cite{isralewitz2001steered} were performed on the top compounds. Bharath et al. \cite{bharath2020silico} and Choudhary et al. \cite{choudhary2020identification} screened 4015 and 1280 small compounds, respectively, and predicted that fytic acid and GR hydrochloride had the highest energies of -10.296 kcal/mol and -11.23 kcal/mol. Fantini et al. \cite{fantini2020nsynergistic} and Kalathiya et al. \cite{kalathiya2020highly} also evaluated some potential small molecules against the S protein.

Since the RBD of S protein is relatively large, the small-molecule drugs may not efficiently block the entire RBD. The entire RBD of S protein needs to be blocked by peptides \cite{wan2020receptor}. Basit et al. \cite{basit2020truncated} designed a truncated version of ACE2 (tACE2) receptor covering the binding residues; they performed protein-protein docking and MD simulations to analyze its binding affinity to RBD and complex stability; since the tACE2 was predicted to have a higher binding score, it could compete with the wild type of ACE2 for binding to SARS-CoV-2. Baig et al. \cite{baig2020identification} also identified a potential peptide inhibitor against S protein through docking and MD simulations. Souza et al. investigated some synthetic peptides using the same method \cite{souza2020molecular}. 

\subparagraph{3. Targeting the RdRp.} 
RdRp is another important target of SARS-CoV-2.  Pokhrel et al. \cite{pokhrel2020potential} screened 1930 FDA-approved drugs,  optimized the protein structures by MD simulations, and predicted poses and binding affinities by molecular docking. As a result, quinupristin was identified as the most potent drug with a docking score of -12.3 kcal/mol. 

\subparagraph{4. Targeting the PLpro.} 
Bosken et al. \cite{bosken2020insights} ran MD simulations to investigate the conformational change of PLpro and docked  inhibitors to the PLpro to reveal binding behaviors.

\subparagraph{5. Other targets.} 
Following the workflow of docking, optimizing the top predictions by MD simulations, and redocking, Sharma et al. \cite{sharma2020computational} repurposed 3000 FDA-approved and experimental drugs against the 2'-O-methyltransferase in NSP16 of SARS-CoV-2; dihydroergotamine and irinotecan were predicted to be the best drugs with binding affinities of -9.3 kcal/mol for both. Menezes et al. \cite{de2020identification} used MD simulation to investigate the dynamics of NSP1. More importantly, they screened 8694 approved and experimental drugs from DrugBank against NSP1 and predicted that tirilazad was the most potent one. Tazikeh-Lemeski et al. \cite{tazikeh2020targeting} docked 1516 FDA-approved drugs from DrugBank to the NSP16 and investigated the drug-protein interactions by MD simulation, finding that raltegravir had the best predicted binding affinity (-10.4 kcal/mol). Following a similar scheme, Selvaraj et al. \cite{selvaraj2020structure} screened 22122 Chinese traditional medicines from TCM Database@Taiwan against the NSP14; the best one, TCM57025, had a docking score of -11.486 kcal/mol. Tatar et al. \cite{tatar2020investigation} also studied the potency of 34 drugs against the N protein.

\subparagraph{6. Multiple targets.}
In many reports, the same drugs were tested against multiple targets. Dwarka et al.  \cite{dwarka2020identification} studied the potency of 14 South African medicinal plants against the main protease, RdRp, and S-protein RBD using docking. They also investigated the dynamics and interactions inside the complexes using MD simulations. However, no potent inhibitors were found. Through a similar procedure,  Adeoye et al.\cite{adeoye2020repurposing} evaluated some clinically approved antiviral drugs against the main protease, PLpro, and S protein. Sinha et al. \cite{sinha2020identification} identified some bioactive natural products from glycyrrhiza glabra targeting the S protein and NendoU. Ortega et al. \cite{ortega2020class} and Aouidate et al. \cite{aouidate2020identification} repurposed some drugs and small molecules in the category of histamine type 2 receptor antagonists to inhibit the main protease and RdRp. Ahmed \cite{ahmed2020destabilizing} predicted the potency of the Caulerpin and its derivatives against the main protease and S protein. Khan et al. \cite{khan2020targeting} utilized the virtual drug repurposing approach to test some antiviral drugs against the main protease and 2/'-O-ribose methyltransferase in NSP16. Some works involved the structural proteins of SARS-CoV-2. Bhowmik et al. \cite{bhowmik2020identification} virtually screened more than 200 anti-viral natural compounds and 348 anti-viral drugs targeting the E, M, and N proteins responsible for envelope formation and virion assembly. Gentile et al. \cite{gentile2020new} simulated chloroquine and hydroxychloroquine against the E Protein, NSP10/NSP14 complex, and NSP10/NSP16 complex.

\subparagraph{7. The human ACE2 target with controversy.} 
Khelfaoui et al.\cite{khelfaoui2020molecular}, Marciniec et al. \cite{marciniec2020phosphate}, and Gutierrez et al. \cite{gutierrez2020alkamides} performed docking and MD studies targeting the human ACE2. However, as mentioned in Section \ref{ace2_targeting}, it is quite controversial to design anti-SARS-CoV-2 drugs targeting ACE2.

\paragraph{MD based MM/PBSA or MM/GBSA binding free energy calculations.}

\begin{figure}[ht!]
	\centering
	\includegraphics[width=0.8\textwidth]{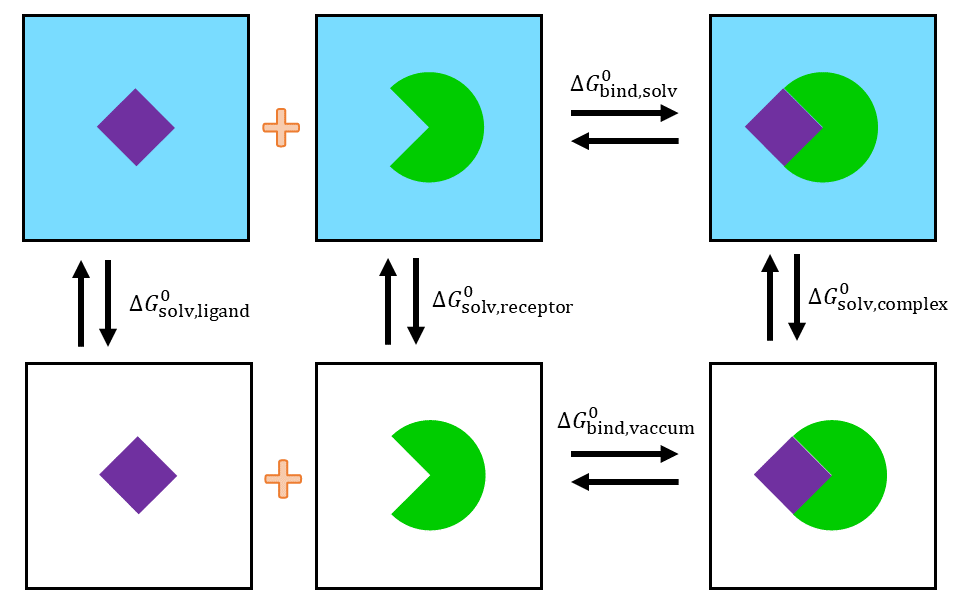}
	\caption{The thermodynamic cycle of the MM/PB(GB)SA calculation.}
	\label{fig:mm-pbsa}
\end{figure}

To obtain more accurate free energies, after docking and MD simulation, many works calculated binding free energies based on the MM/PBSA or MM/GBSA methods \cite{kumari2014g_mmpbsa}.  The basic idea of MM/PBSA and MM/GBSA is to divide up the calculation according to the thermodynamic cycle in Figure \ref{fig:mm-pbsa}, then evidently, the binding free energy $\Delta G_{\text{bind,solv}}$ can be calculated by:

\begin{equation}\label{eq:mm-pbsa}
\Delta G_{\text{bind,solv}}^0=\Delta G_{\text{bind,vacuum}}^0+\Delta G_{\text{solv,complex}}^0-(\Delta G_{\text{solv,ligand}}^0+\Delta G_{\text{solv,receptor}}^0).
\end{equation}

Polar solvation free energies for each of the three states are calculated by either solving the linearized Poisson Boltzmann (PB) equation in MM/PBSA or generalized Born (GB) equation in MM/GBSA. More detail about the PB and GB models can be found in Section \ref{sect:PB_GB}. Nonpolar solvation free energy is obtained by solvent accessible area (SA). $\Delta G_{\text{bind,vacuum}}$ is from calculating the average interaction energy between receptor and ligand and taking the entropy change upon binding into account if necessary \cite{massova1999computational}.

\subparagraph{1. Targeting the main protease.} Florucci et al. \cite{fiorucci2020computational} used docking, MD simulations, and MM/PBSA binding free energy calculations to investigate around 10000 compounds from the DrugBank \cite{wishart2018drugbank}. These compounds were first screened by docking, and then MD-based MM/PBSA binding free energy calculations were performed on the best 36 compounds. The reported binding data were the consensus of docking and MM/PBSA prediction and leuprolide was the best one. A very similar work by Ibrahim et al. \cite{ibrahim2020silico} also screened thousands of compounds from the DrugBank by docking and MM/GBSA MD simulations.  Gahlawat \cite{gahlawat2020structure} implemented an MM/GBSA procedure on 2454 FDA-approved and experimental drugs, 138 natural products, and 144 other inhibitors to predict the MM/GBSA binding free energy of top ones screened by docking. Their reveal the highest one, lithospermic acid B's, with an MM/GBSA energy of -118.69 kcal/mol. Sharma \cite{sharma2020identification} also applied docking and MM/PBSA calculation to screen 2100 drugs as well as 400 other compounds and reported that cobicistat had the highest binding affinity of -11.42 kcal/mol. Wang \cite{wang2020fast} calculated the MM-PBSA-WSAS binding free energies of 2201 drugs, suggesting flavin adenine dinucleotide to be the most potent. Rahman \cite{rahman2020virtual} studied 1615 FDA-approved drugs by docking and MM/PBSA, predicting simeprevir to be highly effective. Khattab Al-Khafaji et al. \cite{al2020using}  evaluated some FDA-approved drugs forming covalent bonds with the main protease through covalent docking and MM-GBSA MD simulations. Other assessments based on docking and MM/PB(GB)SA calculations focused on specific drugs such as lopinavir, ritonavir, saquinavir, anti-HIV drugs, chloroquine, hydroxychloroquine, noscapine, and their derivatives \cite{nutho2020lopinavir, martiniano2020identification, sang2020anti, ancy2020possibility, khan2020combined, mukherjee2020structural, beura2020silico, kumar2020understanding, dash2020drug, nayeem2020target}.

MM/PBSA or MM/GBSA methods were also applied to test natural products against the main protease. Ibrahim et al. \cite{ibrahim2020natural} virtually screened the MolPort database containing 113,756 natural or natural-like products (https://www.molport.com) by docking. The top 5,000 compounds were selected and subjected to MD simulations combined with MM/GBSA binding affinity calculations, and the compound MolPort-004-849-765 was predicted to have the highest binding free energy of -58.4 kcal/mol. Kapusta et al. \cite{kapusta2020protein} also performed docking on 13,496 natural or natural-like products from MolPort and the top 15 were chosen to rescore by MM/GBSA calculations, which reported MolPort-039-338-330 as the most potent one (-57.39 kcal/mol). Mahmud et al. \cite{mahmud2020molecular} screened 1480 natural plant products from literature initially by docking scores and then the best 10\% were rescored by MM/GBSA. Other natural product sources screened by docking and MM/PB(GB)SA against the main protease were flavonoid-based phytochemical constituents of calendula officinalis, tea plant products, Withania somnifera (Ashwagandha) products, lichen compounds, Curcuma longa products, and polyphenols from Broussonetia papyrifera \cite{das2020silico, bhardwaj2020identification, tripathi2020identification, kumar2020identification, ghosh2020evaluation, joshi2020structure, gupta2020identification, ghosh2020identification}.

Other compounds screened by docking and MM/PBSA or MM/GBSA against the main protease are below. Andrianov et al. \cite{andrianov2020computational} first virtually screened over 213.5 million chemical structures from \url{http://pharmit.csb.pitt.edu/} to select the ones satisfying the pharmacophore model from the known X77 potent main protease inhibitor (PDB ID: 6W63). Then they docked them to the main protease, and ran MM/GBSA simulations of the docking complexes to calculate binding free energy. Through this procedure, they reported some potent inhibitors such as Pub-chem-22029441 with a MM/GBSA energy of -47.66 kcal/mol. The pharmacophore procedure was also performed by Arun et al. \cite{arun2020drug} and the top 3 by docking scores were subjected to MM/GBSA calculations, reporting Macimorelin acetate (-50.03 kcal/mol) as the best one. Choudhary et al. \cite{choudhary2020silico} docked the 15754 compounds in their in-house dataset to the main protease and rescored them by MM/GBSA calculations, reporting the most potent one, dimethyl lithospermate (-81.82 kcal/mol). Khan et al. \cite{khan2020identification_1} used docking to screen approximately 8000 compounds in their in-house database and applied MM/GBSA MD simulations to calculate the binding affinities of the top 5 inhibitors, with remdesivir being the best. Fakhar et al. \cite{fakhar2020anthocyanin} screened 3435 anthocyanin substructure compounds by docking and MM/GBSA calculations, reporting the best compound 44256921 with the energy -77.04 kcal/mol. Some other compounds, such as oxazine substituted 9-anilinoacridines, nitric oxide (NO) donor furoxan, withanone caffeic acid phenethyl ester, tetracycline, peptides, etc. were also screened by docking and MM/PB(GB)SA simulations \cite{rajagopal2020identification, al2020potential, kumar2020withanone, pant2020peptide, sk2020elucidating, al2020tackling, zhao2020tetracycline, gimeno2020prediction, seo2020supercomputer}.

\subparagraph{2. Targeting the S protein.} Both SARS-CoV and SARS-CoV-2 infect humans through the S protein binding to the human ACE2. Therefore, many investigations focused on the interaction between the S protein and the ACE2. Hassanzadeh et al.\cite{hassanzadeh2020considerations} used MM/PBSA, and He et al. \cite{he2020molecular} used MM/GBSA calculations to compare the binding affinities of the S proteins from SARS-CoV and SARS-CoV-2 to the human ACE2. Their calculations indicated that the S protein of SARS-CoV-2 bound to ACE2 much more tightly than that of SARS-CoV. Spinello et al. \cite{spinello2020rigidity}, Bhattacharyay et al. \cite{bhattacharyay2020impact}, and Armijos et al. \cite{armijos2020sars} investigated the mechanism of tight binding of the SARS-CoV-2 S-protein through MD simulations and MM/GBSA or MM/PBSA calculations. Shah et al. \cite{shah2020sequence} and Ou et al. \cite{ou2020emergence} ran some MM/PBSA calculations and found that some mutations on the S protein could facilitate stronger interactions with human ACE2. 

Other two interesting studies by Piplani et al.\cite{piplani2020silico} and by Shen et al.\cite{shen2020sars} performed MM/PBSA or MM/GBSA calculations to reveal the binding affinities of the SARS-CoV-2 S-protein to the ACE2s from different species. Their results showed that chimpanzee's binding affinity was even higher than human, cat, pangolin, dog, monkey, and chimpanzee had a similar affinity to human, which suggested some mammals were also vulnerable to SARS-CoV-2. 

Drug repurposing against the S protein was also implemented by MM/GBSA or MM/PBSA. De Oliveira et al. \cite{de2020repurposing} docked 9091 approved or experimental drugs to the S protein and selected the top 3 to perform MM/PBSA calculation, which led to suramin sodium having the highest affinity of -51.07 kcal/mol. Following a similar scheme, Romeo et al. \cite{romeo2020targeting} repurposed 8770 approved or experimental drugs by docking and MM/GBSA, reporting 31h-phthalocyanine as the most potent (-84.8 kcal/mol). Padhi et al. \cite{padhi2020does} performed docking and MM/PBSA calculations studies on the inhibition of Arbidol to the RBD/ACE2 complex.

Some MM/GBSA or MM/PBSA investigations were about the use of natural products against the S protein. Pandey et al. \cite{pandey2020targeting} used docking and MM/PBSA calculations on Among 11 phytochemicals, suggesting that quercetin has the highest affinity (-22.17 kcal/mol).

MM/GBSA or MM/PBSA approaches were applied  to other compounds against S protein. Sethi et al.  \cite{sethi2020understanding} performed docking to 330 galectin inhibitors against the S protein and ran MM/GBSA calculations to some active ones, revealing that ligand No.213 had the highest binding free energy of -54.11 kcal/mol. Rane \cite{rane2020targeting} and Li et al. \cite{li2020mers} calculated the potency of all diaryl pyrimidine derivatives and the MERS-CoV receptor DPP4, respectively.

\subparagraph{3. Targeting the RdRp.} Ruan et al. \cite{ruan2020sars} studied 7496 approved or experimental drugs against both SARS-CoV-2 and SARS-CoV RdRp. They screened by docking and ran MM/GBSA calculations to the top ones, suggesting lonafarnib, tegobuvir, olysio, filibuvir, and cepharanthine's potency to both SARS-CoV-2 and SARS-CoV RdRp. Khan et al. \cite{khan2020phylogenetic} screened 6842 South African natural products against RdRp using docking, and selected the top 4 to further investigate by MD simulations and MM/GBSA calculations. Their most potent one was Genkwanin 8-C-beta-glucopyranoside with a MM/GBSA binding free energy of -63.695 kcal/mol. By contrast, such energy of an approved SARS-CoV-2 drug, remdesivir, against RdRP was reported to be -54.406 kcal/mol. Singh et al.\cite{singh2020plant} also studied 100 natural polyphenols by docking. The leading 8 compounds were used in MD simulations and MM/GBSA calculations, which shows the compound TF3 being the best (-42.27 kcal/mol). Aktacs et al. \cite{aktacs2020new} used MM/PBSA calculations to show that CID294642 was a potent RdRp inhibitor. Venkateshan et al. \cite{venkateshan2020azafluorene} and Ahmad et al. \cite{ahmad2020molecular} also carried out MM/GBSA or MM/PBSA calculations.

\subparagraph{4. Targeting the PLpro.} Kandee et al. \cite{kandeel2020repurposing} repurposed 1697 approved drugs against the PLpro by docking, and their top 10 were studied by MD simulations and MM/GBSA, with the drug phenformin being their best one (-56.5 kcal/mol). Bosken et al. \cite{bosken2020insights} assessed the potential effectiveness of one naphthalene-based inhibitor, 3k, and one thiopurine inhibitor, 6MP, through docking, MD simulations, and MM/PBSA calculations.

\subparagraph{5. Other targets.} Many MM/GBSA or MM/PBSA investigations focused on the SARS-CoV-2 N protein. For instance, Khan et al. \cite{khan2020structural} studied the mechanism of RNA recognition by the N-terminal RNA-binding domain of the SARS-CoV-2 N protein as well as mutation-induced binding affinity changes by docking, MD simulations, and MM/GBSA calculations. Yadav et al. \cite{yadav2020virtual} docked 8987 compounds from Asinex and PubChem databases against the N protein and assessed the potency of the top 10 by MM/GBSA. TMPRSS2 is another attractive target. Some natural products and compounds were repurposed through docking, MD simulations, and MM/GBSA calculations against  TMPRSS2 \cite{chikhale2020identification, kumar2020withanone}, suggesting the compound neohesperidin as the most potent. Khan et al. \cite{khan2020identification_n} docked 123 antiviral drugs to NendoU and found simeprevir had the highest binding energy. Their MM/PBSA calculations also confirmed this finding. Encinar et al. \cite{encinar2020potential} used docking to screen 8696 approved or experimental drugs against NSP16/NSP10 protein complex and, through MM/PBSA calculations, discovered that the presence of NSP10 strengthens the ligand binding to NSP16. Fulbabu Sk et al.\cite{sk2020computational} performed MM/PBSA simulations to study the binding interactions inside NSP16/NSP10 complex. Vijayan et al. \cite{vijayan2020identification} carried out repurposing against NSP16 involved 4200 drugs or compounds. their best one predicted from MM-PBSA was Carba-nicotinamide-adenine-dinucleotide. Chandra et al. \cite{chandra2020identification} studied 2895 approved or experimental drugs against the NendoU and  selected the top 3 compounds from docking results and ran MM/PBSA calculations to these three, finding glisoxepide, with a MM/PBSA binding free energy of -141 kcal/mol, being the most potent one. Dhankhar et al\cite{dhankhar2020computational} and Parida et al.\cite{parida2020natures}  used MM/PBSA calculations to target   6 different non-structural proteins of SARS-CoV-2.

\subparagraph{6. Multiple targets.} Some researchers screened drugs or compounds against multiple targets of SARS-CoV-2.  Gupta et al.\cite{sen2020binding} docked drug famotidine to the twelve targets of SARS-COV-2 including four structural targets: M protein, E protein, S protein, N protein, and eight non-structural targets: the main protease and PLpro, NendoU, Helicase, RdRp, NSP14, NSP16, and NSP10. They found famotidine had the highest docking binding energy of -7.9 kcal/mol with the PLpro, and the MM/PBSA energy was -59.72 kcal/mol. 

 Naik et al. \cite{naik2020high} screened  3963 natural compounds from the NPASS database (http://bidd.group/NPASS/index.php)  against 6 different SARS-CoV-2 targets, namely, the main protease, RdRp, NendoU, helicase, exoribonuclease in NSP14, and methyltransferase in NSP16. They docked these compounds to these targets and calculated MM/GBSA binding free energies of the top ones. Quimque et al.\cite{quimque2020virtual} studied 97 secondary metabolites from marine and terrestrial fungi. They screened these compounds against 5 different SARS-CoV-2 targets, i.e., the main protease, S protein, RdRp, PLpro, and NendoU, by docking and MM/PBSA calculations. Murugan \cite{murugan2020computational} investigated four compounds in A. paniculata targeting four different proteins in SARS-CoV-2, i.e., the main protease, S protein-ACE2 complex, RdRp, and PLpro through docking, MD simulations and MM/GBSA, finding that  AGP-3 had the potency to all four targets. Alazmi et al. \cite{alazmi2020silico} assessed around 100,000 natural compounds against RdRp, NSP4, NSP14, and human ACE2 by docking. Their top compounds were further investigated by MD simulations and MM/PBSA, reporting that Baicalin was potent against RdRp, NSP4, and NendoU. Kar et al.\cite{kar2020natural}   studied  main protease, S protein, and RdRp. Their ligands were natural products from Clerodendrum spp. After docking and rescoring the top ones by MM/GBSA, these authors found taraxerol being effective to all three targets.  Using docking and MM/GBSA or MM/PBSA calculations, Alajmi et al. \cite{alajmi2020antiviral} and Sasidharan et al. \cite{sasidharan2020bacterial} evaluated the potency of around 40 compounds, including some existing drugs and protein azurin secreted by the bacterium pseudomonas aeruginosa as well as its derived peptides,  against the main protease, PLpro, and S protein. In a similar way, Mirza et al. \cite{mirza2020structural} repurposed some compounds against the main protease, RdRp, and helicase. Maroli et al. \cite{maroli2020potential} investigated procyanidin inhibiting the main protease, S protein, and human ACE2.

Panda et al. \cite{panda2020structure} targeted the main protease and S protein; they screened 640 compounds through docking and MD simulations and identified PC786 had high docking scores both to the S protein (-11.3 kcal/mol) and main protease (-9.3 kcal/mol). Moreover, their MD simulations and  MM/PBSA calculations revealed the binding of PC786 could change the conformation of the S-protein and weaken the S-protein's binding interactions to ACE2. Also targeting the main protease and S protein, Prasanth et al. \cite{prasanth2020silico} studied 48 isolated compounds from cinnamon by docking and MD-simulation based MM/PBSA calculations, suggesting the compounds tenufolin and pavetannin C1 were potent to both the main protease and S protein. Parida et al.\cite{parida2020natural} also predicted some potential inhibitors against the SARS-CoV-2 main protease and S protein from Indian medicinal phytochemicals by MM/PBSA.

Gul et al.\cite{gul2020silico} and Ahmed et al.\cite{ahmed2020investigating} used docking, MD simulations, and MM/GBSA calculations to suggest the potency of current drugs against the main protease and RdRp. Targeting the main protease and PLpro, Mitra et al. \cite{mitra2020dual}'s pharmacophore-based virtual screening yielded 6 existing FDA-approved drugs and 12 natural products with promising pharmacophoric features, and through docking, MD simulations, and MM/PBSA calculations, lopinavir and tipranavir being predicted to be the best inhibitors against the two proteases. Similarly, Naidoo et al. \cite{naidoo2020cyanobacterial} investigated the potency of cyanobacterial metabolites against these two proteases. Chikhale et al. \cite{chikhale2020sars} repurposed Indian ginseng against the S protein and NendoU by docking and MM/GBSA calculation. Borkotoky et al. \cite{borkotoky2020computational} focused on the M protein and E protein. Docking and MM/PBSA calculations were performed on the inhibitors from Azadirachta indica (Neem).


\paragraph{Other MD-based binding free energy calculation methods.}
Besides MM/PBSA or MM/GBSA, other binding free energy calculation methods such as FEP and metadynamics were also applied to evaluate the binding affinities of inhibitors to the main protease. 

\subparagraph{1. FEP.} Wang et al. \cite{wang2020enhanced} applied MD simulations and FEP free energy calculations to uncover the mechanism of the stronger binding of SARS-CoV-2 S protein to ACE2 than that of SARS-CoV S protein. They compared hydrogen-bonding and hydrophobic interaction networks of SARS-CoV-2 S protein and SARS-CoV S protein to ACE2 and calculated the free energy contribution of each residue mutation from SARS-CoV to SARS-CoV-2. Ngo et al. \cite{ngo2020computational} first docked about 4600 drugs or compounds to the main protease and then, used pull work obtained from the fast pulling of ligand simulation \cite{ngo2016fast} to rescore the top 35 compounds. They reevaluated the top 3 using FEP free energy calculations, which suggested that the inhibitor 11b was the most potent. Zhang et al. \cite{zhang2020structural} docked remdesivir and ATP to RdRp, and used FEP to calculate the binding free energy, which indicated the binding of remdesivir was about 100 times stronger than that of ATP, so it could inhibit the ATP polymerization process.

\subparagraph{2. Metadynamics.} Namsan et al. \cite{namsani2020potential} docked 16 artificial-intelligence generated compounds by Bung \cite{bung2020novo} to the main protease, then ran metadynamics to calculate their binding affinity, and predicted some potential inhibitors.

\paragraph{Coarse grained MD simulations.} De Sacho et al. \cite{de2020coarse} used the G{\=o} coarse-grained MD model to simulate the process of the SARS-CoV-2 S protein RBD binding to the human ACE2, characterized the free energy landscape, and obtained the free energy barrier between bound and unbound states was about 15 kcal/mol. Nguyen et al. \cite{nguyen2020does} also applied the G{\=o} coarse-grained model and replica-exchange umbrella sampling MD simulations to compare the binding of SARS-CoV-2 S protein and SARS-CoV S protein to the human ACE2, revealing SARS-CoV-2 binding to human ACE2 to be stronger than that of SARS-CoV. Giron et al. \cite{giron2020interactions} studied the interactions between SARS-CoV-2 S protein and some antibodies by coarse-grained MD simulations.

\paragraph{MD simulations combining with deep learning.}
Gupta \cite{gupta2020profiling} first selected 92 potential main-protease inhibitors from FDA-approved drugs by docking, then further evaluated their potency by MD simulations and MM/PBSA calculations using the hybrid of the ANI deep learning force field \cite{smith2019approaching} and a conventional molecular mechanics force field. Their results suggested that targretin was the most potent drug against the main protease. Joshi et al.\cite{joshi2020predictive}  screened compounds by first making use of a deep neural model, and then performed docking and MD simulations to further evaluate them.

\paragraph{MD simulations combining with experiments.} Turovnova et al. \cite{turovnova2020situ}'s MD simulations revealed three  flexible  hinges  within  the  stalk,  coined  hip,  knee,  and  ankle of the S protein, which were consistent with their tomographic experiments.

\paragraph{MD simulation studies on mutation.}
Through MD simulations, Qiao et al. \cite{qiao2020enhanced} found that the mutation on the distal polybasic cleavage sites of the S protein could weaken the binding between the S protein and human ACE2, which means these distal polybasic cleavage sites were critical to the binding. Zou et al. \cite{zou2020computational} also performed virtual alanine scanning mutagenesis by FEP MD simulations to uncover key S protein residues in binding to ACE2.
Similarly, Ou et al. \cite{ou2020rbd} investigated the interaction impact of S protein mutation through MM/PBSA calculations. Dehury et al. \cite{dehury2020effect} compared the interactions of mutated S proteins and the wild type S protein to ACE2. Haidi \cite{hadi2020studying} mutated the human ACE2 and assessed the impact on interactions with S protein by MM/GBSA calculations. Sheik et al. \cite{sheik2020impact} studied the impact of main protease mutations on its 3D conformation.

\paragraph{MD simulation studies on vaccine.} 
Grant et al.\cite{grant2020analysis} ran MD simulations and evaluated the extent to which glycan microheterogeneity could impact epitope exposure of the S protein. Their studies indicated that glycans shield approximately 40\% of the underlying protein surface of the S protein from epitope exposure. Peele et al. \cite{peele2020design} generated more than 30 epitode vaccine candidates originating from the S protein by the online servers NetCTL, IEDB, and FNepitope. These epitodes' tertiary structures were predicted and also docked to the toll-like receptor 3. MD simulations were run for these complexes and the immune reactions were simulated. Lizbeth et al. \cite{lizbeth2020immunoinformatics} not only predicted some epitodes, but also used MM/PBSA MD simulations to calculate the binding affinity of the MCH II-epitope complexes, with the highest one being -1810.9855 kcal/mol. Through docking and MD simulations, De Moura et al.  \cite{de2020immunoinformatic} identified epitopes from the S protein that were able to elicit an immune response mediated by the most frequent MHC-I alleles in the Brazilian population. Other similar reports include Refs. \cite{bhattacharya2020sars, samad2020designing, sanami2020design, pourseif2020prophylactic, rahman2020vaccine}.

Not only epitodes from the S-proteins were investigated, but also epitodes from other targets were studied. Rahman et al. \cite{rahman2020epitope} also researched some epitodes from the S, M, and E proteins. Chauhan et al. \cite{chauhan2020excavating}, Kalita et al. \cite{kalita2020design}, Ranga et al. \cite{ranga2020immunogenic}, and \cite{sarkar2020immunoinformatics} studied multiple targets.

\paragraph{MD simulation data analysis methods.} 
One of the popular methods to analyze dynamics characteristic in MD simulation is principal component analysis (PCA), which can extract the principal modes of motion from MD simulations \cite{skjaerven2011principal, amadei1993essential}. Towards SARS-CoV-2, Kumar et al.\cite{kumar2020molecular}, Mahmud et al.\cite{mahmud2020molecular}, Islam et al.\cite{islam2020molecular}, and Sk et al.\cite{sk2020elucidating} applied PCA to reveal the internal motions of the main protease. Rane et al. \cite{rane2020targeting} and Dehury et al. \cite{dehury2020effect} used such analysis to investigate the dynamics of the S protein. Henderson et al. \cite{henderson2020assessment} and Chandra et al.\cite{chandra2020identification} used PCA to elucidate the motions of the PLpro and NendoU, respectively.

Normal mode analysis (NMA) \cite{case1994normal} is another way to study protein fluctuations. Bhattacharya et al. \cite{bhattacharya2020sars}  applied NMA to display the mobility of the human TLR4/5 protein and SARS-CoV-2 vaccine component complex.

\subsubsection{Density-functional theory (DFT).}
DFT is a computational quantum-mechanics modeling method widely used in computational physics, computational chemistry, and computational materials science to investigate the electronic structure of atoms, molecules, and condensed phases \cite{parr1980density,lee1988development}. Using this theory, the properties of a many-electron system are represented by functionals (functions of another function) of the spatially dependent electron density \cite{parr1980density}. Because of the development of DFT, Walter Kohn won the Nobel Prize in Chemistry in 1998 \cite{kohn1999nobel}.

Bui et al. \cite{bui2020density} applied DFT calculations to optimize the silver/bis-silver-lighter tetrylene complex and study the molecular orbits, and docked the optimized compounds to the main protease and ACE2, reporting NHC-Ag-bis has a -16.8 kcal/mol binding affinity to the main protease.
Gatfaoui et al. \cite{gatfaoui2020synthesis}, and Hagar et al. \cite{hagar2020investigation} also optimized the conformation of 1-Ethylpiperazine-1,4-diium Bis(Nitrate) and some antiviral N-heterocycles, investigated that partial charge distribution as well as hydrogen bond strength by DFT. They docked these compounds to the main protease and studied the interactions between them.

\subsubsection{Quantum mechanics/molecular mechanics (QM/MM).}  
The QM/MM approach is a molecular simulation method that combines the accuracy of QM and the speed of MM: the region of the system in which the chemical process takes place is treated at an appropriate level of QM. The remainder is described by a MM force field \cite{warshel1976theoretical}. This approach can be used  to study  chemical processes in solution and proteins.  The Nobel Prize in Chemistry in 2013 was awarded to Arieh Warshel and Michael Levitt for the introduction of QM/MM.

Khrenova et al. \cite{khrenova2020dynamical} used QM/MM to simulate the covalent bonds forming between the substrate and Cys145 residue of the main protease based on the crystal structure with a PDB ID 6LU7 \cite{jin2020structure}, revealing the inhibition mechanism of covalent inhibitors. Swiderek et al. \cite{swiderek2020revealing} studied the covalent bonding of the polypeptide Ac-Val-Lys-Leu-Gln-ACC (ACC is the 7-amino-4-carbamoylmethylcoumarin fluorescent tag) to the main protease by QM/MM, suggesting that the free energy barrier is 22.8  kcal/mol. Ramos et al. \cite{ramos2020unraveling} also performed a QM/MM simulation on the covalent complex of Ac-Ser-Ala-Val-Leu-Gln-Ser-Gly-Phe-NMe and the main protease.

\subsubsection{Poisson-Boltzmann and generalized Born models} \label{sect:PB_GB}

\begin{figure}[ht!]
	\centering
	\includegraphics[width=0.8\textwidth]{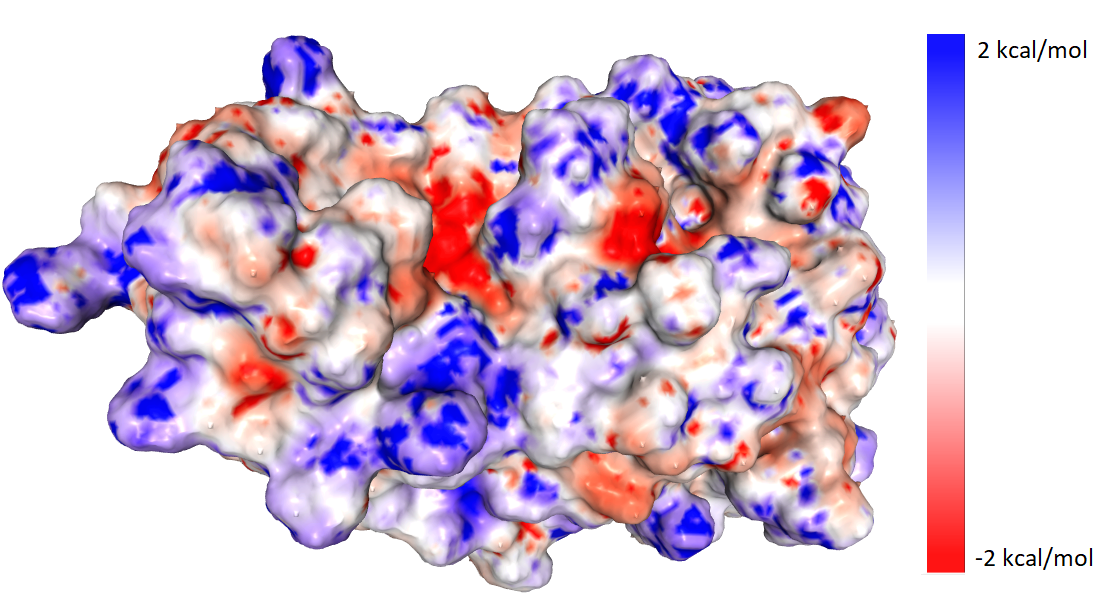}
	\caption{A electrostatic potential of the SARS-CoV-2 main protease based on the PB model.}
	\label{fig:esp}
\end{figure}

In biomolecular studies, electrostatic interactions are of paramount importance due to their ubiquitous existence in the protein-protein interactions, protein-ligand interactions, amino acid interactions, et al. Electrostatics potentials can be calculated using explicit or implicit solvent models as shown in  Figure \ref{fig:esp}. However, including explicit solvent models in free energy calculation is computationally expensive due to its detailed description of solvent effect. Implicit solvent models describe the solvent as a dielectric continuum, while the solute molecule is modeled by an atomic description \cite{davis1990electrostatics,sharp1990electrostatic,honig1995classical,roux1999implicit,rocchia2001extending}. A wide variety of two-scale implicit solvent models has been developed for electrostatic analysis, including Poisson-Boltzmann (PB) \cite{sharp1990electrostatic,fogolari2002poisson}, generalized Born (GB) \cite{dominy1999development,bashford2000generalized,onufriev2002effective,mongan2007generalized}, and polarized continuum \cite{tomasi2005quantum,cossi1996ab} models. GB models are approximations of PB models. GB models are faster but provide only heuristic estimates for electrostatic energies, while PB methods offer more accurate methods for electrostatic analysis \cite{bashford2000generalized,onufriev2000modification, jurrus2018improvements,chen2011mibpb,zhou2006high,chen2010differential,rocchia2001extending}.

\begin{figure}[htb]
	\setlength{\unitlength}{1cm}
	\begin{center}
		\includegraphics[width=2.in]{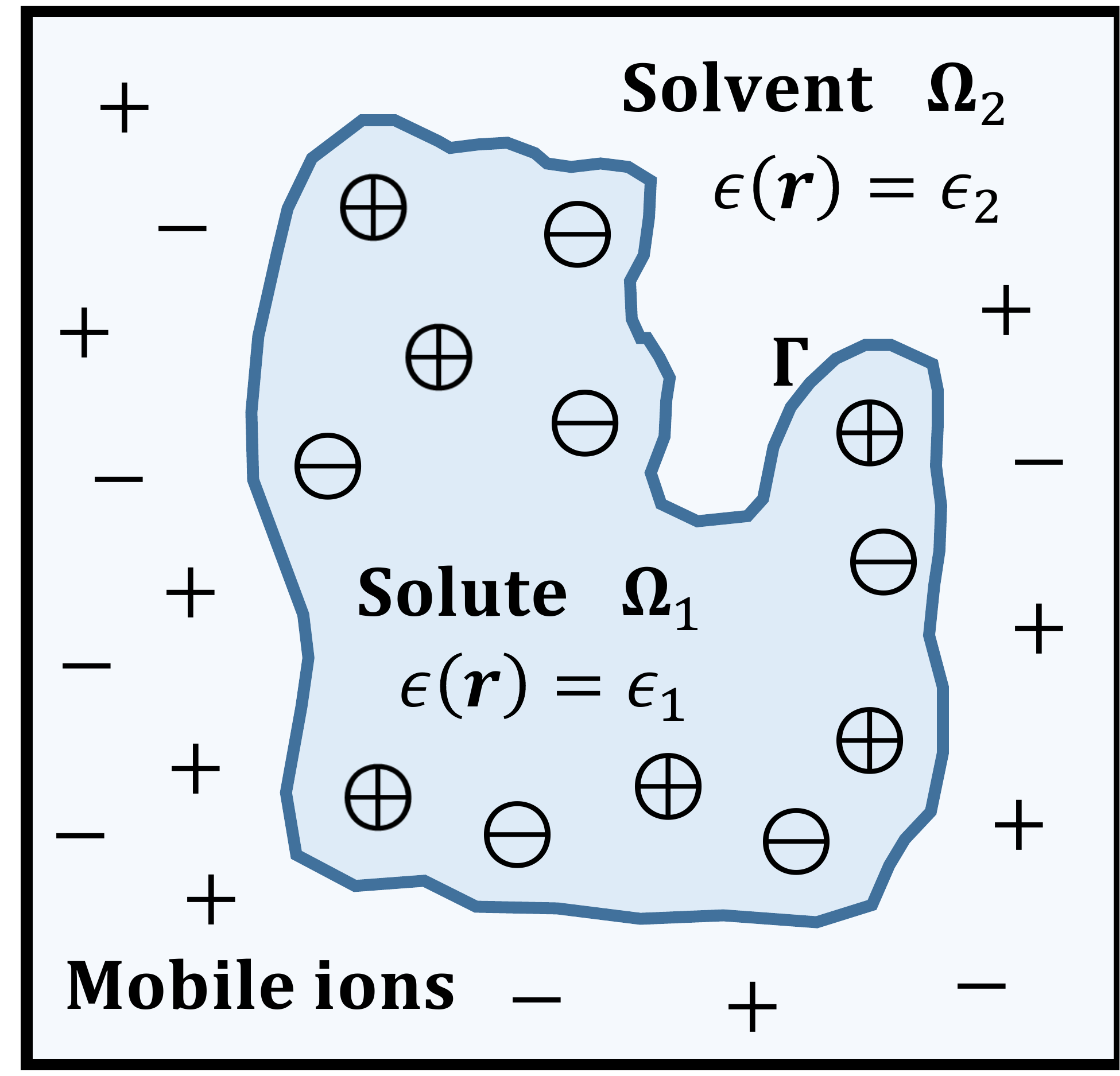}
		\caption{An illustration of  the Poisson-Boltzmann (PB) model, in which the molecular surface $\Gamma$ separates the computational domain into the solute region $\Omega_1$ and solvent region $\Omega_2$.}
		\label{fig_1}
	\end{center}
\end{figure}

As demonstrated in Fig. \ref{fig_1}, the PB model describes the two-scale treatment of the electrostatics within the interior domain $\Omega_1$ containing the solute biomolecule with fixed charges and the exterior domain $\Omega_2$ containing the solvent and dissolved ions.  The interface $\Gamma$ separates the biomolecular domain and solvent domain. While various surface models are available, the most commonly used ones are the solvent excluded surface or molecular surface.

A biomolecule in domain $\Omega_1$ consists of a set of atomic charges $q_k$ located at atomic centers ${\bf r}_k$ for $k=1,...,N_c$, with $N_c$ as the total number of charges. In domain $\Omega_2$, the charge source density of mobile ions is approximated by the Boltzmann distribution. For simplicity, a linearized PB model is always applied: 
\begin{equation}
-\nabla \cdot \epsilon({\bf r}) \nabla \phi({\bf r}) + \epsilon_2 \kappa^2\phi({\bf r}) =
\sum_{k=1}^{N_c} q_k \delta({\bf r}-{\bf r}_k),
\label{eqNPBE}
\end{equation}
where  $\phi({\bf r})$ is the electrostatic potential, $\epsilon({\bf r})$ is a dielectric constant given by $\epsilon({\bf r})=\epsilon_1$ for ${\bf r} \in \Omega_1$ and $\epsilon({\bf r})=\epsilon_2$ for ${\bf r} \in \Omega_2$, and $\kappa$ is the inverse Debye length representing the ionic effective length.
For the PB equation to be well-posed, interface conditions on the molecular surface are needed:

\begin{equation}
\phi_1({\bf r}) = \phi_2({\bf r}),
\quad
\epsilon_1 \frac{\partial\phi_1({\bf r})}{\partial \bf{n}}=\epsilon_2 \frac{\partial \phi_2 ({\bf r})}{\partial {\bf n}},
\quad
{\bf r} \in \Gamma,
\label{eqInterface}
\end{equation}
where $\phi_1$ and $\phi_2$ are the limit values when approaching the interface from inside or outside the solute domain, and $\bf{n}$ is the outward unit normal vector on $\Gamma$. The  far-field boundary condition for the PB model is $\lim_{|{\bf r}|\rightarrow\infty}\phi({\bf r}) = 0$. The electrostatic solvation free energy can be obtained from  the PB model by 
\begin{equation}
\Delta G= \frac{1}{2}\sum_{k=1}^{N_c}q_k (\phi({\bf r}_k) -\phi_0({\bf r}_k) )
\label{solvationEnergy}
\end{equation}
where $\phi_0({\bf r}_k)$ is the solution of the PB equation as if there were no solvent-solute interface.

The GB model is devised to offer a relatively simple and efficient approach to calculate electrostatic solvation free energy. However, with an appropriate parametrization, a GB solver can be as accurate as a PB solver \cite{forouzesh2017grid}. The GB approximation of electrostatic solvation free energy can be expressed as,
\begin{equation}
\label{GB_eqn}
\Delta G^{\rm GB} \approx \sum_{ij}\Delta G^{\rm GB}_{ij} = -\frac{1}{2} \Big( \frac{1}{\epsilon_1}-\frac{1}{\epsilon_2} \Big) \frac{1}{1+\alpha\beta} \sum_{ij} q_i q_j \Big( \frac{1}{f_{ij}(r_{ij}, R_{i}, R_j)} + \frac{\alpha\beta}{A} \Big),
\end{equation}
where $R_i$ is the effective Born radius  of atom $i$, $r_{ij}$  is the distance between atoms $i$ and $j$, $\beta = \epsilon_1/\epsilon_2$, $\alpha = 0.571412$, and $A$ is the electrostatic size of the molecule. The function $f_{ij}$ is given as
\begin{equation}
\label{f_ij}
f_{ij} = \sqrt{r^2_{ij}+R_iR_j {\rm exp}\Big( -\frac{r^2_{ij}}{4R_iR_j} \Big)}.
\end{equation}
The effective Born radii $R_i$ is calculated by the following boundary integral:
\begin{equation}
\label{eqn:Born_radii}
R^{-1}_i = {\Big( -\frac{1}{4\pi} \oint_{\Gamma } \frac{{\bf r}-{\bf r}_i}{|{\bf r}-{\bf r}_i|^6} \cdot \text{d}{\bf S} \Big)}^{1/3}. 
\end{equation}

Due to its success in describing biomolecular systems, the PB and GB models have attracted a wide attention in both mathematical and biophysical communities \cite{gilson1988calculating,zhou2008highly,ren2012biomolecular}. Meanwhile much effort has been given to the development of accurate, efficient, reliable, and robust PB solvers. A large number of methods have been proposed in the literature, including the finite difference method (FDM)\cite{jo2008pbeq}, finite element method (FEM)\cite{baker2001adaptive}, and boundary element method (BEM)\cite{juffer1991electric}. The emblematic solvers in this category are MM-PBSA\cite{wang2008poisson}, Delphi\cite{li2012delphi,rocchia2002rapid}, ABPS\cite{baker2001electrostatics,jurrus2018improvements}, MPIBP\cite{zhou2008highly,geng2007treatment,chen2011mibpb}, CHARM PBEQ\cite{jo2008pbeq}, and TABIPB\cite{geng2013treecode,chen2018preconditioning}.

PB and GB models have been applied to the SARS-CoV-2 studies including protein-ligand binding and protein-protein binding energetics. Adaptive Poisson-Boltzmann solver \cite{baker2001electrostatics,jurrus2018improvements} is one of the most popular PB solvers. By using it, Su et al.\cite{shu2020potential}, Rosario et al. \cite{rosario2020computational}, Amin et al.\cite{amin2020comparing}, and Morton et al. \cite{morton2020electrostatic} calculated electrostatic potentials of the S protein and ACE2 complex. Su et al.\cite{shu2020potential} also calculated that of the main protease. Jin et al.\cite{jin2020virus} and Xi et al.\cite{xi2020virus} studied human Furin, which is involved in the SARS-CoV-2 binding. Zhang et al.\cite{zhang2020whole} calculated the electrostatic potential of SARS-CoV-2 SUD-core dimer. Nerli et al.\cite{nerli2020structure} studied the electrostatic surface potentials of some SARS-CoV-2 antigens. Lu et al.\cite{lu2020genetic} investigated multiple other targets. Another popular software is DELPHI \cite{rocchia2002rapid}. Ali et al.\cite{ali2020ace2} solved the S protein and ACE2 complex using it. Moreover, AQUASOL\cite{koehl2010aquasol} can solve the dipolar nonlinear Poisson-Boltzmann-Langevin equation, Smaoui et al.\cite{smaoui2020unraveling} used AQUASOL to study the S protein RBD.

\subsubsection{Gibbs-Helmholtz equation}
The Gibbs-Helmholtz equation describes the thermodynamics calculating changes in the Gibbs energy of a system as a function of temperature. It is a separable differential equation which is given as 
\begin{equation}
\bigg(\frac{\partial (\Delta G/T)}{\partial T}\bigg)_{p} = \frac{-\Delta H}{T^2}
\label{eq:gibbs-helmholtz}
\end{equation}
where $\Delta G$ is the change in Gibbs free energy, $\Delta H$ is the enthalpy change, $T$ is the absolute temperature, and $p$ is the constant pressure. 

In the study of nucelocapsid protein (N-protein) of SARS, it was shown that the N-protein's maximum conformational stability near pH 9.0, and the oligomer dissociation and protein unfolding occur simultaneously \cite{luo2004vitro}. In the denaturation of the N-protein by chemicals, the free energy changes ($\Delta G$) of unfolding at temperature ($T$) is calculated by the solution of Gibbs-Helmholtz equation \cite{brumano2000thermodynamics}
\begin{equation}
\Delta G(T) = \Delta H_m(1-T/T_m)-\Delta C_p[(T_m-T)+T\ln(T/T_m)]
\end{equation}
where $T_m$ is the transition temperature, $\Delta H_m$ is the enthalpy of unfolding at $T_m$, and $\Delta C_p$ is the heat capacity change. The Gibbs free energy, $\Delta G(T)$ of unfolding is applied to estimated the protein stability in enduring the denaturants, where high $\Delta G(T)$ means the protein might be more stable against denaturant \cite{deshpande2003equilibrium}.

\subsection{Monte Carlo (MC) methods.}
MC methods are a broad class of computational algorithms that rely on repeated random sampling to obtain optimized numerical results. In principle, Monte Carlo methods can be used to solve any problem having a probabilistic distribution \cite{kroese2014monte}. When the probability distribution of the variable is parametrized, mathematicians often use a Markov chain Monte Carlo (MCMC) sampler \cite{hastings1970monte}: the central idea is to design a judicious Markov chain model with a prescribed stationary probability distribution. By the ergodic theorem, the stationary distribution is approximated by the empirical measures of the random states of the MCMC sampler. 

Moreover, Metropolis Monte Carlo methods \cite{metropolis1953equation} are a branch of MC methods popularly used in molecular modeling. The essential idea is that, if the energy of a trial conformation is lower than or equal to the current energy, it will always be accepted; if the energy of a trial conformation is higher than that of the current energy, then it will be accepted with a probability determined by the Boltzmann (energy) distribution,
\[
P_{\rm accept}(m \rightarrow n)= 
\begin{cases}
{\rm exp}\left(-\frac{\Delta U_{nm}}{kT}\right),& \text{if } \Delta U_{nm} > 1\\
1,              & \text{if } \Delta U_{nm} \leq 1,\\
\end{cases}
\] 
where $m$ is the current conformation, $n$ is the new conformation, $P_{\rm accept}(m \rightarrow n)$ is the probability to accept the new conformation, $U_{nm}$ is the energy difference between $n$ and $m$, $k$ is the Boltzmann constant, and $T$ is temperature. Therefore, the evolution of molecular conformations can be simulated. 

There are several aspects of MC applications regarding SARS-CoV-2.

\subsubsection{Applications to molecular modeling.}

\subparagraph{1. The whole virus.} 

Francis et al. \cite{francis2020monte} performed a MC simulation of ionizing radiation damage to the SARS-CoV-2 and found that $\gamma$-rays  produced  significant  S protein  damage,  but much less  membrane  damage. Thus,  they proposed  $\gamma$-rays as  a  new effective  tool  to develop inactivated  vaccines. Cojutti et al. \cite{cojutti2020comparative} used a Metropolis MC sampling process to simulate a pharmacokinetic model of HIV drug darunavir against SARS-CoV-2. 

\subparagraph{2. The main protease.} 
Amamuddy et al.\cite{amamuddy2020impact} performed coarse-grained MC simulations of the SARS-CoV-2 main protease in the free and ligand-bound forms.  Toropov et al. \cite{toropov2020sars} used MC simulations to search weights of the QSAR model about the main protease inhibitory activity of aromatic disulfide compounds. Sheik et al. \cite{sheik2020impact} investigated mutation-induced conformational changes of the main protease by coarse-grained MC models.

\subparagraph{3. The S protein.} 
Othman et al. \cite{othman2020interaction} ran MC simulations to predict the flexibility of the S protein. Wong et al. \cite{wong2020assessing} assessed the mutation impact on the structure of the S protein via a sequential MC model. Polydorides et al. \cite{polydorides2020computational} used MC simulations to collect the S protein mutations that could enhance its binding to ACE2. Amin et al. \cite{amin2020comparing} used  MC simulations to sample the protonation states of the amino acids. The generated conformer occupancies based on Boltzmann distributions were used to calculate the electrostatic and van der Waals interactions between the SARS-CoV-2 and the ACE2. Bai et al. \cite{bai2020critical} used a MC proton transfer (MCPT) method to determine the charge configuration of all ionizable residues so that the coarse-grained free energy of each protein configuration could be obtained. Giron et al. \cite{giron2020interactions} performed coarse-grained MC simulations to calculate antibodies 80R, CR3022, m396, and F26G19's binding processes and binding free energies to the S protein. Becerra et al. \cite{becerra2020sars} applied MC simulations to optimize the structures of mutated S proteins. Huang et al. \cite{huang2020novo} performed MC simulations to generate some potential peptide-based inhibitors against the S protein.

\subparagraph{4. Other targets.} 
Amin et al. \cite{amin2020chemical} built a MC-based QSAR model to study the potency of some in-house molecules against the PLpro. Cubuk et al. \cite{cubuk2020sars} sampled different conformations of the N protein. Pandey et al. \cite{pandey2020thermal} used coarse-grained MC simulations to investigate the structural dynamics of the E protein at multiple temperatures.  Wu et al. \cite{wu2020analysis} applied  MC simulations to optimize the structures of the main protease, PLpro, and RdRp.

\subsubsection{Applications to gene evolution.} 
Most investigations focused on the full genome of the SARS-CoV-2. Lai et al. \cite{lai2020early}, Nie et al.\cite{nie2020phylogenetic}, Li et al. \cite{li2020bayesian}, Koyama et al.\cite{koyama2020variant}, and Khan et al.\cite{khan2020phylogenetic} aimed to investigate the temporal origin, the rate of viral evolution, and the population dynamics of the virus globally using a Bayesian MCMC simulation. Li et al. \cite{li2020evolutionary} implemented cross-species gene analysis using the MCMC method, revealing that human SARS-CoV-2 is close to Bat CoV. Nabil et al. \cite{nabil2020transmission} employed  MC-based phylogenetic analysis to deduce the SARS-CoV-2 gene transmission route among China, Italy, and Spain. Castells et al.\cite{castells2020evidence} performed a MCMC analysis of complete genome sequences of SAR strains recently isolated in different regions of the world (i.e., Europe, North America, South America, and Southeast Asia). Using MC models, Zehender et al. \cite{zehender2020genomic} and Xavier et al. \cite{xavier2020ongoing} studied  SARS-CoV-2 in Italy and the city Minas Gerais of Brazil, respectively. Rice et al. \cite{rice2020evidence} used MC simulations to calculate mutation matrices. Andonegui et al. \cite{andonegui2020molecular} used MC sampling to estimate relative samples of RNA transcripts. Makhoul et al. \cite{makhoul2020analyzing} carried out MC sampling to simulate random polymerase chain reaction (PCR) test.

Other works consider the genome for some proteins in SARS-CoV-2. Flores et al.\cite{flores2020receptor} conducted MC-based phylogenetic analysis, suggesting that SARS-CoV-2 S protein resulted from ancestral recombination between the bat-CoV RaTG13 and the pangolin-CoV MP789. Sarkar et al.\cite{sarkar2020structural} used a MC procedure to test homogeneity among the sequence of the E proteins.

\subsubsection{Applications on virus transmission.}
The serial interval is defined as the duration between the symptom-onset time of the infector and that of the infectee. Ali et al. \cite{ali2020serial} built a MCMC model to estimate the serial interval of SARS-CoV-2, which concluded that this serial interval was shortened over time by nonpharmaceutical interventions. Peak et al. \cite{peak2020individual} also focused on the serial interval. They studied sequential MC models with different serial intervals. Miller et al. \cite{miller2020transmission} applied an MC simulation to predict the emission rate in one super spreading event. Kucharski et al. \cite{kucharski2020early}  ran sequential MC simulations to  infer the transmission rate over time in Wuhan, China. Silverman et al. \cite{silverman2020using} and Lavezzo et al. \cite{lavezzo2020suppression} used MC models to investigate the prevalence of SARS-CoV-2 in the US and the Italian municipality Vo'. Fu et al. \cite{fu2020simulating} performed a MC-based approach to estimate daily cumulative numbers of confirmed cases of SARS-CoV-2. Mizumoto et al. \cite{mizumoto2020estimating} used the Hamiltonian MC algorithm to calculate the probability of being asymptomatic given infection  as well as  the  infection  time  of  each  individual. Yang et al. \cite{yang2020analysis} built a MC infection model considering age structures. 

The role of suppression strategies in SARS-CoV-2 transmission was also studied by MC models. Yang et al. \cite{yang2020impact} focused on the impact of household quarantine on SARS-Cov-2 infection. Chu et al. \cite{chu2020physical} simulated the effect of physical distancing, face masks, and eye protection to prevent person-to-person transmission. Girona et al. \cite{girona2020confinement}'s MC model tried to answer the question: ``how long should suppression strategies last to be effective to avoid quick rebounds in the transmission once interventions are relaxed".

MC models were applied to simulate other aspects of SARS-CoV-2 transmission as well. Ahmed et al. \cite{ahmed2020first} used MC simulations to deduce the number of infected individuals from the viral RNA copy numbers observed in the waste water. Khalil et al. \cite{khalil2020sars} built an MC model to reveal SARS-CoV-2 infection in pregnancy. Vuorinen et al. \cite{vuorinen2020modelling} modeled aerosol transport and virus exposure with MC simulations in different public indoor environments.

\subsubsection{Miscellaneous.} 
Pascual et al. \cite{pascualmonte} employed a MC model to simulate the SARS-CoV-2 virus replication cycle. Jain et al. \cite{jain2020sars} used MC simulations to study SARS-CoV-2's impact on the backlog of orthopedic surgery, which concluded that it would take over 2 years to end the cumulative backlog.

\subsection{Structural bioinformatics}

\subsubsection{Protein pocket detection.}

The  detection  and  characterization  of protein pockets and cavities is a critical issue in molecular biology studies. Pocket detection algorithms can be classified as grid-based and grid-free approaches \cite{le2009fpocket,zhao2018protein}.  Grid-based approaches embed  proteins in  3D grids and then search for grid points that satisfy  some  conditions. Grid-free ones include methods based on probe (sphere) or the concepts of Voronoi diagrams. Zhao et al developed differential geometry and algebraic topology based protein pocket detections using convex hull surface evolution and associated Reeb graph \cite{zhao2018protein}.

Gervasoni et al. \cite{gervasoni2020comprehensive} evaluated the performance of the pocket-detecting algorithms Fpocket \cite{le2009fpocket} and PLANTS \cite{korb2009empirical} on 12 different SARS-CoV-2 proteins with an accuracy of 0.97. Manfredonia et al. \cite{manfredonia2020genome} predicted the 3D structure of SARS-CoV-2 RNA by coarse-grained modeling and detected potential pockets. Boldrini et al.\cite{boldrini2020identification} constructed a ACE2 folding pathway through ratchet-and-pawl MD (rMD) \cite{du1998transition} and high temperature MD simulations, identified intermediates, and predicted potential druggable pockets on some late intermediates. Sheik et al.\cite{sheik2020impact} predicted potential binding pockets of the main protease.

\subsubsection{Homology-modeling based protein structure prediction}

Homology modeling constructs an atomic-resolution model of the ``target" protein from its amino acid sequence based on experimental 3D structures of related homologous proteins (the ``templates") \cite{schwede2003swiss}. Homology modeling relies on identifying one or more known protein structures likely to resemble the structure of the query sequence, and producing an alignment that maps residues in the query sequence to residues in the template sequence \cite{chothia1986relation}.

Because the 3D experimental structures of SARS-CoV-2 proteins were largely unknown at the early stage of the epidemic, homology modeling was widely applied to predict the 3D structures of SARS-CoV-2 proteins, such as the main protease \cite{abdelrheem2020inhibitory, jimenez2020virtual, calligari2020molecular,chen2020prediction,wang2020virtual,milenkovic2020several,shanker2020whole, khan2020targeting, khan2020targeting, mirza2020structural}, S protein \cite{hall2020search,luan2020spike,hussain2020structural,sigrist2020potential,baig2020elucidation,jaimes2020phylogenetic, ortega2020role, ling2020silico, singh2020structure, choudhary2020identification, zhou2020novel, kharisma2020construction, shanmugarajan2020curcumin, feng2020eltrombopag, li2020delving,park2020spike,liu2020computational,parray2020identification, kumar2020structural, su2020biological}, RdRp \cite{aftab2020analysis, arya2020potential, zhang2020binding, zhang2020structural, lung2020potential, ruan2020potential, venkateshan2020azafluorene, elfiky2020ribavirin, beg2020anti, neogi2020feasibility}, PLpro\cite{hu2020prediction}, E protein \cite{de2020improved,sarkar2020structural,oany2020design,alam2020functional, azeez2020state}, N protein \cite{azeez2020state}, and others \cite{dong2020guideline, martin2020repurposing, wu2020analysis,culletta2020exploring,zhou2020bioaider,hijikata2020knowledge,meng2020insert,chen2020genomics, liu2020potential, selvaraj2020structure, benvenuto2020evolutionary,borgio2020state,mirza2020structural,cardenas2020exclusive,khan2020targeting}.

Some human  proteins that interact with SARS-CoV-2 were also predicted, such as ACE2 \cite{ekins2020deja}, TMPRSS2 \cite{rensi2020homology, sonawane2020homology, huggins2020structural}, and CD147 \cite{xia2020tumor}. Some 3D structures of vaccine proteins  \cite{peele2020design,samad2020designing} were also built by homology modeling.

\subsubsection{Quantitative structure-activity relationship models (QSAR)}

QSAR models refer to regression or classification models to predict the physicochemical, biological, and environmental properties of compounds from the knowledge of their chemical structure \cite{gramatica2007principles}. In QSAR modeling, the predictors consist of physico-chemical properties and theoretical molecular descriptors of chemicals; the QSAR response-variable could be a biological activity of the chemicals. Building a QSAR model includes two steps: first, summarizing a supposed relationship between chemical structures and biological activity in a data-set of chemicals and then, using QSAR models to predict the activities of new chemicals \cite{roy2015primer}.

To identify potential main protease inhibitors, Ghaleb et al. \cite{ghaleb2020silico} and Acharya et al.\cite{acharya2020kinase} applied 3D QSAR models: Ghaleb et al.'s 3D model was based on comparative molecular similarity indices analysis (CoMSIA); Acharya et al.'s 3D model was based on pharmacophores. More works used 2D QSAR models: Alves et al. \cite{alves2020qsar} used the random forest algorithm to build the model; Kumar et al.\cite{kumar2020development} and Masand et al.\cite{masand2020extending, masand2020structure} used genetic algorithms; Basu et al.\cite{basu2020novel} used support vector machine; Islam et al. \cite{islam2020molecular}, De et al. \cite{de2020silico1}, and As et al. \cite{as2020theoretical} used multiple linear regression; Toropov et al. \cite{toropov2020sars} used linear models. Moreover, Ghost et al.  \cite{ghosh2020chemical} built a Monte Carlo-based classification model.    

Inhibitors against other targets were also investigated. QSAR models based on multiple linear regression and Monte Carlo classification were constructed by Laskar et al.\cite{laskar2020search} and Amin et al.\cite{amin2020chemical}  against PLpro. Against the main protease and RdRp, Ahmed et al. \cite{ahmed2020investigating} built a QSAR model followed partial-least-square regression. Borquaye et al. \cite{borquaye2020alkaloids} used multiple linear regression.

\subsection{Machine learning and deep learning}

\subsubsection{Linear regression}

The linear regression is one of the basic  algorithms in machine learning. It can be used to solve the regression problem. We assume the training set is $\{(\mathbf{x}_i, y_i) | \mathbf{x}_i\in \mathbb{R}^m, y_i \in \mathbb{R}\}_{i=1}^{n}$. The predictor is  defined as 
\begin{equation}
\hat{\mathbf{y}}(\mathbf{x}) = \mathbf{w}^T\mathbf{x}+\mathbf{b},
\end{equation}
where $\mathbf{w}$ is the weights, $\mathbf{b}$ is the bias, and $\mathbf{w}^T$ represents the transpose of $\mathbf{w}$. The aim of the linear regression is to minimize the loss function, which can be defined as 
\begin{equation}
L = \frac{1}{2n}\sum_{i=1}^{n}( \hat{\mathbf{y}}_i - \mathbf{y}_i)^2.
\end{equation}
As a basic machine learning algorithm, linear regression can be commonly found in the literature related to COVID-19 research. In the framework of QSAR \cite{roy2015primer}, to identify potential SARS-CoV-2 main protease inhibitors, Toropov et al. \cite{toropov2020sars}, Islam et al. \cite{islam2020molecular}, De et al. \cite{de2020silico1}, and As et al. \cite{as2020theoretical} used linear or multiple linear regression methods to build prediction models. Other SARS-CoV-2 targets were also investigated: Against papain-like protease, QSAR models based on multiple linear regression were constructed by Laskar et al. \cite{laskar2020search} against the main protease and RdRp. Ahmed et al. \cite{ahmed2020investigating}'s QSAR model was based on partial-least-square regression. Borquaye et al. \cite{borquaye2020alkaloids} used multiple linear regression.

More examples involve applying linear regression to calculate the correlation coefficient and predict different types of dependent variable values. Becerra et al. \cite{becerra2020sars} and Korber et al. \cite{korber2020tracking} calculated infectivity of the SARS-CoV-2 S protein D614G mutation percentage by employing linear regression. In Mckay et al.'s work about vaccine candidates \cite{mckay2020self}, a linear regression was performed between  SARS-CoV-2 IgG and viral neutralization. Bunyavanich et al. \cite{bunyavanich2020nasal} built linear regression models between ACE2 gene expression and age. Ma et al.\cite{ma2020effect} used linear regression to analyze the  relationship  between  serum T:LH  ratio and  the  clinical  characteristics  of  COVID-19 patients. Jary et al. \cite{jary2020evolution} applied linear models to study the gene evolution of SARS-CoV-2.

\subsubsection{Logistic regression}
Logistic regression is an algorithm designed  for solving classification problems. Assume the training set is $\{(\mathbf{x}_i, y_i) | \mathbf{x}_i\in \mathbb{R}^m, y_i \in\mathbb{Z} \}_{i=1}^{n}$. The predictor of the logistic regression is 
\begin{equation}
\hat{\mathbf{y}}(\mathbf{x}) = \frac{1}{1+e^{-\mathbf{w}^T\mathbf{x}+\mathbf{b}}},
\end{equation}
where $\mathbf{w}$ is the weights and $\mathbf{b}$ is the bias. The loss function can be defined by
\begin{equation}
L = -\frac{1}{n}\sum_{i=1}^{n}[-\mathbf{y}_i \text{log}(\hat{\mathbf{y}}_i) - (1-\mathbf{y}_i)\text{log}(1-\hat{\mathbf{y}}_i)].
\end{equation}

Ayouba et al. \cite{ayouba2020multiplex} used logistic regression to represent the dynamics of IgG response to the S protein, N protein, or both antigens at the same time since symptoms onset. In addition, researchers stated that the publicly shared CD8$^{+}$ might be used as a potential biomarker of SARS-CoV-2 infection at high specificity and sensitivity by applying the logistic regression in \cite{levantovsky2020shared}.

\subsubsection{\texorpdfstring{$k$}{k}-nearest neighbors}
The $k$-nearest neighbors algorithm ($k$-NN) is a non-parametric technique proposed by Thomas Cover and P. E. Hart in 1967 \cite{cover1967nearest}. $k$-NN can be used for solving both regression and classification problems \cite{altman1992introduction}, and it is sensitive to the local structure of the data. The \autoref{alg:KNN} below shows the pseudo code for the $k$-NN. Different distance metrics can be employed in the $k$-NN algorithm such as Euclidean distance, Manhattan distance, Minkowski distance, Chebyshev distance, natural log distance, generalized exponential distance, generalized Lorentzian distance, Camberra distance, quadratic distance, and mahalanobis distance.

\begin{algorithm}[ht!]
	\SetKwData{Left}{left}\SetKwData{This}{this}\SetKwData{Up}{up}
	\SetKwFunction{Union}{Union}\SetKwFunction{FindCompress}{FindCompress}
	\SetKwInOut{Input}{Input}\SetKwInOut{Output}{Output}
	\SetAlgoLined
	
	\Input{$k$: The nearest data points;\\$\mathbf{x}$: The feature of the training set with shape $\mathbb{R}^{n\times m}$;\\ $\mathbf{y}$: labels with shape $\mathbb{R}^{n\times 1}$;\\ $\mathbf{x^{\prime}}$: unknown samples with shape $\mathbb{R}^{s\times m}$.}
	\Output{The predicted labels of $\mathbf{x^{\prime}}$ with shape $\mathbb{R}^{s\times 1}$.}
	\For{$i=1$ \KwTo $s$}{
		Compute the distance $d(\mathbf{x},\mathbf{x}^{\prime}_i)$ with shape $n\times 1$;\\Sort the distance values in ascending order;\\Choose the top $k$ rows from the sorted array\;
		\If{Classification}{
			Assign the label of $\mathbf{x}^{\prime}_i$ based on the most frequent label of these $k$ rows;} 
		\If{Regression}{
			Assign the label of $\mathbf{x}^{\prime}_i$ based on the average label(value) of these $k$ rows.}
	}
	\caption{$k$-NN algorithm}
	\label{alg:KNN}
\end{algorithm}

The classifier can be built by using the $k$-NN algorithm. Granholm et al.\cite{granholm2020metagenomic} used $k$-NN as a classifier to distinguish the SARS-CoV-2 virus genome from other viruses, bacteria, and eukaryotes. Similarly, Naeem et al.\cite{naeem2020diagnostic} trained a $k$-NN model to distinguish the SARS-CoV-2 genome from SARS-CoV genome and MERS genome.  Moreover, AllerTOP v.2.0 classified allergens and non-allergens based on the $k$-NN method with an accuracy of 88.7\% \cite{mukherjee2020immunoinformatics}.  Furthermore, the $k$-NN algorithm can be employed to classify the human protein sequences of COVID-19 according to country \cite{alkady2020computational}. Stanley et al.\cite{stanley2020coronavirus} classified cells  using the $k$-NN algorithm.

\subsubsection{Support vector machine} 
The support vector machine (SVM) was developed by Vapnik and his colleagues, which can be used for both classification and regression analysis \cite{cortes1995support,drucker1997support}. For the classification problem, assume the training set is $\mathbf{x} = \{\mathbf{x}_1, \mathbf{x}_2, \cdots,\mathbf{x}_i,\cdots, \mathbf{x}_n\}$ with $\mathbf{x}_i \in \mathbb{R}^{1\times m}$, and the label of the training set is $\mathbf{y} = \{y_1, y_2, \cdots, y_i, \cdots, y_n\}\in \mathbb{R}^{n\times 1}$ with $y_i \in \{-1, 1\}$. The predictor of the SVM will be $\hat{\mathbf{y}} = \mathbf{w}^{T}\mathbf{x}+ \mathbf{b}$. Here, $\mathbf{w}$ is the weights and $\mathbf{b}$ is the bias. If the training set is linearly separable, the aim is to minimize $\|\mathbf{w}\|$ subject to $y_i(\mathbf{w}^{T}\mathbf{x}_i - \mathbf{b}) \geq 1$. If the training set is not linearly separable, then the hinge loss function $\max(0, 1 - y_i(\mathbf{w}^{T}\mathbf{x}_i - \mathbf{b}))$ will be involved. The aim of SVM is to minimize
\begin{equation}
\|\mathbf{w}\| + \lambda \sum_{i=1}^{n}\max(0, 1 - y_i(\mathbf{w}^{T}\mathbf{x}_i - \mathbf{b}),
\end{equation}
where $\lambda$ is the regularization term. For the regression problem, the aim is to minimize $\|\mathbf{w}\|$ subject to $|y_i - \langle \mathbf{w}, \mathbf{x}_i \rangle - b |\leq \epsilon$.

The SVM mentioned above is a linear classifier. To design a non-linear classifier, the kernel trick is employed to maximize margin hyper-planes. The feature of the kernel SVM will become $k(\mathbf{x},\mathbf{x})$, where the commonly used kernels are the linear kernel $k(\mathbf{x}, \mathbf{z}) = \mathbf{x}^T\mathbf{z}$, the polynomial kernel defined by $k(\mathbf{x}, \mathbf{z}) =( \alpha\mathbf{x}^T\mathbf{z}+r)^d$, the radial basis function kernel (RBF) $k(\mathbf{x}, \mathbf{z}) = e^{-(\frac{\|\mathbf{x}-\mathbf{z}\|}{\sigma})^{\mu}}$, and the sigmoid kernel denoted as $k(\mathbf{x}, \mathbf{z}) = \frac{1}{1+e^{-\gamma\mathbf{x}^T\mathbf{z}}}$.

Through SVM models, Basu et al.\cite{basu2020novel} and Ghosh et al.\cite{ghosh2020chemical} identified potential main protease inhibitors. Kowalewski et al. \cite{kowalewski2020predicting} screened near 100,000 FDA-registered chemicals and approved drugs as well as about 14 million other purchasable chemicals against multiple SARS-CoV-2 targets. 

Additionally, Dutta et al.\cite{dutta2020novel} predicted a novel peptide analogue of S protein using SVM models implemented by the  AVPred antiviral peptide prediction server.  Yadav et al. \cite{yadav2020full}, Peele et al.\cite{peele2020design},  Rahman et al.\cite{rahman2020vaccine}, Martin et al.\cite{martin2020rational}, and  \cite{yang2020silico} used an SVM-based online server to identify epitopes. In Beg's work \cite{beg2020computational}, the Ease-MM web server based on the SVM algorithm was applied to predict protein stability. Kumar et al.\cite{kumar2020exploring}, Kibria et al.\cite{kibria2020multi}, Rajput et al.\cite{rajput2020engineering}, and Chauhan et al.\cite{chauhan2020excavating} used SVM-based web servers to predict allergenicity of proposed epitopes or vaccines. 

\subsubsection{Decision trees}

Decision trees (DTs) are a basic machine method, which is used to perform both classification and regression model by representing the attribute of the data using a flowchart-like structure. Decision trees were used commonly in diagnosis of COVID-19. In Ref. \cite{kadry2020development,yoo2020deep,nour2020novel,ardakani2020covidiag,warman2020interpretable}, authors used decision trees in conjunction with a feature detection model to diagnose COVID-19 from CT scans \cite{kadry2020development,ardakani2020covidiag,warman2020interpretable} or chest X-rays \cite{yoo2020deep,nour2020novel}. Diagnosis was also done using physical symptom information \cite{otoom2020iot,zimmerman2020proposed} and demographic information \cite{zimmerman2020proposed,bhatnagar2020descriptive} with decision tree models.
Decision trees were also used to make predictors of case severity of COVID-19, using physical data from infected individuals \cite{pourhomayoun2020predicting,assaf2020utilization,muhammad2020predictive} and using physical data in conjunction with demographic data \cite{chassagnon2020ai} with the intent that the predictors will prove useful for hospitals' allocation of resources.
Other predictors of case severity were constructed using the clarity of the decision tree algorithm to make conclusions on which factors most influenced case severity \cite{gong2020tool,petrilli2020factors,elshazli2020diagnostic,toraih2020association,he2020comment,chakraborty2020real,chen2020characteristics,massie2020identifying}. Age was found to be a significant factor in predicting an individual case's outcome \cite{gong2020tool,petrilli2020factors,toraih2020association}, as was obesity \cite{petrilli2020factors,he2020comment}. Blood data like oxygenation \cite{petrilli2020factors}, troponin \cite{petrilli2020factors,toraih2020association}, aspartate aminotransferase levels (AST) \cite{toraih2020association}, lymphocyte and neutrophil count \cite{elshazli2020diagnostic,he2020comment,chen2020characteristics}, procalcitonin \cite{petrilli2020factors,elshazli2020diagnostic}, C-reactive protein \cite{petrilli2020factors,elshazli2020diagnostic}, D-dimer levels \cite{elshazli2020diagnostic}, and white blood cell count \cite{elshazli2020diagnostic} were also linked to case severity. In Ref. \cite{massie2020identifying}, authors used a decision tree model to investigate the benefit or harm of kidney transplantation during the pandemic, concluding that despite the risks, kidney transplantation provided survival benefit in most scenarios. Authors in \cite{chakraborty2020real} created a case fatality rate regression tree, from which it was concluded that total number of cases, percentage of people older than 65 years, total population, doctors per 1000 people, lockdown period, and hospital beds per 1000 people were significant predictors is case fatality rate.
Decision tree models were constructed in order to understand environmental effects on COVID-19 spread \cite{keshavarzi2020coronavirus,gupta2020developing} and severity \cite{malki2020association}.
In Refs. \cite{erraissi2020machine,ribeiro2020short,kumar2020preparedness}, authors used decision trees to create short term predictors of disease spread and fatalities.
The disease genome was classified using decision trees, one used to conclude the zoonotic source of Pangolin in \cite{el2020study} and one created to provide a reliable option for taxonomic classification \cite{randhawa2020machine}.
Others used decision trees to evaluate the effectiveness of lockdowns and shelter-in-place orders \cite{al2020measurement,utama2020optimizing}.
In \cite{loey2020hybrid}, authors used a decision tree model in conjunction with a feature detection model to determine if a person is wearing a mask. Decision trees were also used for language processing to classify sentiments \cite{sethi2020sentiment} or information quality \cite{hussein2020real} in social media posts about COVID-19.

\subsubsection{Random forest}
Random Forest (RF) \cite{ho1995random} is an ensemble learning method, which is designed to reduce the over-fitting in the original decision trees. Both classification and regression problems are suitable for  random forest models. In Refs. \cite{pourhomayoun2020predicting,sarkar2020machine,mei2020artificial,iwendi2020covid,shi2020large,tang2020severity,chen2020hypertension,farid2020novel,di2020common,cheng2020using,gong2020tool,assaf2020utilization,muhammad2020predictive,fernandes2020multipurpose,yao2020severity,chassagnon2020ai,bae2020predicting,navlakha2020projecting,oniani2020constructing}, authors used Random Forest to predict severity of individual cases of COVID-19 from physical, demographic, or geographic data. From CT images in particular, a number of sources used RF algorithms in order to diagnose a COVID-19 infection more rapidly than the available testing procedures \cite{mei2020artificial,shi2020large,tang2020severity,farid2020novel,wu2020rapid,sun2020adaptive,alqudah2020covid,chen2020early,kumar2020accurate,kadry2020development,guo2020improved}.
Many authors also used the nature of the Random Forest algorithm to determine which of the provided data fields were most important in determining severity or mortality of COVID-19. Age was found to be a highly relevant factor in \cite{sarkar2020machine,chen2020hypertension,di2020common,gong2020tool,chen2020early,salas2020data,heldt2020early,sainaghi2020fatality}. Male gender was associated with higher risk of mortality in \cite{sainaghi2020fatality,chen2020hypertension}. In \cite{di2020common,gong2020tool,chen2020early}, C-reactive protein level was determined to be a major factor in predicting recovery. Climate was determined to be a factor in COVID-19 spread and severity in Russia in \cite{pramanik2020climatic}, while weather factors were also associated with severity in \cite{malki2020association}. Other notable factors were high population density in \cite{mathur2020explainable}, D-dimer levels in \cite{li2020dynamic,chen2020early}, and other illnesses such as arterial hypertension, history of coronary artery disease (CAD), active cancer, atrial fibrillation, dementia and chronic kidney disease in \cite{sainaghi2020fatality}. 
Authors in \cite{batra2020screening} used RF to predict effectiveness of drugs and other therapeutic agents for treatment of COVID-19.
Short term prediction models of disease spread were created using Random Forest on environmental predictors in \cite{singh2020kalman,rao2020contextualizing,keshavarzi2020coronavirus}.
RF was used to predict cases and deaths in different geographic regions; in Russia in \cite{pramanik2020climatic}, in the United States in \cite{watson2020fusing}, in a region of Spain in \cite{benitez2020short}, in Iran in \cite{pourghasemi2020spatial}, in Morocco in \cite{erraissi2020machine}, and worldwide in \cite{yecsilkanat2020spatio}.
The authors in \cite{cobb2020examining,al2020measurement,haug2020ranking} used Random Forest to determine the effectiveness of social distancing and shelter-in-place orders at containing the spread of the virus.
RF was also used in language processing to determine public opinion and track the propagation of information on social media, specifically the Chinese Sina-Weibo in \cite{han2020using,li2020characterizing} and Twitter in \cite{sethi2020sentiment}.
The effects on the US stock market by COVID-19 were predicted by RF in \cite{dey2020quantifying}. Random Forest was also used to fill out datasets in \cite{zhang2020association,zhang2020hospital}.

\subsubsection{Gradient boost decision tree (GBDT)}
GBDT is a machine learning technique for regression and classification problems, which produces a prediction model in the form of an ensemble of decision trees \cite{mason2000boosting}. This ensemble of decision trees is built in a stage-wise fashion like other boosting methods. That is, algorithms optimize a cost function over function space by iteratively choosing a function that points in the negative gradient direction.

Many sources used GBDT in order to diagnose COVID-19 from CT images \cite{zhang2020clinically,kassani2020automatic}, chest X-rays \cite{kumar2020accurate}, blood test data \cite{de2020covid,soltan2020artificial,kukar2020covid}, or clinical/physical symptoms \cite{zoabi2020covid,shoer2020should}.
In Refs. \cite{fernandes2020multipurpose,yao2020severity,chassagnon2020ai,decaprio2020building,khanday2020machine,haimovich2020development,burdick2020prediction,estiri2020individualized,quiroz2020severity,souza2020predicting,vaid2020machine,zhou2020identifying,carr2020supplementing,wang2020clinical,barda2020developing,li2020development,eva2020development,karthikeyan2020machine}, authors used GBDT to create predictors of case severity/mortality, with the intent that the predictors will be used to aid in the decision making for distributing health resources.
Some authors trained GBDTs to predict case severity and used the nature of the algorithm to determine which data fields most impact the prediction \cite{mathur2020explainable,salas2020data,heldt2020early,vaid2020machine,zhou2020identifying,skorka2020macroecology,paul2020socio,yang2020dies,yan2020interpretable,wollenstein2020personalized,chowdhury2020early,zeng2020enhanced}. In  Refs.\cite{salas2020data,heldt2020early,vaid2020machine,zhou2020identifying,yang2020dies,wollenstein2020personalized,chowdhury2020early,zeng2020enhanced} it was found that age is one of the most important parameters for predicting case severity, with \cite{wollenstein2020personalized} finding age as a significant predictor in hospitalization, mortality, and ventilator need. Preexisting issues including hypertension, diabetes, immunosuppression, and respiratory illness were linked to case severity in \cite{zhou2020identifying,wollenstein2020personalized}. In refs. \cite{vaid2020machine,yan2020interpretable,chowdhury2020early}, lactic dehydrogenase (LDH) as well as C-reactive protein were important in prediction models. Other parameters included population density in \cite{mathur2020explainable,skorka2020macroecology}, measures of oxygenation status in \cite{heldt2020early}, coagulation parameters in \cite{vaid2020machine}, male sex in \cite{zhou2020identifying}, a country's number of tourists and gross domestic product in \cite{skorka2020macroecology}, socio-economic standing in \cite{paul2020socio}, neutrophils and lymphocye percentage in \cite{chowdhury2020early}, country-wise research sentiment and local weather conditions in \cite{zeng2020enhanced}.
Weather conditions were also linked to COVID-19 spread using a GBDT model in \cite{malki2020association}.
Gradient boosting was also used to predict cases and deaths in Para-Brazil \cite{torres2020quest} and worldwide \cite{liu2020boosting}.
In \cite{ong2020covid}, authors used GBDT to predict which proteins would likely make up an effective vaccine for COVID-19.
The effectiveness of social distancing measures were studied using GBDT in \cite{delen2020no}. In \cite{mowery2020improved}, the current spread and landscape of COVID-19 were assessed by integrating GBDT with lateral flow assays. A predictor for infection risk in nursing homes was constructed with gradient boosting in \cite{sun2020predicting}.
GBDT was employed for language processing in \cite{sethi2020sentiment,muthusami2020covid} to classify topics/sentiments of social media posts relating to COVID-19. Pandemic inspired lockdowns' effects on pollution were studied using gradient boosting in \cite{petetin2020meteorology,keller2020global}. Effects on psychological state among Chinese undergraduates were studied employing GBDT as well \cite{ge2020predicting}. Gao et al.\cite{gao2020repositioning}'s GBDT model repurposed 8565 approved or experimental drugs targeting the main protease, suggesting some existing drugs could be effective. Wang et al.\cite{wang2020decoding} used topology-based features and GBDT models to predict the NSP6 protein stability upon mutation.



\subsubsection{Artificial neural network (ANN)}
Artificial neural network (ANN) is a computational model inspired by the biological neural network that constitutes animal brains \cite{chen2019design}. ANN can be viewed as a weighted directed graph in which artificial neurons can be considered as nodes, and weights can be considered as the links between input and output nodes. ANN is designed for both regression and classification problems. We assume the training set is $\mathbf{x} = \{\mathbf{x}_1, \mathbf{x}_2, \cdots,\mathbf{x}_i,\cdots, \mathbf{x}_n\}$ with $\mathbf{x}_i \in \mathbb{R}^{1\times m}$. Here, $n$ is the number of samples, and $m$ represents the number of features. The label of the training set is $\mathbf{y} = \{y_1, y_2, \cdots, y_i, \cdots, y_n\}\in \mathbb{R}^{n\times 1}$. There are two main procedures in the ANN algorithm, the feed-forward and the back-propagation procedures. The feed-forward starts from the input layer to the first hidden layer. We define
\begin{equation}
\mathbf{z}_1 = f(\mathbf{x}\mathbf{W}_1+ \mathbf{b}_1),
\end{equation}
where $\mathbf{W}_1 \in \mathbb{R}^{m\times h_1}$ represents the weights from the input layer to the first hidden layer,  $\mathbf{b}_1 \in \mathbb{R}^{1\times h_1}$ represents the bias from the input layer to the first hidden layer; $h_1$ is the number of the neurons in the first hidden layer, and function $f$ represents the activation functions such as ReLu and Sigmoid function. Next, from the first hidden layer to the second hidden layer, we apply a similar function defined as: 
\begin{equation}
\mathbf{z}_2 = f(\mathbf{z_1}\mathbf{W}_2+ \mathbf{b}_2),
\end{equation}
where $\mathbf{W}_2 \in \mathbb{R}^{h_1\times h_2}$ and $\mathbf{b}_2 \in \mathbb{R}^{1\times h_2}$. Here, $h_2$ is the number of neurons in the second hidden layer. We repeat a similar procedure until we get to the output layer. Our predictor from the last hidden layer to the output layer is:
\begin{equation}
\hat{\mathbf{y}} = \mathbf{z_j}\mathbf{W}_j+ \mathbf{b}_j.
\end{equation}
where $\mathbf{W}_i \in \mathbb{R}^{h_j\times 1}$ and $\mathbf{b}_j \in \mathbb{R}^{1\times 1}$. $h_j$ is the number of neurons in the last hidden layer In the ANN. We use the cross-entropy loss to describe the cost function, which is defined as
\begin{equation}
L = -\sum_{i=1}^{n}\mathbf{y}_i \text{log}(\hat{\mathbf{y}}_i) .
\end{equation}
The ANN algorithm obtains the prediction via the feed-forward procedure and then minimizes the cross-entropy loss through the back-propagation procedure.

The typical ANN application to SARS-CoV-2 is to repurpose existing drugs and compounds or even generate new ones to treat SARS-CoV-2. Ton et al. \cite{ton2020rapid} developed a deep docking (DD) model, which provides fast prediction of docking scores from Glide or any other docking program, hence, enabling structure-based virtual screening of billions of purchasable molecules in a short time. The DD model relies on a deep neural network trained with docking scores of small random samples of molecules extracted from a large database to predict the scores of remaining molecules. Karki et al. \cite{karki2020predicting} predicted potential SARS-COV-2 drugs using a deep neural network framework, scale selection network (SSnet). SSnet was trained to predict drug binding affinity. Beck et al. \cite{beck2020predicting} used a pre-trained deep learning-based drug-target interaction model called molecule transformer-drug target interaction (MT-DTI) to identify commercially available drugs that could act on viral proteins of SARS-CoV-2. 

\subsubsection{Convolutional neural network (CNN)}
\begin{figure}[ht!]
	\centering
	\includegraphics[width=0.8\textwidth]{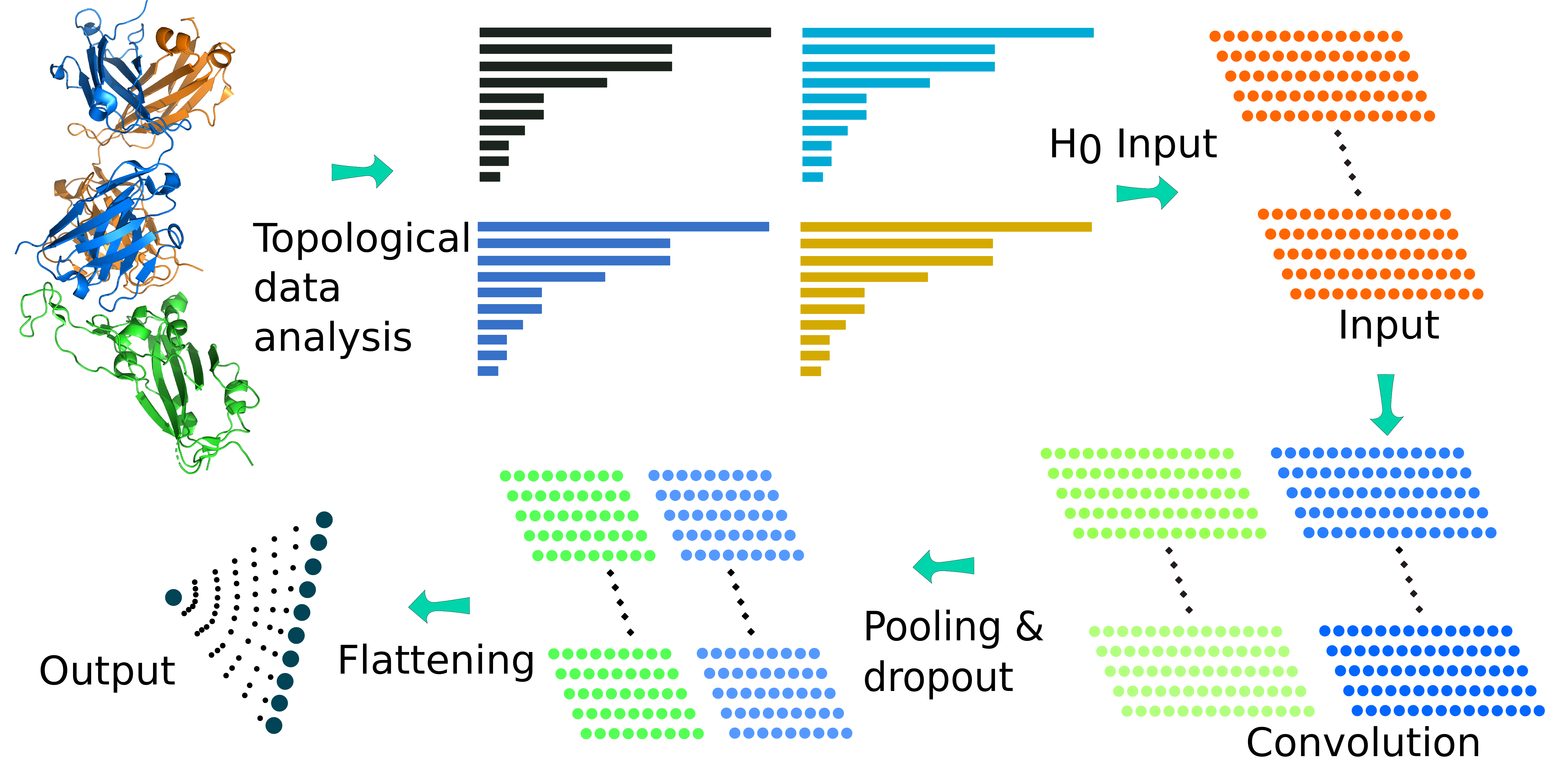}
	\caption{The CNN model from Ref. \cite{chen2020review}.}
	\label{fig:cnn}
\end{figure}

The CNN \cite{krizhevsky2017imagenet} is a specialized type of neural network model originally designed to analyze visual imagery, but it can be also applied to lots of areas. CNN is a superstar of neural networks, since the first successful CNN was developed in the late 1990s, it has achieved much success in image and video recognition, natural language processing, etc., even in biophysics areas such as protein structure prediction and protein-ligand binding \cite{cang2017topologynet,cang2018representability}. The core of CNN is the convolutional layer where its name comes from (see Figure \ref{fig:cnn}). In the context of CNN, convolution is a linear operation that involves the multiplication of a set of weights with the input. This multiplication is always called a filter or a kernel. Using a filter smaller than the input is intentional as it allows the same filter to be multiplied by the input array multiple times at different points on the input. Specifically, the filter is applied systematically to each overlapping part or filter-sized patch of the input data, left to right, top to bottom, which allows the filter an opportunity to discover that feature anywhere in the input. 

\begin{figure}[ht!]
	\centering
	\includegraphics[width=0.9\textwidth]{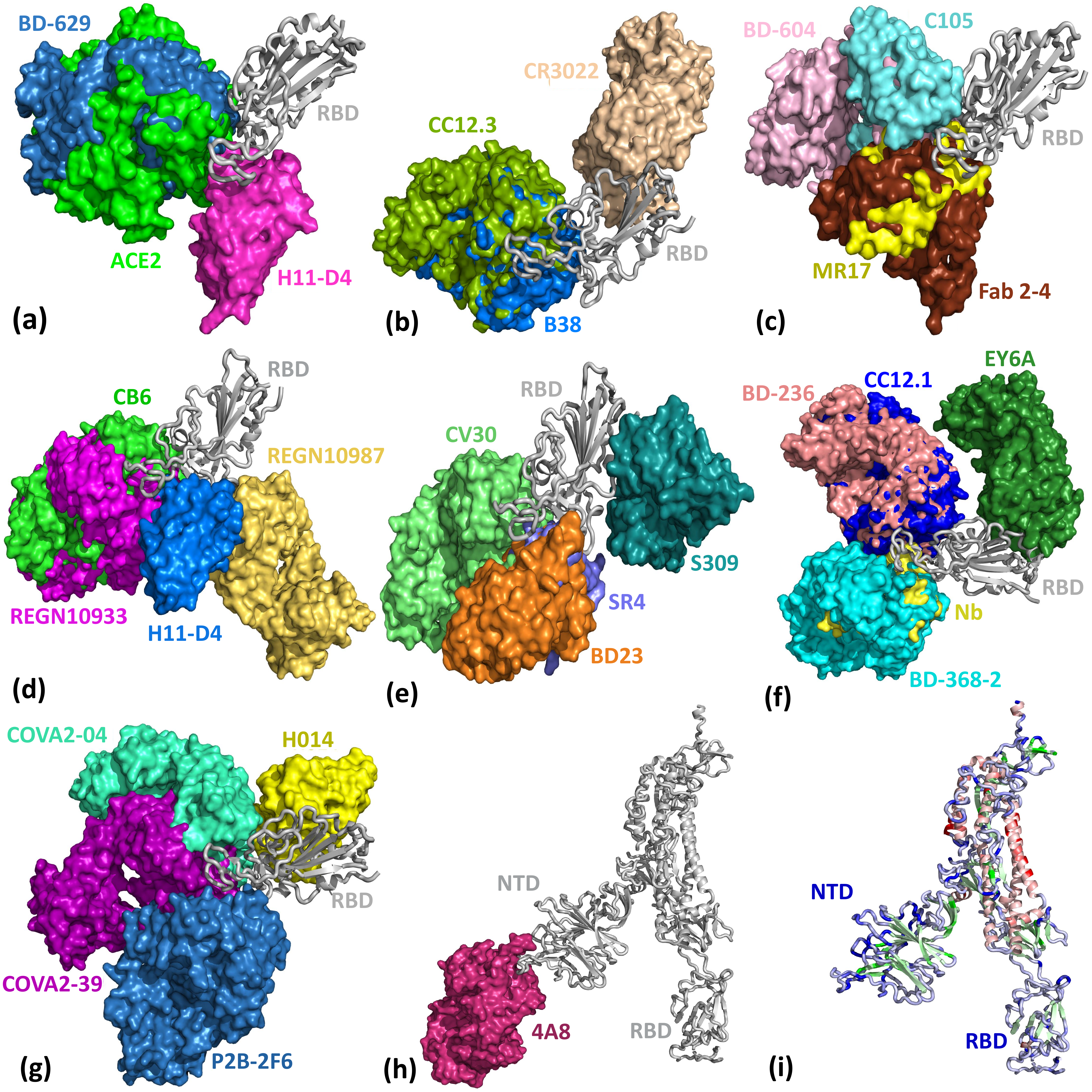}
	\caption{The 3D alignment of the available unique 3D structures of SARS-CoV-2 S protein RBD in binding complexes with 27 antibodies as well as ACE2 in Ref. \cite{chen2020review}.}
	\label{fig:sars2-antibody}
\end{figure}

In the antibody and vaccine research, Chen et al. used  algebraic-topology based features to build a CNN-GBT hybrid model for predicting mutation-induced binding affinity change, investigating the impact of S protein mutations on the ACE2 \cite{chen2020mutations, wang2020characterizing} 27 antibodies (see Figure \ref{fig:sars2-antibody})  \cite{chen2020review}, as well as suggesting some highly risky ones to vaccine design \cite{chen2020prediction}. In the inhibitor research, Nguyen et al. \cite{nguyen2020unveiling} used algebraic-topology based features and CNN models to predict the potency of ligands from the 137 crystal structures of the main protease.

Critical Assessment of Protein Structure Prediction (CASP) also proved a domain for application of powerful CNN methods in protein structure prediction. For example, CNN-based AlphaFold by Google Deepmind obtained the highest accuracy in CASP13 \cite{senior2020improved,wei2019protein}. During this epidemic, Deepmind applied the AlphaFold to predict the 3D structures of SARS-CoV-2 M protein, PLpro, NSP2, NSP4, NSP6 \cite{deepmind_predict}. Meanwhile, the CNN-based C-I-TASSER algorithm developed by the Zhang Lab was implemented to predict as many as 24 SARS-CoV-2 proteins \cite{zhang_predict}.

\subsubsection{RNN, LSTM, and GRU} 
The recurrent neural network (RNN) is a class of artificial neural network where connections between nodes form a directed graph along a temporal sequence \cite{rumelhart1986learning}, which allows it to exhibit temporal dynamic behavior. Derived from the feed-forward neural network, RNN can use its internal state (memory) to process variable length sequences of inputs. RNN was originally designed for language processing tasks, but it can also be applied to other circumstances.

\begin{figure}[ht!]
    \centering
	\includegraphics[width=1\textwidth]{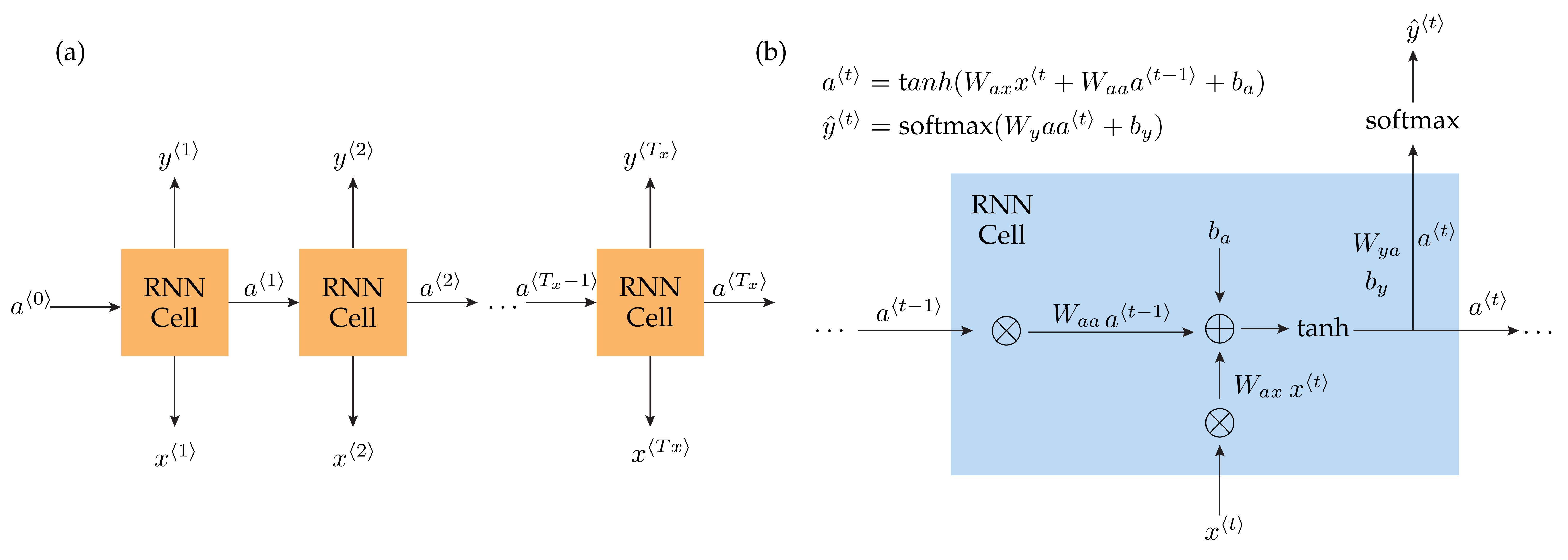}
	\caption{The workflow of RNN (See \autoref{fig:RNN}). $\langle t \rangle$ represents the an object at time-step $t$. $x^{\langle t \rangle}$, $y^{\langle t \rangle}$, and $a^{\langle t \rangle}$ denote the input $x$, output $y$, and activation at time-step $t$, respectively. $\hat{y}^{\langle t \rangle}$ represents the prediction at time-step $t$. (a) The forward propagation of RNN. (b) The operations for a single time-step of a RNN cell. $W$ and $b$ represent weights and bias at a specific state. }
	\label{fig:RNN}
\end{figure}

\begin{figure}[ht!]
    \centering
	\includegraphics[width=1\textwidth]{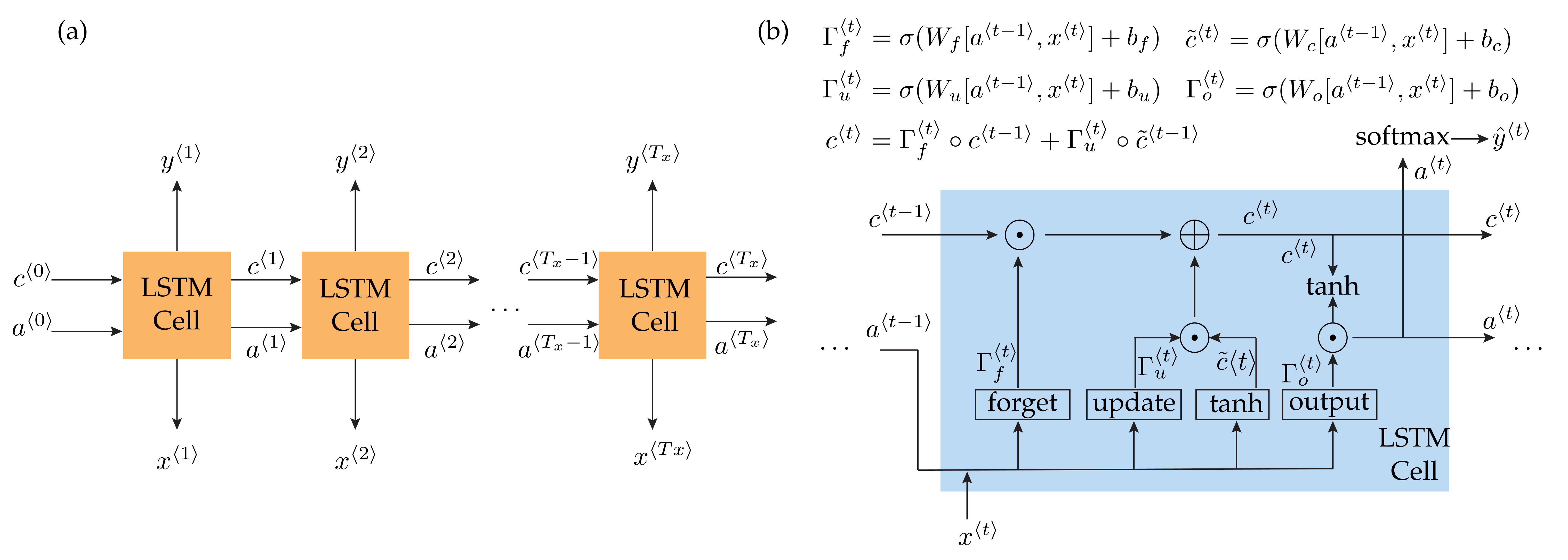}
	\caption{The workflow of LSTM. $\langle t \rangle$ represents an object at time-step $t$. $x^{\langle t \rangle}$, $y^{\langle t \rangle}$, $a^{\langle t \rangle}$, and $c^{\langle t \rangle}$ denote the input $x$, output $y$, activation, and cell state at time-step $t$. respectively. $\hat{y}^{\langle t \rangle}$ represents the prediction at time-step $t$. (a) The forward propagation of LSTM. (b) The operations for a single time-step of a LSTM cell. $\Gamma_{f}^{\langle t \rangle}$, $\Gamma_{i}^{\langle t \rangle}$, $\Gamma_{o}^{\langle t \rangle}$, $c^{\langle t \rangle}$, and $\tilde{c}^{\langle t \rangle}$ denote the forget gate state, update gate state, output gate state, cell state, and previous cell state at time-step $t$. $W$ and $b$ represent weights and bias at a specific state, and $\sigma$ is the activation function such as tanh.}
	\label{fig:LSTM}
\end{figure}

The long short-term memory (LSTM) shown in \autoref{fig:LSTM}) and gated recurrent unit (GRU) are two popular variants of RNN. LSTM \cite{hochreiter1997long} is designed to avoid the vanishing gradient problem. LSTM is normally augmented by recurrent gates called ``forget gates'', and so errors can flow backwards through unlimited numbers of virtual layers unfolded in space. GRU is a gating mechanism in recurrent neural networks introduced in 2014 \cite{cho2014learning}. Its performance was found to be similar to that of LSTM. However, as it lacks an output gate, its parameters are fewer than LSTM, so it is easier and faster to train.

\begin{figure}[ht!]
    \centering
	\includegraphics[width=0.8\textwidth]{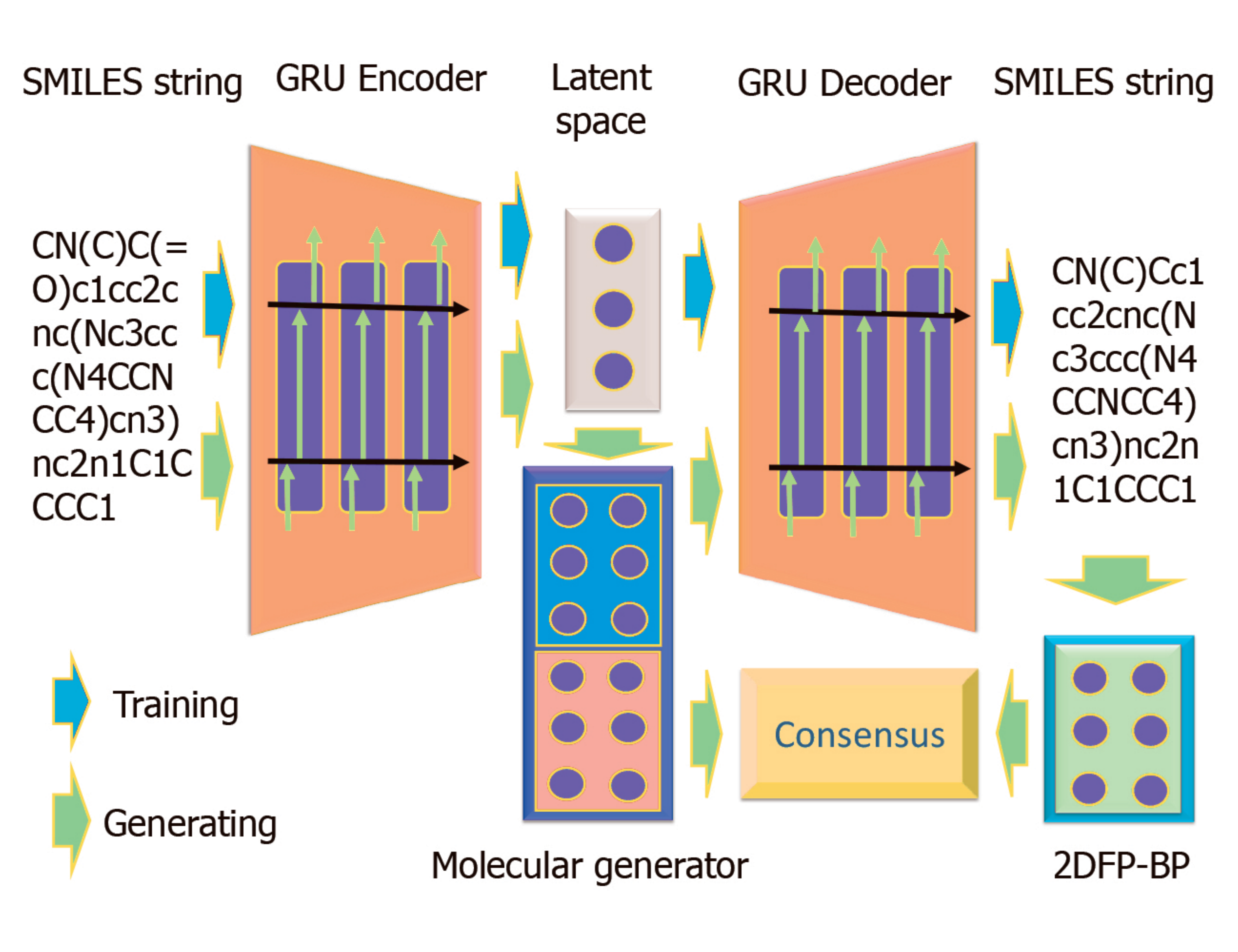}
	\caption{Illustration of the generative network complex \cite{gao2020machine}. SMILES strings are encoded into latent vector space through a gated recurrent neural network (GRU)-based encoder.}
	\label{fig:GRU}
\end{figure}

Their application to SARS-CoV-2 includes the following: Hofmarcher et al.\cite{hofmarcher2020large} utilized ``ChemAI'' to screen and rank around one billion molecules from the ZINC database for favourable effects against CoV-2; in more detail, the network is of the type Smiles LSTM \cite{mayr2018large}. Bung et al. \cite{bung2020novo} employed RNN-based generative and predictive models for de novo design of new small molecules capable of inhibiting the main protease of SARS-CoV-2. The generative network complex\cite{gao2020generative} is a GRU-based generative model. Gao et al. \cite{gao2020machine} used this AI technology to generate some potential main protease inhibitors as illustrated in \autoref{fig:GRU}.

\subsubsection{Machine learning and viral mutations}
In Refs. \cite{sardar2020integrative,gussow2020genomic,sardar2020comparative,dehury2020effect,sagkan2020structural,beg2020computational}, authors used a variety of different machine learning approaches in order to predict the effect of given mutations on disease stability or severity.
Others utilized machine learning to aid in the phylogenic analysis and geographic modeling \cite{villmann2020analysis,kandpal2020identification,nagpal2020if,ramazzotti2020quantification}.
Alam et al. \cite{alam2020functional} used machine learning approaches to predict Gene Ontology and extrapolate on the features most important in the ontology prediction. Wang et al. use $K$-means clustering to cluster the SARS-CoV-2 sequences into different groups based on the single nucleotide polymorphisms (SNP) profiles \cite{wang2020decoding}. Moreover, Huzumi et al. compared the performances of various dimensional reduction algorithms such as PCA, t-SNE, and UMAP, which aims to find a best-suited, stable, and efficient technique to improve the clustering accuracy of SARS-CoV-2 sequences \cite{hozumi2020umap}.

\subsection{Mathematical approaches}

\subsubsection{Network analysis}
Networks represent interactions between pairs of units in biomolecular or other systems, such as atomic interactions, protein-protein interactions, drug-target interactions, disease-protein associations, and drug-disease treatments. 
The unique characteristics of these networks can be quantified for descriptions and comparisons of different networks. If considering protein-protein interactions as networks, each descriptor evaluates the network properties and measures how proteins connect.  

{\bf Network heterogeneity.} The network heterogeneity is an index that evaluates the heterogeneity of a network on different distributions \cite{estrada2010quantifying}. The heterogeneity can reflect the distribution of a network on different impacts or compare the heterogeneity of two networks, which is defined as: 
\begin{equation}
\rho=\sum_{i=1}^{N_e}\sum_{j=i+1}^{N_e}(k_i^{-1/2}-k_j^{-1/2})^2,
\end{equation}
where $N_e$ is the number of edges of the network, and  $k_i$ is the degree of the $i$-th node, which is the number of connections that the $i$-th node has with other nodes.

{\bf Edge density.} The edge density is defined as 
\begin{equation}
D=\frac{2N_e}{N_v(N_v-1)},
\end{equation}
where $N_e$ is the number of edges and $N_v$ is the number of vertices. The edge density is also called the average degree centrality. For a complete network in which each pair of network vertices is connected, the edge density is equal to one. A non-complete network has an edge density smaller than one.  

{\bf Path length.} The characteristic path length studied the typical separation between two vertices in the network. It was used to study infectious diseases spread in so-called ''small-world'' networks \cite{watts1998collective}. The shortest path distance $d(i,j)$ was defined as the shortest path between the corresponding pairs of vertex $i$ and $j$.  The average path length was defined as: 

\begin{equation}
\label{eq:ave_path_length}
\langle L\rangle = \frac{1}{N_v(N_v-1)} \sum_{i=1}^{N_v}\sum_{j=i+1}^{N_v} d(i,j).
\end{equation}

For instance, in protein-protein interactions, the path length between two atoms reflects how ACE2 or antibodies connect to RBD.

{\bf Betweenness centrality} The concept of betweenness centrality illustrates communications in a network \cite{freeman1978centrality}. The betweenness centrality of a vertex $v_k$ is given as:
 
\begin{equation}
C_b(v_k) = \sum_{i=1}^{N_v}\sum_{j=i+1}^{N_v} g_{ij}(v_k)/g_{ij},
\end{equation}

and the average betweenness centrality is given as:
\begin{equation}
\langle C_b \rangle = \frac{1}{N_v}\sum_{k=1}^{N_v} C_b(v_k),
\end{equation}
where $g_{ij}(v_k)$ is defined as the number of geodesics linking vertex $v_i$ and $v_j$ that passes $v_k$, and $g_{ij}$ considers all the paths between $v_i$ and $v_j$.

{\bf Eigencentrality.} The eigenvector centrality is the elements of the eigenvector $V_\text{max}$ with respect to the largest eigenvalue of the adjacency matrix $A$ of networks \cite{bonacich1987power}. It describes the probability of starting at and returning to the same point for infinite length walks. Thus, the average eigenvector centrality is, 
\begin{equation}
\langle C_e \rangle = \frac{1}{N_v}\sum_{i=1}^{N_v}e_i,
\end{equation}
where $e_i$ are elements of $V_\text{max}$, which stands for the average impact spread of vertices beyond its neighborhood for an infinite walk.

{\bf Subgraph centrality.} The following descriptors are built on the exponential of the adjacency matrix, $E = e^A$. The average subgraph centrality is defined as 
\begin{equation}
\langle C_s \rangle = \frac{1}{N_v}\sum_{k=1}^{N_v} E(k,k),
\end{equation} 
which indicates the vertex participating in all subgraphs of the graphs \cite{estrada2005subgraph, estrada2020topological}. Subgraph centrality is the summation of weighted closed walks of all lengths starting and ending at the same node. The long path length has a small contribution to the subgraph  centrality. 

{\bf Communicability.} Finally, the last two descriptors are average communicability, given as
\begin{equation}
\langle M \rangle = \frac{2}{N_v(N_v-1)}\sum_{i=1}^{N_v}\sum_{j=i+1}^{N_v} E(i,j),
\end{equation}
and average communicability angle, given as
\begin{equation}
\langle \Theta \rangle = \frac{2}{N_v(N_v-1)}\sum_{i=1}^{N_v}\sum_{j=i+1}^{N_v} \theta(i,j),
\end{equation}
where $\theta(i,j) = \arccos \Big( \frac{E(i,j)}{\sqrt{E(i,i),E(j,j)}} \Big)$. The average communicability measures how much two vertices can communicate by using all the possible paths in the network, where the shorter paths have more weight than the longer paths \cite{estrada2008communicability}. The average communicability angle evaluates the efficiency of two vertices passing impacts to each other in the network with all possible paths \cite{estrada2016communicability,estrada2020topological}.

\paragraph{Network based biomolecular structure analysis.}

\begin{figure}[ht!]
	\centering
	\includegraphics[width=0.9\textwidth]{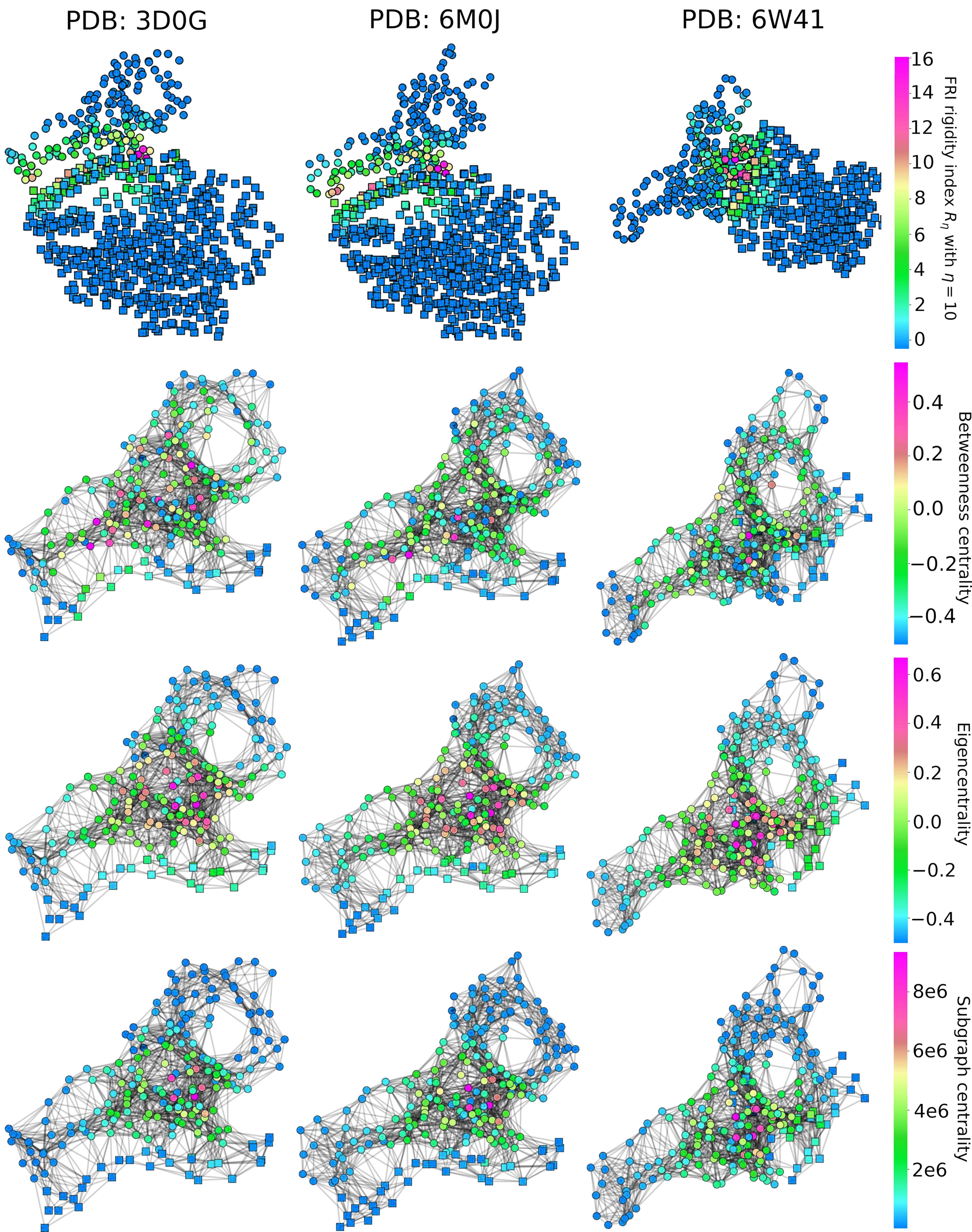}
	\caption{C$\alpha$ network analysis of three antibody-antigen complexes. Here, circle markers represent antigen (spike protein RBD), and cube markers represent antibody or ACE2.  The PDB ID of the three antibody-antigen complexes are 2D0G, 6M0J, and 6W41. The rows represent the FRI rigidity index, betweenness centrality, and subgraph centrality \cite{chen2020review}.}
	\label{fig:combine1}
\end{figure}

Using networks to analyze the structural similarities is important to drug repurposing and functional mechanisms. Estrada applied the aforementioned network indices to analyze the interaction networks between the SARS-CoV-2 main protease and various inhibitors \cite{estrada2020topological}. Chen et al.\cite{chen2020mutations} applied a similar strategy to predict binding affinity changes induced by mutations. A variety of studies using the network indexes on protein residue/atom networks followed the same path \cite{wang2020decoding,chen2020review,wang2020characterizing,griffin2020sars,diaz2020sars,karathanou2020graph}. Moreover, Chen et al. employed the network analysis of antibody-antigen complexes on C$\alpha$ atoms in \cite{chen2020review} as illustrated in \autoref{fig:combine1}.

\paragraph{Network-based drug repurposing.}

Drug repurposing methods require comparing the unique features, such as chemical components, or proteomic, metabolomic, or transcriptomic data, of a drug candidate with existing drugs, diseases, or clinical phenotypes. One idea of drug repurposing is that one drug currently working for one disease may also work for other diseases if these diseases share some similar protein targets \cite{chiang2009systematic,keiser2007relating}. Thus, integrated disease-human-drug interactions could form a network with nodes as drugs, diseases, and proteins, weighted edges referring to interactions between them, e.g., the number of drugs with a certain treatment. Novel drug usage can be discovered based on shared treatment profiles from any disease connections, and the weight between two disease connections determines the possibility of repurposing drugs \cite{chiang2009systematic}. Common pathways between different viruses or diseases are already identified on a large scale\cite{smith2012identification}. Meanwhile, another way to define drug repurposing is based on the structural similarities of two drugs: two drugs may work on the same therapeutic target if the two drugs have similar structures. 
 
Network-based drug repurposing studies have already been performed on SARS-CoV-2. Gordon et al.\cite{gordon2020sars} investigated the protein-protein interaction (PPI) network between SARS-CoV-2 and humans, and identified 332 high-confidence PPIs between SARS-CoV-2 and human proteins; based on that and considering the features of drugs such as drug status, drug selectivity, drug availability, and the statistical calculations of the protein interactions, they screened drugs targeting the human proteins in the SARS-CoV-2 human interactome. Consequently, 29 drugs already approved by USDA, 12 investigational new drugs, and 28 preclinical compounds were identified according to their studies. Zhou et al.\cite{zhou2020network} studied the antiviral drug repurposing methodology targeting SARS-CoV-2; a systematic pharmacology-based network medicine platform was implemented to identify the interplay between the virus-host interactome and drug targets where they investigated the network proximity of SARS-CoV-2 host and drug targets interaction. Based on that, they reported three potential drug combinations. In the study by Sadegh et al.\cite{sadegh2020exploring}, CoVex was developed for SARS-CoV-2 host interactome exploration and drug (target) identification, which also explored the virus-host interactome and potential drug target; the network was constructed based on PPIs, drug-protein-protein interactions, etc. for repurposing drug candidates. Additionally, Srinivasan et al.\cite{srinivasan2020structural} developed a network of the comprehensive structural gene and interactome of SARS-CoV-2. Messina et al.\cite{messina2020covid} investigated host-pathogen interaction model through the PPI network.

\subsubsection{Flexibility-rigidity index (FRI)}

The flexibility-rigidityindex (FRI) is a geometric graph-based method that utilizes weighted graph edges to molecular interactions\cite{nguyen2016generalized, xia2013multiscale}. Multiscale FRI \cite{opron2015communication},  colored (i.e., element-specific) FRI \cite{bramer2018multiscale} and their algebraic graph counterpart \cite{xia2015multiscale} have been also proposed.   The atomic rigidity index at position ${\bf r}_i$ is defined as a summation of all the weighted edges around it:  
\begin{equation}
\label{eq:rigidity}
\nu_i (\eta) =  \sum_{j=1}^{N_{c'}}  e^{-\big(\frac{\|\textbf{r}_i-\textbf{r}_j\|}{\eta}\big)^2},
\end{equation}
where $\textbf{r}_j$ are atom positions and $N_{c'}$ is the number of atoms in the neighborhood of $\textbf{r}_i$. Here, $\eta$ is a characteristic scale. Element-specific rigidity \cite{bramer2018multiscale} and molecular rigidity \cite{xia2013multiscale} can be obtained by appropriate collection of atomic rigidity indices. 

FRI has been applied for 	protein and  nucleic acid flexibility and fluctuation analysis \cite{xia2013multiscale} and protein-ligand binding affinity prediction \cite{nguyen2017rigidity}. Protein-protein interactions, such as the elasticity between antibody and antigen, especially long-range impacts, are studied by calculating the FRI index of the network consisting of $\text{C}_\alpha$ atoms.  FRI rigidity index is an important feature for machine learning models to predict the binding affinity changes on mutations \cite{wang2020topology} and the protein folding energy changes on mutations \cite{cang2017analysis}. 

\begin{figure}[ht!]
	\centering
	\includegraphics[width=0.8\textwidth]{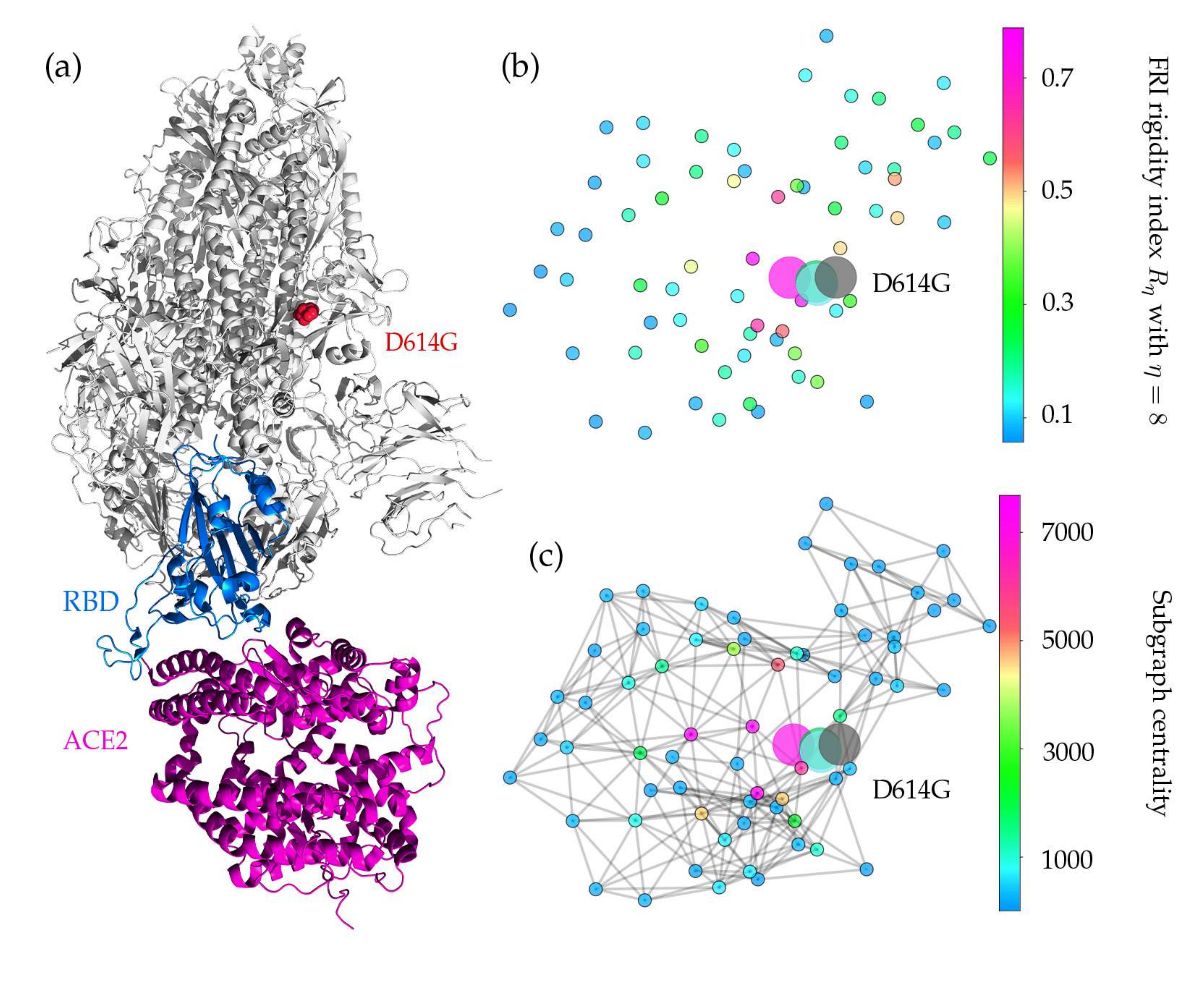}
	\caption{The FRI rigidity index of the SARS-CoV-2 S protein. (a) Illustration of S protein and ACE2 interaction. The RBD is displayed in blue, the ACE2 is given in pink, and mutation D614G is highlighted in red. (b) The difference of FRI rigidity index of the S protein between the network with wild type and the network with mutant type. (c) The difference of the subgraph centrality between the network with wild type and the network with mutant type in Ref. \cite{wang2020characterizing}.}
	\label{fig:s-pro-sta-change}
\end{figure}

Some studies already applied the machine learning models based on FRI index to study the SARS-CoV-2 proteins: combining with network analysis, Wang et al. \cite{wang2020decoding,wang2020characterizing} calculated FRI rigidity index and investigated the folding stability changes of the S protein (see Figure \ref{fig:s-pro-sta-change}, the definition of subgraph centrality is in the next section) and other proteins caused by mutations. FRI-based binding affinity change between the S protein and human ACE2 due to mutations was also calculated by Chen et al. \cite{chen2020mutations,chen2020review}.

\subsubsection{Topological data analysis (TDA)}

Recent years have witnessed a rapid increase in topological data analysis (TDA) and its applications to a wide variety of scientific and engineering problems \cite{kaczynski2006computational,carlsson2009topology}. The main workhorse of TDA is persistent homology \cite{zomorodian2005computing,edelsbrunner2008persistent}, a new branch of algebraic topology. This approach has been applied to characterize biomolecular systems \cite{xia2014persistent,kovacev2016using,xia2015multiresolution}. 
More powerful methods that provide simultaneous  topological persistence and spectral analysis have been proposed \cite{wang2020persistent,chen2019evolutionary,meng2020persistent}.
In algebraic topology, molecular atoms can be treated as $k\!+\!1$ affinely independent points $v_0$, $v_1$, $...$, $v_k$. A simplicial complex, the essential building block, is a finite collection of sets of points $K=\{\sigma_i\}$, and $\sigma_i$ is a linear combination of these points in $\mathbb{R}^n$ ($n\ge k$).
A simplicial complex $K$ is valid if a face $\tau$ of a $k$-simplex $\sigma_i$ of $K$ is also in $K$, such that $\tau \subseteq \sigma_i$ and $\sigma_i\in K$ imply $\tau \in K$ and the non-empty intersection of any two simplices is a face for both. Given a simplicial complex $K$, a $k$-chain is a finite formal sum of $k$-simplices; that is, $\sum_{i}\alpha_i\sigma^k_i$. The set of all $k$-chains of the simplicial complex $K$ equipped with an algebraic field (typically, $\mathbb{Z}_2$) forms an abelian group $C_k(K, \mathbb{Z}_2)$. A boundary operator $\partial_k: C_k\!\rightarrow\!C_{k-1}$ for a $k$-simplex $\sigma^k=\{v_0,v_1,\cdots,v_k\}$ are homomorphisms defined as $\partial_k \sigma^k = \sum^{k}_{i=0} (-1)^i\{ v_0, v_1, \cdots, \hat{v_i}, \cdots, v_k \},$where $\{v_0, v_1, \cdots ,\hat{v_i}, \cdots, v_k\}$ is a $(k\!-\!1)$-simplex excluding $v_i$ from the vertex set. Consequently, an important property of boundary operator, $\partial_{k-1}\partial_k= \emptyset$, follows from that boundaries are boundaryless. Moreover, the $k$th cycle group $Z_k={\rm ker} \partial_k=\{c\in C_k \mid \partial_k c=\emptyset\}$ is defined to be the kernel of $\partial_k$, whose elements are called $k$-cycles; and the $k$th boundary group is the image of $\partial_{k+1}$, denoted as $B_k={\rm im} ~\partial_{k+1}= \{ \partial_{k+1} c \mid c\in C_{k+1}\}$. The algebraic construction to connect a sequence of complexes by boundary maps is called a chain complex,
\[
\cdots \stackrel{\partial_{i+1}}\longrightarrow C_i(X) \stackrel{\partial_i}\longrightarrow C_{i-1}(X) \stackrel{\partial_{i-1}}\longrightarrow \cdots \stackrel{\partial_2} \longrightarrow C_{1}(X) \stackrel{\partial_{1}}\longrightarrow C_0(X) \stackrel{\partial_0} \longrightarrow 0,
\]
and the $k$th homology group is the quotient group defined by
$H_k = Z_k / B_k$.
The key property of boundary operators implies $B_k\subseteq Z_k \subseteq C_k$. The Betti numbers are defined by the ranks of $k$th homology group $H_k$ which counts $k$-dimensional holes. Especially, $\beta_0\!=\!{\rm rank}(H_0)$ reflects the number of connected components, $\beta_1\!=\!{\rm rank}(H_1)$ reflects the number of loops, and $\beta_2\!=\!{\rm rank}(H_2)$ reveals the number of voids or cavities. Together, the set of Betti numbers $\{\beta_0,\beta_1,\beta_2,\cdots \}$ indicates the intrinsic topological property of a system. 

Persistent homology is devised to track the multiscale topological information over different scales along a filtration \cite{edelsbrunner2000topological}. A filtration of a topology space $K$ is a nested sequence of subspaces $\{K^t\}_{t=0,...,m}$ of $K$ such that $
\emptyset = K^0 \subseteq K^1 \subseteq K^2 \subseteq \cdots \subseteq K^m = K$. Moreover, on this complex sequence, we obtain a sequence of chain complexes by homomorphisms: $C_*(K^0) \to C_*(K^1) \to \cdots \to C_*(K^m)$ and a homology sequence: $H_*(K^0) \to H_*(K^1) \to \cdots \to H_*(K^m)$, correspondingly. The $p$-persistent $k$th homology group of $K^t$ is defined as $H_k^{t,p} = Z^t_k/(B_k^{t+p}\bigcap Z^t_k)$, where $B_k^{t+p} = {\rm im} \partial_{k+1}(K^{t+p})$. Intuitively, this homology group records the homology classes of $K^t$ that are persistent at least until $K^{t+p}$. Under the filtration process, the persistent homology barcodes can be generated. 
Then, the feature vectors can be constructed from these sets of intervals for machine learning models \cite{cang2018representability}. 
 
Since the first integration of persistent homology and machine learning \cite{cang2015topological}, topology-based approaches have found much success in biomolecular modeling and prediction \cite{nguyen2020review,cang2017topologynet,cang2018integration,cang2018representability}. Combined with large datasets and machine learning algorithms, TDA is a powerful tool in predicting biomolecular properties such as protein-ligand binding affinity \cite{cang2017topologynet,cang2018representability} and drug discovery \cite{nguyen2019mathematical}. Features are generated by constructing complexes on protein atoms. According to the biomolecular properties, complexes are constructed as an atomic-specific strategy or bipartition graph. For instance, when studying the protein folding energy of the ACE2 and SARS-CoV-2 S protein, one can use element-specific and/or site-specific persistent homology to simplify the structural complexity of protein structure and encode vital biological information into topological invariants \cite{cang2018representability,wang2020decoding}. Wang et al. \cite{wang2020decoding} applied topology features on the studying of protein folding studies on the energy changes on mutations of SARS-CoV-2 NSP6 protein. Moreover, in the complex forms in a bipartite graph, the features of protein-protein interaction can be studied where the atoms of antibody and antigen consist of two disjoint and independent sets. Chen et al. \cite{chen2020review} used this idea to predict the binding free energy changes on mutations of the protein-protein interactions between the S protein and antibodies. Nguyen et al. \cite{nguyen2020unveiling} studied the potency and molecular mechanism of main protease inhibition from 137 crystal structures by integrating mathematics, deep learning methods, and applied persistent homology. Topological data analysis is not only applied to study protein-protein interactions. Chen et al.\cite{chen2020mutations} studied the mutations that strengthened SARS-CoV-2 infectivity where persistent homology played a key role in analyzing the interactions between the S protein and human ACE2.

\section{Discussion}

Since the outbreak of the COVID-19 epidemics in December 2019, enormous  effort has been devoted to the scientific research relating to SARS-CoV-2, leading to significant breakthroughs, such as the development of vaccines and experimental determination of protein structures. However, effective drugs and therapies are still absent. Notably, thanks to the rapid development of high-performance computers, biophysical methods, and AI algorithms in recent decades,  plenty of theoretical and computational studies were carried out  against SARS-CoV-2. Theoretical and computational studies are significant to combat urgent epidemics such as this COVID-19 because they lead to important understandings faster and cheaper \cite{estrada2020covid}. This review strives  to summarize the existing SARS-CoV-2 theoretical \& computational works and enlighten future ones. 

Most of the researches covered by this review are about repurposing current drugs or inhibitors to target SARS-CoV-2 because drug development has been one of the most urgent issues in combating COVID-19. A variety of drug-repurposing approaches has been applied, from molecular docking and MD simulation to machine learning \& deep learning, as summarized below. (1) The most straightforward approach is molecular docking, which provides both binding poses and corresponding scores. (2) In many studies, docking poses were further optimized by MD simulations, and these optimized poses were rescored by docking programs. (3) More accurate binding free energies can be achieved by MD simulation-based or even QM-based calculations, such as MM/PB(GB)SA, free energy perturbation, metadynamics, QM/MM, and DFT. (4) Other than traditional molecular docking and MD simulations, thanks to the development of AI, machine learning \& deep learning technologies such as GBDT, DNN, and CNN open a new trail to discover SARS-CoV-2 drugs. With existing drugs as training sets, machine learning \& deep learning could predict the potency of a large number of potential SARS-CoV-2 inhibitors in a short time \cite{gao2020repositioning}. 3D models also provided binding poses \cite{nguyen2020unveiling}.  (5) Network-based drug repurposing was also performed to hunt SARS-CoV-2 drugs. The basic idea is that one drug currently curing one disease may also work for other diseases if sharing some similar protein targets \cite{chiang2009systematic,keiser2007relating}. Thus, integrated disease-human-drug interactions form a network connecting drugs, diseases, and targets. Novel drug usage can be discovered based on shared treatment profiles from disease connections. (6) Traditional QSAR approaches were implemented in many calculations for drug discovery.

The magic of AIs is not limited to the repurposing of  existing drugs or inhibitors. They also have the potential to create new drugs \cite{winter2019efficient,gao2020generative} to combat COVID-19. For example, Bung et al. \cite{bung2020novo} employed RNN-based networks and Gao et al. \cite{gao2020machine} used GRU-based generative networks to design new potential main protease inhibitors.  

Since SARS-CoV-2 is an RNA virus, it is quite vulnerable to gene mutation. New variants have already been spotted in some places in the world. Mutations are potentially harmful to the efficacy of vaccines, drugs, etc. Mutation studies collected in this review include MD based and deep-learning based approaches. MD-based mutation studies mainly investigated mutation-induced conformation and binding affinity changes such as that between the S protein and human ACE2. The deep-learning models were designated to predict mutation-induced binding affinity changes, applied to reveal the mutation impacts on the ACE2 and/or antibody binding with the S protein. These impacts are significant to SARS-CoV-2 infectivity \cite{chen2020mutations} and antibody therapies \cite{chen2020review}.    

Some computational investigations were devoted to vaccine design. MD simulations were employed to simulate vaccine-related immune reaction, such as the binding of the MCH II-epitope complexes \cite{lizbeth2020immunoinformatics}. Deep learning was also applied to study mutation impacts on vaccine efficiency. Based on predicted mutation-induced binding affinity changes by their CNN model and frequency of mutations, they suggested some hazardous ones \cite{chen2020prediction}.

SARS-CoV-2 protein structure prediction also plays an important role, especially at the early stage of the epidemics when experimental structures were largely unavailable. At this point, besides traditional homology modeling, a more fancy solution is the high-level deep learning based models such as Alphafold \cite{deepmind_predict} and C-I-TASSER \cite{zhang_predict}, both making use of deep CNN.

\section{Conclusion and Perspective}

Since the first COVID-19 case was reported in December 2019, this pandemic has gone out of control worldwide. Although scientists around the world have already placed a top priority on SARS-CoV-2 related researches, there is still no effective and specific anti-virus therapies at this point. Moreover, despite the exciting progress on vaccine development, the reasons that caused the side effects, such as allergy reactions to COVID-19 vaccines, are unknown. Furthermore, whether the newly emergent variants of SARS-CoV-2 could make the virus more transmissible, infectious and deadly are also unclear, indicating that our understanding of the infectivity, transmission, and evolution of SARS-CoV-2 is still quite poor. Therefore, providing a literature review for the study of the molecular modeling, simulation, and prediction of SARS-CoV-2 is needed. Since the related literature is huge and varies in quality, we cannot collect all of existing literature for the topic. However, we try to put forward a methodology-centered review where we emphasize the methods used in various studies. To this end,  we gather the existing theoretical and computational biophysics studies of SARS-CoV-2 with respect to the aspects such as molecular modeling, machine learning \& deep learning, and mathematical approaches, aiming to provide a comprehensive, systematic, and indispensable component for the understudying of the molecular mechanism of SARS-CoV-2. Our review provides a methodology-centered  description of the status on molecular model, simulation, and prediction of SARS-CoV-2. 


Although the U.S. Food and Drug Administration (FDA) has approved the emergency use of vaccines from Pfizer and Moderna in December 2020, the vaccination rate is still quite low. Even with the promising news of the vaccines, COVID-19 as a global health crisis may still last for years before it is fully stopped globally. 

The research on SARS-CoV-2 will  also last for many years. It will take researchers many more years to fully understand the molecular mechanism of coronaviruses, such as RNA proofreading, virus-host cell interactions, antibody-antigen interactions, protein-protein interactions, protein-drug interactions, and viral regulation of host cell functions. Even if we could control the transmission of SARS-CoV-2 one day in the near future, newly emergent conronaviruses may still cause similar pandemic outbreaks. Therefore, the conronavirus studies will continue even after the current pandemic is fully under control. 

Currently, epidemiologists, virologists, biologists, medical scientists, pharmacists,  pharmacologists, chemists, biophysicists, mathematicians, computer scientists, and many others are called to the investigation of various aspects of COVID-19 and SARS-CoV-2. This trend of joint effort on  COVID-19 investigations will continue and be kept the present pandemic. 

The urgent need for molecular mechanistic understanding of SARS-CoV-2 and COVID-19 will further stimulate the development of computational biophysical, artificial intelligent, and advanced mathematical methods. The theoretical, computational, and mathematical communities will benefit from this endeavor against the pandemic.

Year 2020 has witnessed the birth of human mRNA vaccines for the first time --- a remarkable accomplishment in science and technology. Although there are more dark days ahead us, humanity will prevail in a post-COVID-19 world. Science will emerge stronger against all  pathogens and diseases in the future.

\section*{Acknowledgments}
This work was supported in part by NIH grant  GM126189, NSF Grants DMS-1721024,  DMS-1761320, and IIS1900473, NASA grant 80NSSC21M0023, Michigan Economic Development Corporation,  George Mason University award PD45722,  Bristol Myers Squibb, and Pfizer.
The authors thank The IBM TJ Watson Research Center, The COVID-19 High Performance Computing Consortium, NVIDIA, and MSU HPCC for computational assistance.


\end{document}